\documentclass[11pt,a4paper]{article}
\usepackage{amsmath}
\usepackage{amssymb}
\usepackage{amsfonts}
\usepackage{amsthm,amsxtra}
\usepackage{nicefrac}
\usepackage{float}
\usepackage[all,web]{xy}
\usepackage{epsfig,pstricks}
\usepackage{color}
\usepackage{array}
\usepackage{pifont}
\usepackage{rotating}
\usepackage{multirow}
\usepackage[vcentermath]{youngtab}

\definecolor{purple}{rgb}{1,0,1}
\definecolor{orange}{rgb}{1,.5,0}
\definecolor{yellow}{rgb}{1,1,0}
\definecolor{lightblue}{rgb}{.4,1,1}
\definecolor{lightlightblue}{rgb}{1,1,.8}
\definecolor{lightlightgreen}{rgb}{.85,1,.85}
\definecolor{lightlightred}{rgb}{1,.85,1}
\definecolor{midblue}{rgb}{.2,.5,.8}
\definecolor{darkgreen}{rgb}{.1,.6,.3}
\definecolor{grey}{rgb}{.2,.2,.2}
\def\be{\begin{equation}}
\def\ee{\end{equation}}
\def\bea{\begin{eqnarray}}
\def\eea{\end{eqnarray}}

\def\mybox#1{\mbox{$\displaystyle #1$}}
\def\half {\mbox{$\textstyle {1 \over 2}$}}
\def\Re{\mathop{\rm Re}}
\def\re{\mathop{\rm Re}}
\newcommand{\im}{\mbox{Im}}

\theoremstyle{plain}

\theoremstyle{definition}

\theoremstyle{remark}

\renewcommand{\baselinestretch}{1.1}
\renewcommand{\theequation}{\thesection.\arabic{equation}}

\newcommand{\ZZ}{\mathbb Z}
\newcommand{\calL}{\mathcal{L}}

\newcommand{\A}{\mathcal{A}}
\newcommand{\C}{\mathcal{C}}
\newcommand{\F}{\mathcal{F}}
\newcommand{\pv}{\diagup\hspace{-4mm}\int_{-\infty}^{\infty}}
\newcommand{\mn}{$(\m,\n)$}
\newcommand{\m}{\boldsymbol{m}}
\newcommand{\n}{\boldsymbol{n}}
\def\DD{\vec D}
\def\dd{\vec d}
\def\ge{\geqslant}
\def\le{\leqslant}
\newcommand{\td}{\tilde{d}}
\newcommand{\tD}{\tilde{D}}

\newcommand{\hy}{\hat{y}}
\newcommand{\hd}{\varepsilon}
\newcommand{\floor}[1]{\lfloor #1\rfloor}
\def\vec#1{\mbox {\boldmath $#1$}}
\def\svec#1{\mbox {\scriptsize\boldmath $#1$}}

\renewcommand{\title}[1]{{\Large\bf\mbox{}\\\medskip#1\bigskip\medskip\\}}
\renewcommand{\author}[1]{{\large #1\smallskip\\}}
\newcommand{\address}[1]{{\em #1\medskip\\}}

\numberwithin{equation}{section}
\def\qbin#1#2#3{{\Big[{#1\atop #2}\Big]}_{#3}}
\def\sqbin#1#2#3{\mbox{$[{#1\atop #2}]$}_{#3}}

\begin{document}
%\begin{titlepage}
%\vspace*{\fill}
\hfill LAPTH--1293/08

\renewcommand{\baselinestretch}{1.1}
\begin{center}
\title{Physical Combinatorics and Quasiparticles}
\medskip
\author{Giovanni Feverati}
\address{Laboratoire de Physique Th\'eorique LAPTH\\
CNRS, UMR 5108, associ\'e \`a l'Universit\'e de Savoie\\
9, Chemin de Bellevue, BP 110, 74941, Annecy-le-Vieux Cedex, France}
\medskip
\author{Paul A.~Pearce and Nicholas S.~Witte}
\address{
Department of Mathematics and Statistics\\
University of Melbourne, Parkville, Victoria 3010, Australia}
\bigskip\medskip

\begin{abstract}
\noindent  We consider the physical combinatorics of critical lattice models and their associated conformal field theories arising in the continuum scaling limit. As examples, we consider $A$-type unitary minimal models and the level-$1$ $s\ell(2)$ Wess-Zumino-Witten (WZW) model. The Hamiltonian of the WZW model is the $U_q(s\ell(2))$ invariant XXX quantum spin chain. For simplicity, we consider these theories only in their vacuum sectors on the strip. Combinatorially, fermionic particles are introduced as certain features of RSOS paths. They are composites of dual-particles and exhibit the properties of quasiparticles. The particles and dual-particles are identified, through a conjectured energy preserving bijection, with patterns of zeros of the eigenvalues of the fused transfer matrices in their analyticity strips. The associated $(m,n)$ systems arise as geometric packing constraints on the particles. The analyticity encoded in the patterns of zeros is the key to the analytic calculation of the excitation energies through the Thermodynamic Bethe Ansatz (TBA). As a by-product of our study, in the case of the WZW or XXX model, we find a relation between the location of the Bethe root strings and the location of the transfer matrix 2-strings.
\end{abstract}
\end{center}

%%%%%%%%%%%%%%%%%%%%%%%%%%%%%%%%%%%%%%%%%%%%%%%%%%%%
\newpage
\tableofcontents
\newpage

\section{Introduction}
\label{intro}
The phrase {\it physical combinatorics}~\cite{PhysComb,HatayamaETAl} was termed in 2000 to refer to a new body of results appearing at the intersection of solvable lattice models~\cite{BaxterBook82}, representation theory of Conformal Field Theory (CFT) and combinatorics. The term {\it quasiparticle} takes somewhat different meanings according to the field, but we use (fermionic) quasiparticles in the sense described by McCoy and his collaborators~\cite{KlassenEtAl93} in the early nineties. Quasiparticles have properties usually associated with particles but may not correspond to real physical particles. More specifically, {\it physical combinatorics} is the combinatorial realization of {\it finitized} conformal characters~\cite{DateEtAl,Melzer94, Berkovich94} (or more generally conformal partition functions built from such characters) as  generating functions of diagrammatic objects weighted by an energy statistic $E\in{\Bbb N}$ for system size $N$
\bea
\chi^{(N)}_{\Delta}(q)=q^{-c/24+\Delta} \sum_{object} q^{E(object)},\qquad 
\chi_{\Delta}(q)=\lim_{N\to\infty} \chi^{(N)}_{\Delta}(q)
\eea
where $c$ is the central charge. In the thermodynamic limit $N\to\infty$, this leads to {\it fermionic} expressions for the conformal characters $\chi_{\Delta}(q)$ as positive term series in the modular nome $q=e^{\pi i\tau}$ where $\tau$ is the modular parameter. In the limit $q\to 1$, the finitized character $\chi^{(N)}_{\Delta}(q)$ counts the number of states for system of size $N$.  The diagrammatic objects can be strings, lattice paths, rigged configurations, Young tableaux or other suitable combinatorial construct. Ultimately, these combinatorial objects should manifest the nature of the constituent fermionic quasiparticles.

Historically, physical combinatorics first appeared in the context of the {\it string hypothesis}~\cite{Takahashi} of Bethe's solution~\cite{bethe} of the Heisenberg (spin-$\half$ XXX) quantum spin chain. 
For each allowed eigenvalue of the commuting family $\vec T(u)$ of finite-size transfer matrices, Takahashi's string hypothesis posits the allowed patterns of zeros in the complex plane of the spectral parameter $u$. More correctly, the hypothesis specifies  the zero patterns of the eigenvalues of Baxter's auxiliary matrices 
$\vec Q(u)$ but the eigenvalues $T(u)$ are easily reconstructed from the eigenvalues $Q(u)$ through the Bethe ansatz~\cite{bethe}, or equivalently, Baxter's $T$-$Q$ relation~\cite{BaxterTQ}. 
In the intervening years, there have many works~\cite{Ki85,KR86,Ki87,KKR88,KRa88,KRb88, Dasmahapatra95,DasFoda98,Kirillov00,KirillovL01,KirillovEtAl02,Schilling02,Schilling07} developing the enumeration and study of Bethe states in terms of strings, rigged configurations and Young tableaux. Most notably, in the case of the XXZ spin chain with periodic boundary conditions, there is still some debate about whether the set of Bethe states given by the string hypothesis is complete~\cite{KirillovL97,KirillovL97_2,FabMcCoy}.

A major advance in the methods of physical combinatorics came with the advent of Baxter's one-dimensional configurational sums $X^{(N)}(\hat q)$ which derive from off-critical Corner Transfer Matrices (CTMs)~\cite{BaxterBook82}. The one-dimensional configurational sums are generating functions for one-dimensional lattice paths weighted by an energy statistic. In the prototypical example of the minimal models~\cite{BPZ} with central charges
\be
c=1-\frac{6}{L(L+1)}\label{mincentralcharge}
\ee 
the conformal characters exactly coincide~\cite{DateEtAl} with the $N\to\infty$ limit of the one-dimensional configurational sums of the Andrews-Baxter-Forrester (ABF) Restricted-Solid-On-Solid (RSOS) models~\cite{ABF84} in the off-critical regime III. 
This {\em correspondence principle} further extends~\cite{SaleurBauer,Melzer94,Berkovich94} to finitized conformal characters 
\be\label{cmx}
\chi^{(N)}(q)={\hat q}^{-c/24} X^{(N)}(\hat q)\Big|_{\hat q=q}
\ee
In applying physical combinatorics, some form of this {\em correspondence principle} is invariably invoked to introduce RSOS lattice paths through the one-dimensional configurational sums. 
However, substituting $q$ with $\hat q$ is surprising and even logically inconsistent. Specifically, the modular nome $q$ in the finitized characters is related to the aspect ratio and conformal geometry, and only makes sense at criticality, whereas the elliptic nome $\hat q$ in the one-dimensional configurational sums measures the departure from criticality in off-critical models. Reinterpreting $\hat q$ as $q$ does not substitute for understanding how the structure of RSOS paths actually arises in integrable {\em critical\/} lattice models. 
While the approach based on CTMs and the correspondence principle has proved very successful~\cite{PhysComb,HatayamaETAl}, 
it is natural to ask whether this approach to physical combinatorics really is {\it physical}. This approach also has the inherent limitation that it can only be applied with fixed and not periodic boundary conditions. 
There have been many subsequent works studying the eigenvalues and physical combinatorics of the RSOS models both with periodic~\cite{BazhResh89,KP91,KP92} and fixed~\cite{ABBBQ,Sklyanin88,BPO96,Das97,OPW97} boundary conditions but perhaps the RSOS lattice path approach to the physical combinatorics of the RSOS is best typified by the work of Warnaar~\cite{Warnaar96a,Warnaar96b}. For some recent reviews see \cite{Mathieu}.

The aim of this paper is twofold ---
first, to present a conjectured energy-preserving bijection between RSOS paths and string patterns of the transfer matrices at the critical point and, second, to give complete derivations of the conformal energies in the vacuum sectors of the RSOS minimal and XXX models by solving the Thermodynamic Bethe Ansatz (TBA)~\cite{ZamolodchikovTBA,KP92}. 
Since the solution of the TBA equations relies only on the string patterns, it is independent of the conjectured bijection with RSOS paths. 
Following initial work in \cite{FeveratiP03}, the approach we adopt in this paper removes the logical inconsistency of invoking properties of off-critical models to study the physical combinatorics of critical RSOS models. It points the way to a stronger unification of the approaches based on string patterns and RSOS lattice paths. Moreover, our methods can be applied to larger classes of boundary conditions, including models with periodic boundaries which are not tractable by previous RSOS lattice path methods.

In this paper we consider the ABF RSOS $A_L$ models and the $s\ell(2)_1$ WZW lattice model in their vacuum sectors. The WZW model is a six-vertex model with rational Boltzmann weights which reduces, in the Hamiltonian limit, to the $U_q(s\ell(2))$-invariant XXX quantum spin chain. In the sequel, we will always refer to this case as the XXX model. 
The layout of the paper is as follows. Quasiparticles are introduced combinatorially by decomposing one-dimensional RSOS lattice paths into particles decorated by dual-particles in Section~2. It is shown that the geometric packing constraints of these decorated particles encode the $(m,n)$ systems. Moreover, it is argued that the RSOS lattice paths actually encode string patterns through the locations of the quasiparticles (particles and dual-particles). This is argued on the basis of a conjectured energy-preserving bijection between the RSOS paths and the string patterns. The energy statistics are associated with the string patterns through finitized characters (involving integer string quantum numbers and the Cartan matrix) and with RSOS paths through one-dimensional configurational sums (having their origins in off-critical Corner Transfer Matrix calculations). 
In Section~3, we consider the example of the vacuum sector of the minimal RSOS lattice models. Using the commuting family of fused transfer matrices, we show how the eigenvalues are completely classified by string patterns consisting of 1- and 2-strings in $L-2$ analyticity strips. The allowed numbers of 1- and 2-strings in these strips (particle content) is subject to the $(m,n)$ system. The relative positions of the 1- and 2- strings in each strip (integer string quantum numbers) uniquely specifies the analyticity input required to analytically solve the TBA equations. The detailed solution of the TBA presented in this section finally leads to the general conformal energy expression (\ref{energy3}) including the central charge term. This is the first complete derivation of this general result from the TBA. 
In Section~4, we consider the vacuum sector of the XXX model. Although this model has not previously been analysed using RSOS paths, we show that this model is amenable to our new approach. Considering the commuting family of fused transfer matrices again shows that the eigenvalues are completely classified by string patterns consisting of 1- and 2-strings in $L=\lfloor N/2\rfloor+1$ analyticity strips. In this case the fusion hierarchy is infinite and does not truncate. This is the reason that $L$ grows with the system size $N$. The allowed numbers of 1- and 2-strings in these strips (particle content) is again subject to the $(m,n)$ system given in Section~2. The relative positions of the 1- and 2- strings in each strip (integer string quantum numbers) uniquely specifies the analyticity input required to analytically solve the TBA equations. The detailed solution of the TBA finally leads to the general conformal energy expression (\ref{energy_xxx}) including the central charge term. Again this is the first complete derivation of this general result from the TBA.
We conclude with a discussion in Section~5. Details on  dilogarithm identities~\cite{KP92,KirillovDilog95a, KirillovDilog95b} are relegated to an Appendix.

\section{Combinatorial Quasiparticles}
\label{sec:quasi}

In this section we introduce quasiparticles~\cite{KlassenEtAl93} of two kinds in a purely combinatorial manner and refer to them as {\em particles} and {\em dual-particles}  even though they are all in fact quasiparticles. The connection with various lattice models is established in the next section.

\begin{figure}[h]
\setlength{\unitlength}{.6cm}
\psset{unit=.6cm}
\bea
\begin{array}{rcccc}
\mbox{$A_L$ minimal:}
&\mbox{}\hspace{3\unitlength}\mbox{}&
\mbox{\begin{pspicture}(0,-.125)(6,.4)
\psline[linewidth=1pt](0,0)(6,0)
\multiput(0,0)(1,0){7}{\psline[linewidth=1pt](0,-.2)(0,.2)}
\rput[r](-.8,0){$G=A_L$}
\rput[t](0,-.3){\small $1$}
\rput[t](1,-.3){\small $2$}
\rput[t](2,-.3){\small $3$}
\rput[t](4,-.3){$\cdots$}
\rput[t](6,-.3){\small $L$}
\end{pspicture}}
&\mbox{}\hspace{5\unitlength}\mbox{}&
\mbox{\begin{pspicture}(0,-.125)(6,.4)
\psline[linewidth=1pt](0,0)(6,0)
\multiput(0,0)(1,0){7}{\psline[linewidth=1pt](0,-.2)(0,.2)}
\rput[r](-.8,0){$G^*=A_{L-2}$}
\rput[t](0,-.3){\small $1$}
\rput[t](1,-.3){\small $2$}
\rput[t](2,-.3){\small $3$}
\rput[t](4,-.3){$\cdots$}
\rput[t](6,-.3){\small $L\!-\!2$}
\end{pspicture}}\\[.4in]
\mbox{$s\ell(2)_1$ XXX:}
&\mbox{}\hspace{3\unitlength}\mbox{}&
\mbox{\begin{pspicture}(0,-.125)(6,.2)
\psline[linewidth=1pt](0,0)(6,0)
\multiput(0,0)(1,0){7}{\psline[linewidth=1pt](0,-.2)(0,.2)}
\rput[r](-.8,0){$G=T_L$}
\rput[t](0,-.3){\small $1$}
\rput[t](1,-.3){\small $2$}
\rput[t](2,-.3){\small $3$}
\rput[t](4,-.3){$\cdots$}
\rput[t](6,-.3){\small $L$}
\pscircle[linewidth=1pt,fillstyle=solid](0,.3){.3}
\end{pspicture}}
&\mbox{}\hspace{5\unitlength}\mbox{}&
\mbox{\begin{pspicture}(0,-.125)(6,.2)
\psline[linewidth=1pt](0,0)(6,0)
\multiput(0,0)(1,0){7}{\psline[linewidth=1pt](0,-.2)(0,.2)}
\rput[r](-.8,0){$G^*=T'_{L-1}$}
\rput[t](0,-.3){\small $1$}
\rput[t](1,-.3){\small $2$}
\rput[t](2,-.3){\small $3$}
\rput[t](4,-.3){$\cdots$}
\rput[t](6,-.3){\small $L\!-\!1$}
\pscircle[linewidth=1pt,fillstyle=solid](6,.3){.3}
\end{pspicture}}\\[12pt]
\end{array}
\nonumber
\eea
\vspace{-\unitlength}
\caption{Dynkin diagrams $G=A_L, T_L$ and their respective duals $G^*=A_{L-2}, T_{L-1}'$. The particles live on $G$ and the dual-particles live on $G^*$. In the $A_L$ minimal case $L$ is fixed. In the XXX case $L=\lfloor N/2\rfloor+1$ grows with the system size $N$.\label{dynkin}}
\end{figure}
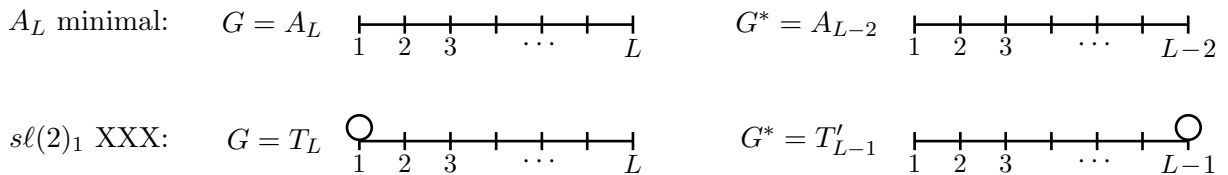

Combinatorially, quasiparticles are introduced via features of Restricted Solid-On-Solid (RSOS) paths. 
In principle, the paths can be arbitrary paths on Dynkin diagrams of $A$-$D$-$E$-$T$ or more general type, but here we only consider $A_L$ (linear) and $T_L$ (tadpole) diagrams with $L$ nodes (Figure~\ref{dynkin}). We denote by $T_L$ the tadpole diagram with a single loop at node $j=1$ and by $T_{L}'$ the tadpole diagram with a single loop at node $j=L$. On the Dynkin diagram $A_L$ with $L$ fixed, an $N$-step path $a=\{a_0,a_1,a_2,\ldots,a_N\}$ satisfies the RSOS constraint $a_{j-1}-a_j=\pm 1$ with $a_{j-1}\in \{1,2,\ldots,L\}$ for $j=1,2,\ldots,N$. On the tadpole diagram $T_L$ paths with steps along the baseline with $a_{j-1}=a_j=1$ are also allowed. In this case, we take $L=\lfloor N/2\rfloor+1$ corresponding to the maximum height occurring in an $N$-step path. We stress that this means, in the XXX case, $L$ grows with the system size $N$. We will only consider the {\em vacuum sector} with boundary conditions given by $a_0=a_N=1$, $a_{N+1}=2$. 

\subsection{Pure particles}

On $A_L$, pure particles~\cite{Warnaar96a,Warnaar96b} of type $a$ correspond to upright two-dimensional {\it pyramids} (triangles) of height $a$ and baseline width $2a$. 
We use the name {\em pyramid} to emphasize that there is a distinguished base (baseline). 
There are thus $L-1$ types of pyramid particles. In addition, on $T_L$, we allow {\it tower} particles consisting of a single step along the baseline with $a_{j-1}=a_j=1$. A typical path can contain many particles. Let $n_a$ with $a=1,2,\ldots,L-1$ be the number of pyramid particles of type $a$ and let $n_0$ denote the number of tower particles. Since there can only be $0$ or $1$ particle of a given type at a given $j$, there is an exclusion principle and the particles are fermionic in nature.

\begin{figure}[htbp]
\setlength{\unitlength}{1pt}
\begin{center}
\mbox{}\hspace{-.7in}
\begin{picture}(360,106)
%\put(-5,-20){\makebox(0,0)[r]{$0$}}
\put(-5,0){\makebox(0,0)[r]{$1$}}
\put(-5,20){\makebox(0,0)[r]{$2$}}
\put(-5,40){\makebox(0,0)[r]{$3$}}
\put(-5,60){\makebox(0,0)[r]{$4$}}
\put(-5,80){\makebox(0,0)[r]{$5$}}
\put(0,-5){\makebox(0,0)[t]{$0$}}
\put(20,-5){\makebox(0,0)[t]{$1$}}
\put(40,-5){\makebox(0,0)[t]{$2$}}
\put(60,-5){\makebox(0,0)[t]{$3$}}
\put(80,-5){\makebox(0,0)[t]{$4$}}
\put(100,-5){\makebox(0,0)[t]{$5$}}
\put(120,-5){\makebox(0,0)[t]{$6$}}
\put(140,-5){\makebox(0,0)[t]{$7$}}
\put(160,-5){\makebox(0,0)[t]{$8$}}
\put(180,-5){\makebox(0,0)[t]{$9$}}
\put(200,-5){\makebox(0,0)[t]{$10$}}
\put(220,-5){\makebox(0,0)[t]{$11$}}
\put(240,-5){\makebox(0,0)[t]{$12$}}
\put(260,-5){\makebox(0,0)[t]{$13$}}
\put(280,-5){\makebox(0,0)[t]{$14$}}
\put(300,-5){\makebox(0,0)[t]{$15$}}
\put(320,-5){\makebox(0,0)[t]{$16$}}
\put(340,-5){\makebox(0,0)[t]{$17$}}
\put(360,-5){\makebox(0,0)[t]{$18$}}
\put(380,-5){\makebox(0,0)[t]{$19$}}
\put(400,-5){\makebox(0,0)[t]{$20$}}
\put(420,-5){\makebox(0,0)[t]{$21$}}
\put(440,-5){\makebox(0,0)[t]{$22$}}
%\put(400,-24.5){\makebox(0,0)[t]{$\ \ \ \ \,20$}}
\put(80,100){\makebox(0,0)[t]{$n_4$}}
\put(220,100){\makebox(0,0)[t]{$n_3$}}
\put(320,100){\makebox(0,0)[t]{$n_2$}}
\put(380,100){\makebox(0,0)[t]{$n_1$}}
\put(420,100){\makebox(0,0)[t]{$n_0$}}
\put(410,108){\makebox(0,0)[t]{$\vdots$}}
\put(430,108){\makebox(0,0)[t]{$\vdots$}}
\multiput(0,0)(0,20){5}{\line(1,0){440}}
\multiput(0,0)(20,0){23}{\line(0,1){80}}
\thicklines
%\put(0,.2){\line(1,0){440}}
%\put(0,-.2){\line(1,0){440}}
%\put(0,.4){\line(1,0){440}}
%\put(0,-.4){\line(1,0){440}}
\put(0,0){\line(1,1){20}}
\put(20,20){\line(1,1){20}}
\put(40,40){\line(1,1){20}}
\put(60,60){\line(1,1){20}}
\put(80,80){\line(1,-1){20}}
\put(100,60){\line(1,-1){20}}
\put(120,40){\line(1,-1){20}}
\put(140,20){\line(1,-1){20}}
\put(160,0){\line(1,1){20}}
\put(180,20){\line(1,1){20}}
\put(200,40){\line(1,1){20}}
\put(220,60){\line(1,-1){20}}
\put(240,40){\line(1,-1){20}}
\put(260,20){\line(1,-1){20}}
\put(280,0){\line(1,1){20}}
\put(300,20){\line(1,1){20}}
\put(320,40){\line(1,-1){20}}
\put(340,20){\line(1,-1){20}}
\put(360,0){\line(1,1){20}}
\put(380,20){\line(1,-1){20}}
\put(400,0){\line(1,0){20}}
\put(420,0){\line(1,0){20}}
\put(380,20){\color{green}\circle*{6}}
\put(60,20){\color{red}\circle*{6}}
\put(20,20){\color{red}\circle*{6}}
\put(40,20){\color{red}\circle*{6}}
\put(60,20){\color{red}\circle*{6}}
\put(100,20){\color{red}\circle*{6}}
\put(120,20){\color{red}\circle*{6}}
\put(140,20){\color{red}\circle*{6}}
\put(180,20){\color{red}\circle*{6}}
\put(200,20){\color{red}\circle*{6}}
\put(240,20){\color{red}\circle*{6}}
\put(260,20){\color{red}\circle*{6}}
\put(300,20){\color{red}\circle*{6}}
\put(340,20){\color{red}\circle*{6}}
\put(320,40){\color{purple}\circle*{6}}
\put(220,60){\color{yellow}\circle*{6}}
\put(40,40){\color{blue}\circle*{6}}
\put(60,40){\color{blue}\circle*{6}}
\put(100,40){\color{blue}\circle*{6}}
\put(120,40){\color{blue}\circle*{6}}
\put(200,40){\color{blue}\circle*{6}}
\put(240,40){\color{blue}\circle*{6}}
\put(60,60){\color{orange}\circle*{6}}
\put(100,60){\color{orange}\circle*{6}}
\put(80,80){\color{lightblue}\circle*{6}}
%%%
\multiput(410,20)(20,0){2}{\color{red}\circle*{6}}
\multiput(410,40)(20,0){2}{\color{blue}\circle*{6}}
\multiput(410,60)(20,0){2}{\color{orange}\circle*{6}}
\multiput(410,80)(20,0){2}{\color{midblue}\circle*{6}}
\end{picture}
\qquad\mbox{}
\end{center}
%\qquad\mbox{}
%\end{center}
\caption{Illustration of the different types of pure particles that can occur in an $N$-step path. There are $L-1$ types of pyramid particles of different heights. Their positions (peaks) are indicated by a solid coloured dot according to their type: \ \mbox{{\color{green}\Large $\bullet$\ }}\ \mbox{{\color{purple}\Large $\bullet$\ }}\ \mbox{{\color{yellow}\Large $\bullet$\ }}\ \mbox{{\color{lightblue}\Large $\bullet$\ }}\ \mbox{{\color{darkgreen}\Large $\bullet$\ }}\ $\cdots$ A pyramid particle of type $a$ is decorated by $a(a-1)$ dual-particles as shown. On $A_L$, there are $L-2$ types of dual-particles with different relative heights above the particle baseline. On $T_L$, there is an additional dual-particle giving a total of $L-1$ different types of dual-particles. 
The dual-particles are also indicated by solid coloured dots according to their type: \ \mbox{{\color{red}\Large $\bullet$\ }}\ \mbox{{\color{blue}\Large $\bullet$\ }}\ \mbox{{\color{orange}\Large $\bullet$\ }}\ \mbox{{\color{midblue}\Large $\bullet$\ }}\ $\cdots$ In addition, on $T_L$, there are {\em tower particles} corresponding to single steps along the baseline with $a_{j-1}=a_j=1$. These particles which are designated to be of type $a=0$ are decorated by a complete tower of $L-1=\lfloor N/2\rfloor$ dual-particles.}
\end{figure}
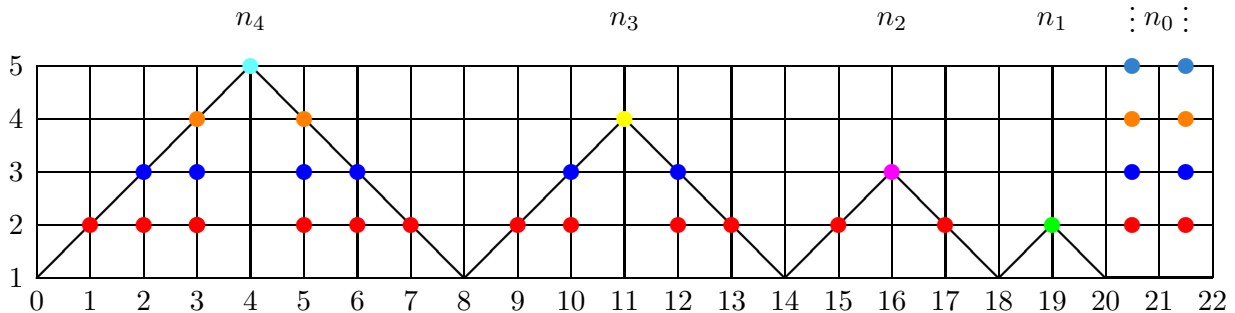

\subsection{Dual particles}

On $A_L$, a particle of type $a$ is decorated by $a(a-1)$ dual-particles as indicated in Figure~2. In this sense particles are composites of dual-particles. On $A_L$ and $T_L$, the type of the dual-particle is fixed by the relative height above the particle baseline. 
Specifically, a dual-particle of type $a$ appears at a height $a$ above the baseline of the particle which contains it. On $A_L$, there are $L-2$ types of dual-particles whereas, on $T_L$, there are $L-1$ different types. 
On $T_L$, the tower particles are decorated by a complete tower of $L-1=\lfloor N/2\rfloor$ dual-particles. A typical path can contain many dual-particles. Let $m_a$ with $a=1,2,\ldots,L-1$ be the number of dual-particles of type $a$. Since there can only be $0$ or $1$ dual-particle of a given type at a given lattice point $j$, there is an exclusion principle and the dual-particles are also fermionic in nature.

\subsection{Geometric packing and $(m,n)$ systems}
\label{sec:geo}

The $(m,n)$ systems~\cite{Melzer94,Berkovich94} describe the possible particle contents. In our combinatorial framework, they arise from the geometric packing constraints of particles and dual-particles along an $N$-step path. The adjacency matrices that appear in the $(m,n)$ systems differ from the adjacency matrices of $G=A_L, T_L$ for the original paths and are in fact those of the dual graphs $G^*$.
Let $A$ denote the adjacency matrix of $G^*=A_{L-2}$ or $T_{L-1}'$ so that the corresponding Cartan matrix is
\bea\label{adjacency}
C=2I-A
\eea

{$A_L$:} The geometric packing constraints to accommodate $n_a$ particles of type $a$ and $m_a$ dual-particles of type $a$ along a path of $N$ steps on $G=A_L$ are
\bea
\begin{array}{rcl}
N&=&2n_1+4n_2+6n_3+8n_4+\cdots+2(L-1)n_{L-1}\\
m_1&=&2n_2+4n_3+6n_4+\cdots+2(L-2)n_{L-1}\\
m_2&=&2n_3+4n_4+\cdots+2(L-3)n_{L-1}\\
m_3&=&2n_4+\cdots+2(L-4)n_{L-1}\\
&\vdots&\\
m_{L-2}&=&2n_{L-1}
\end{array}
\eea
Rearranging yields an $(m,n)$ system of $G^*=A_{L-2}$ type
\bea
\mybox{m_a+n_a=\half N\,\delta(a,1)+\half\sum_{b=1}^{L-2} A_{a,b} m_b,\qquad\quad a=1,2,\ldots,L-2}
\label{Amnsystem}
\eea
where
\bea
n_{L-1}={N-2n_1-4n_2-6n_3-\cdots-2(L-2)n_{L-2}\over 2(L-1)}
\eea
and
\bea
N\ge m_1\ge m_2\ge\cdots\ge m_{L-2}\ge 0
\eea
Here $A$ is the $A_{L-2}$ adjacency matrix. 
Notice that $N$ and $m_a$ must all be even. 

Starting with the $(m,n)$ system, it is possible to eliminate the particle numbers $n_a$ in favour of the dual-particle numbers $m_a$ or vice-versa. It is this fact that underlies the duality between particles and dual-particles. Which is to be regarded as the more fundamental is a matter of choice. 
Explicitly, using column vectors, the $(m,n)$ system can be written in the alternative forms 
\bea
\vec m+\vec n=\half (N\vec e_1+A\vec m),\qquad \vec n=\half(N\vec e_1-C\vec m),\qquad \vec m=C^{-1}(N\vec e_1-2\vec n)
\label{mnalt}
\eea
where
\bea
\vec m=(m_1,m_2,\ldots,m_{L-2})^T,\quad \vec n=(n_1,n_2,\ldots,n_{L-2})^T,\quad \vec e_1=(1,0,0,\ldots,0)^T
\eea
and the inverse Cartan matrix of $A_{L-2}$ is
\bea
C^{-1}_{ab}={1\over L-1}\,\mbox{min}\big(a(L-1-b),b(L-1-a)\big)
\eea

{$T_L$:} Let $L=\lfloor N/2\rfloor+1$ and let $n_0$ be the number of tower particles. Then the geometric packing constraints to accommodate $n_a$ particles of type $a$ and $m_a$ dual-particles of type $a$ along a path of $N$ steps on $G=T_L$ are
\bea\label{geomT}
\begin{array}{rcl}
N&=&n_0+2n_1+4n_2+6n_3+8n_4+\cdots+2(L-1) n_{L-1}\\
m_1&=&n_0+2n_2+4n_3+\cdots+2(L-3)n_{L-2}+2(L-2)n_{L-1}\\
&\vdots&\\
m_{L-3}&=&n_0+2n_{L-2}+4n_{L-1}\\
m_{L-2}&=&n_0+2n_{L-1}\\
m_{L-1}&=&n_0
\end{array}
\eea
Rearranging yields an $(m,n)$ system of $G^*=T_{L-1}'$ type
\bea
\mybox{m_a+n_a=\half N\,\delta(a,1)+\half\sum_{b=1}^{L-1} A_{a,b} m_b,\qquad a=1,2,\ldots,L-1}
\label{Tmnsystem}
\eea
where $A$ is the $T_{L-1}'$ adjacency matrix. Explicitly this yields the inversion of (\ref{geomT})
in the form
\begin{align}
  n_1 & = \tfrac{1}{2}N+\tfrac{1}{2}m_2-m_1 \nonumber\\
  n_a & = \tfrac{1}{2}m_{a-1}+\tfrac{1}{2}m_{a+1}-m_a , \quad 2 \leq a \leq L-2 \\
  n_{L-1} & = \tfrac{1}{2}m_{L-2}-\tfrac{1}{2}m_{L-1} \nonumber
\end{align}
Notice that for each $a$, $m_a=N$ mod 2 and
\bea
N\ge m_1\ge m_2\ge \cdots \ge m_{L-1}\ge 0
\eea
This $(m,n)$ system can also be written in the alternative forms (\ref{mnalt}) where 
\bea
\vec m=(m_1,m_2,\ldots,m_{L-1})^T,\quad \vec n=(n_1,n_2,\ldots,n_{L-1})^T,\quad \vec e_1=(1,0,0,\ldots,0)^T
\eea
and, explicitly, 
\bea
C^{-1}_{ab}=\mbox{min}(a,b),\qquad C^{-1}\vec e_1=\vec e=(1,1,\ldots,1)^T
\eea

\subsection{Decomposition of paths into particles}
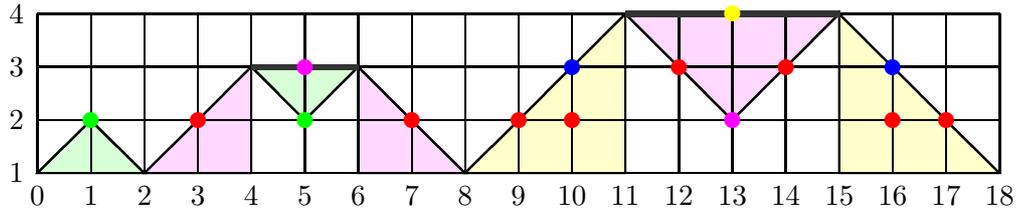
\begin{figure}[htbp]
\setlength{\unitlength}{1pt}
\psset{unit=1pt}
\begin{center}
\begin{pspicture}(360,66)
\pspolygon[fillstyle=solid,fillcolor=lightlightgreen](0,0)(20,20)(40,0)
\pspolygon[fillstyle=solid,fillcolor=lightlightred](40,0)(80,40)(80,0)
\pspolygon[fillstyle=solid,fillcolor=lightlightgreen](80,40)(100,20)(120,40)
\pspolygon[fillstyle=solid,fillcolor=lightlightred](120,0)(120,40)(160,0)
\pspolygon[fillstyle=solid,fillcolor=lightlightblue](160,0)(220,60)(220,0)
\pspolygon[fillstyle=solid,fillcolor=lightlightred](220,60)(260,20)(300,60)
\pspolygon[fillstyle=solid,fillcolor=lightlightblue](300,0)(300,60)(360,0)
\put(-5,0){\makebox(0,0)[r]{$1$}}
\put(-5,20){\makebox(0,0)[r]{$2$}}
\put(-5,40){\makebox(0,0)[r]{$3$}}
\put(-5,60){\makebox(0,0)[r]{$4$}}
\put(0,-5){\makebox(0,0)[t]{$0$}}
\put(20,-5){\makebox(0,0)[t]{$1$}}
\put(40,-5){\makebox(0,0)[t]{$2$}}
\put(60,-5){\makebox(0,0)[t]{$3$}}
\put(80,-5){\makebox(0,0)[t]{$4$}}
\put(100,-5){\makebox(0,0)[t]{$5$}}
\put(120,-5){\makebox(0,0)[t]{$6$}}
\put(140,-5){\makebox(0,0)[t]{$7$}}
\put(160,-5){\makebox(0,0)[t]{$8$}}
\put(180,-5){\makebox(0,0)[t]{$9$}}
\put(200,-5){\makebox(0,0)[t]{$10$}}
\put(220,-5){\makebox(0,0)[t]{$11$}}
\put(240,-5){\makebox(0,0)[t]{$12$}}
\put(260,-5){\makebox(0,0)[t]{$13$}}
\put(280,-5){\makebox(0,0)[t]{$14$}}
\put(300,-5){\makebox(0,0)[t]{$15$}}
\put(320,-5){\makebox(0,0)[t]{$16$}}
\put(340,-5){\makebox(0,0)[t]{$17$}}
\put(360,-5){\makebox(0,0)[t]{$18$}}
\multiput(0,0)(0,20){4}{\line(1,0){360}}
\multiput(0,0)(20,0){19}{\line(0,1){60}}
\thicklines
\put(0,0){\line(1,1){20}}
\put(20,20){\line(1,-1){20}}
\put(40,0){\line(1,1){20}}
\put(60,20){\line(1,1){20}}
\put(80,40){\line(1,-1){20}}
\put(100,20){\line(1,1){20}}
\put(120,40){\line(1,-1){20}}
\put(140,20){\line(1,-1){20}}
\put(160,0){\line(1,1){20}}
\put(180,20){\line(1,1){20}}
\put(200,40){\line(1,1){20}}
\put(220,60){\line(1,-1){20}}
\put(240,40){\line(1,-1){20}}
\put(260,20){\line(1,1){20}}
\put(280,40){\line(1,1){20}}
\put(300,60){\line(1,-1){20}}
\put(320,40){\line(1,-1){20}}
\put(340,20){\line(1,-1){20}}
\put(80,39){\color{grey}\rule{40\unitlength}{2\unitlength}}
\put(220,59){\color{grey}\rule{80\unitlength}{2\unitlength}}
\put(20,20){\color{green}\circle*{6}}
\put(100,20){\color{green}\circle*{6}}
\put(260,20){\color{purple}\circle*{6}}
\put(100,40){\color{purple}\circle*{6}}
\put(260,60){\color{yellow}\circle*{6}}
\put(60,20){\color{red}\circle*{6}}
\put(140,20){\color{red}\circle*{6}}
\put(180,20){\color{red}\circle*{6}}
\put(200,20){\color{red}\circle*{6}}
\put(200,40){\color{blue}\circle*{6}}
\put(320,20){\color{red}\circle*{6}}
\put(240,40){\color{red}\circle*{6}}
\put(280,40){\color{red}\circle*{6}}
\put(320,40){\color{blue}\circle*{6}}
\put(340,20){\color{red}\circle*{6}}
\end{pspicture}
\end{center}
\caption{An arbitrary path decomposes into pure particles and complexes of overlapping particles. 
In this way pure particles not only move (have different locations) but also can be cut apart and turned upside-down. 
Even though a particle can be cut into many pieces with other particles inserted, the area (shown shaded) of the pure particles (pyramids) 
is preserved along with the decorating dual-particles. 
In this example there are particles of type $1,1,2,2,3$ at $j=1,5,5,13,13$ and dual-particles of type 
$1$ at $j=3,7,9,10,12,14,16,17$ and type $2$ at $j=10,16$. Note that particles (dual-particles) of different types can occupy the same lattice site.
%$\mbox{Path}=\mbox{\{pure particles}\}+\mbox{\{complexes of overlapping particles\}}$
}
\end{figure}

In addition to translational motion, pure particles can be cut into many pieces (with other particles inserted) and turned upside-down. 
To see the particle content of an arbitrary path, as in Figure~3, we need an algorithm to decompose the path 
into pure particles and complexes of overlapping particles. Since the particles are composites, the dual-particles must be carried along with the particles. Moreover, the algorithm is conjectured to yield a bijection so that it is possible to return to the original path given the configuration of particles and dual-particles. 

The particle decomposition algorithm we use is based on the bijective decomposition of Warnaar~\cite{Warnaar96a,Warnaar96b} but is modified to be more symmetric. This modification is required to make contact with the various lattice models in the next section. Explicitly, the algorithm to decompose an arbitrary $A_L$ or $T_L$ path to reveal the  particle content and configurations of particles and dual-particles is as follows:
\begin{enumerate}
\item[1.] Identify any flat segments corresponding to tower particles and decorate them with dual-particles. These tower particles automatically separate the path into complexes with respect to the initial baseline at height $1$. A complex with respect to the current baseline is any segment of the path, starting and ending at the baseline, that is not a pyramid.
\item[2.] For each current baseline, separate the pure particles (pyramids with respect to the current baseline) from the complexes and decorate these pyramids with particles and dual-particles.
\item[3.] If there are no remaining complexes then terminate the algorithm. For each remaining complex with respect to the current baseline, identify the left-most and right-most global maxima and connect these with a new baseline. The left-most and right-most global maxima may coincide and in this case no new baseline is drawn. Decorate the new baseline with a particle placed at the midpoint with a type appropriate to its height above the previous baseline.
\item[4.] From each left (right) maxima, follow the profile of the complex moving continuously down and to the left (right), and, where a valley is encountered, draw a further baseline at the height of the highest peak neighbouring the valley. This process thus introduces new baselines at the heights of secondary peaks as they are encountered. Decorate all sloping segments (not valleys or peaks) of the complex, outside or at the endpoints of the new baselines, with dual-particles.
\item[5.] For each of the new baselines generated reflect the path constrained within it about that baseline and identify each as a current baseline. Go to step 2. 
\end{enumerate}
The iterative stages of this particle decomposition algorithm are illustrated in Figure~4.

The pure particles are {\em solitons} in the sense that they can move together from far apart, {\em interact} and then re-emerge as unchanged pure particles. This holds generally for an arbitrary number of particles but is illustrated for particles of types $1$ and $3$ and particles of types $2$ and $3$ moving through each other in Figures~5 and 6 respectively. 
\begin{figure}[htbp]
\centerline {
\includegraphics[width=6.5in]{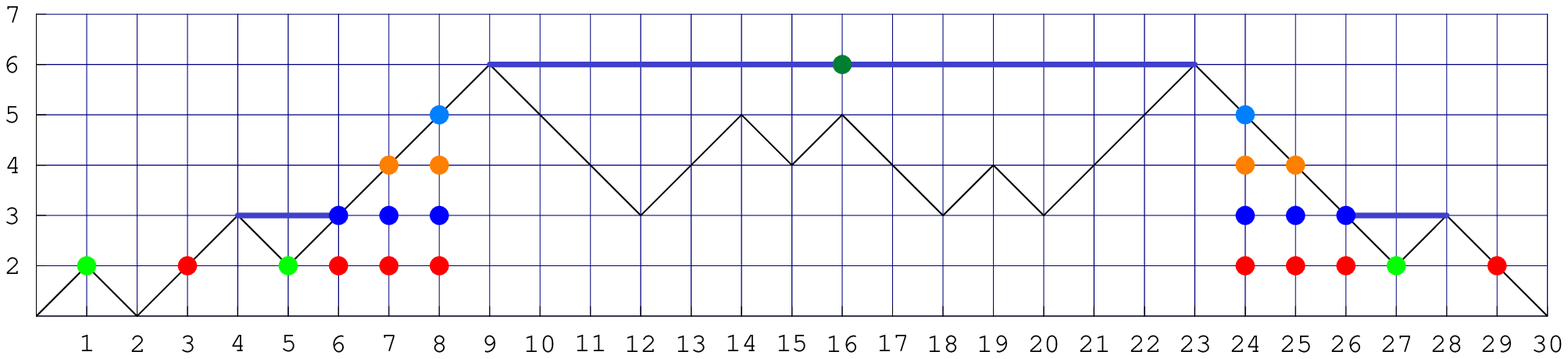}}
\vspace{.1in}
\centerline {
\includegraphics[width=6.5in]{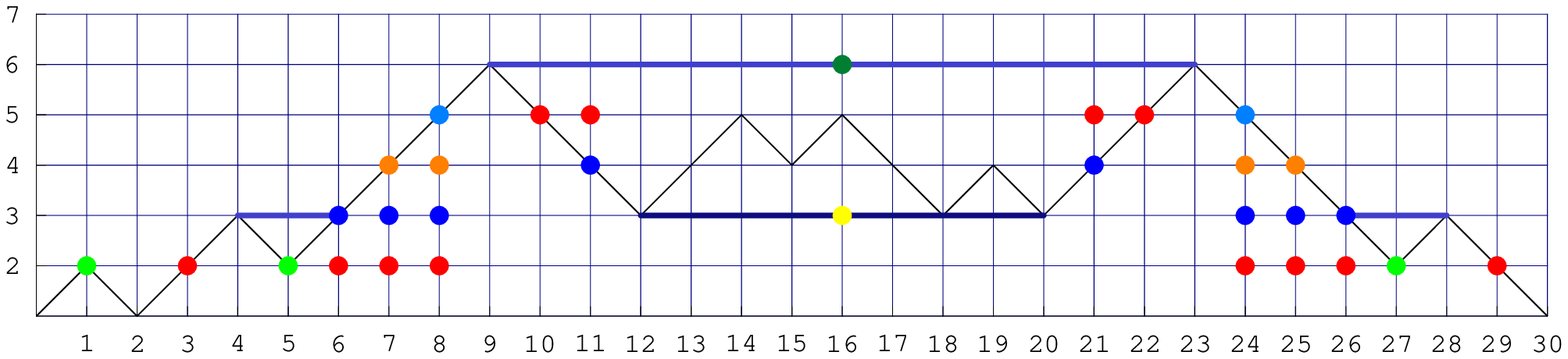}}
\vspace{.1in}
\centerline {
\includegraphics[width=6.5in]{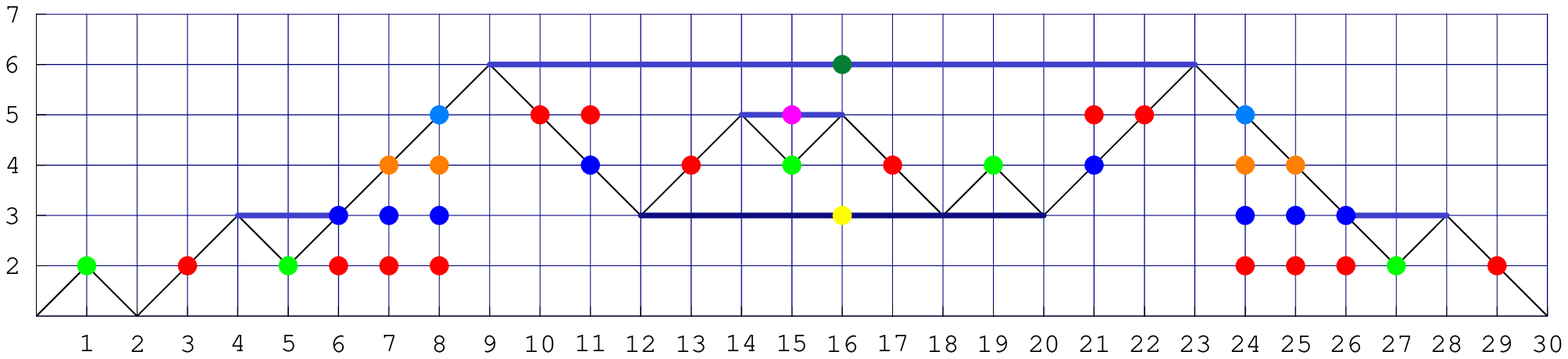}}
\caption{Illustrative example showing three iterative stages of the algorithm for decomposing a particular $A_L$ path to reveal the particle content and configurations of particles and dual-particles. In the first stage, a new baseline is drawn at height 6. The profile is traced down and to the left and down and to the right. This introduces (upside down) particles of type 1 at positions 5 and 27 and the indicated decorations. In the second stage, a new baseline is drawn at height 3. The profile of the complex between sites 9 and 23 is traced out adding the indicated dual-particle decorations. In the third stage, a new baseline is added at height 5. The profile is traced out and the final decorations added.}
\end{figure}

\begin{figure}[hp]
\includegraphics[width=6.2in]{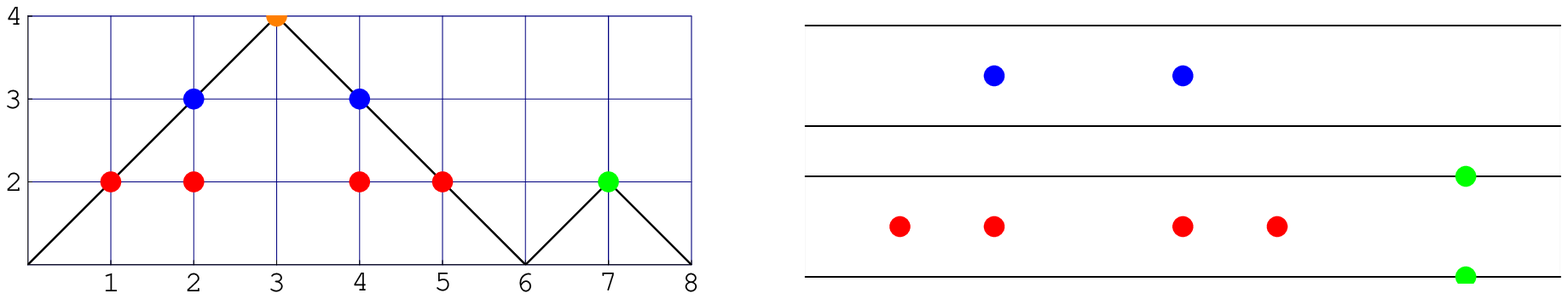}\\[-26pt]
\includegraphics[width=6.2in]{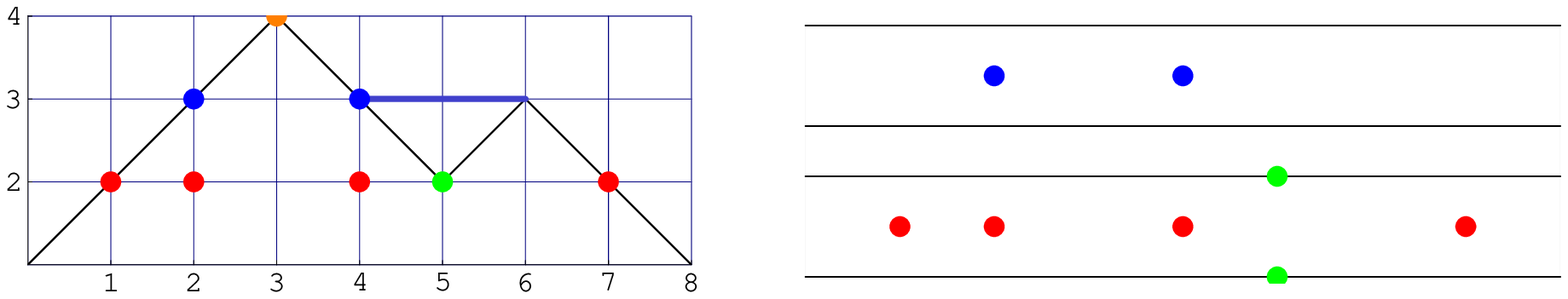}\\[-26pt]
\includegraphics[width=6.2in]{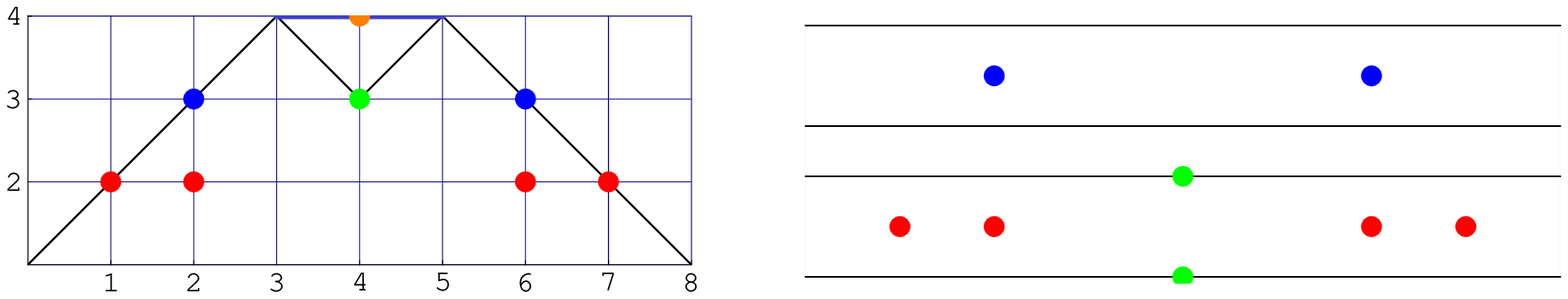}\\[-32pt]
\caption{A particle of type $1$ moving from right to left through a particle of type $3$. The configurations as the particle of type $1$ progresses further to the left are obtained by reflecting in the vertical. Like solitons, after the interaction, the particles maintain their integrity. On the right, the type $1$ and $2$ particle and dual-particle content is shown as a string pattern in two strips.}
%\end{figure}
%\begin{figure}[hp]
\includegraphics[width=6.5in]{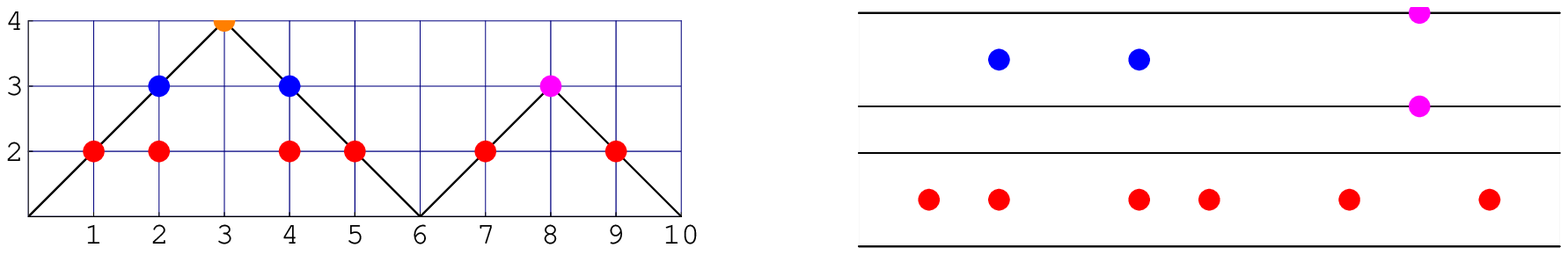}\\[-30pt]
\includegraphics[width=6.5in]{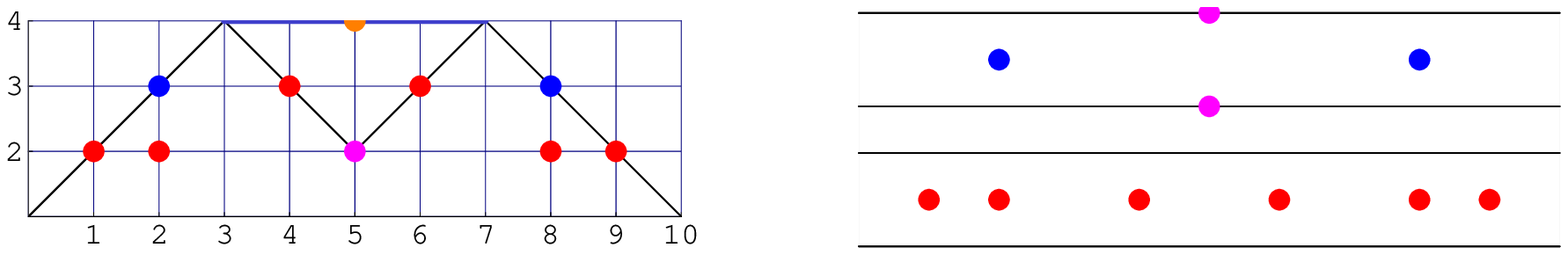}\\[-30pt]
\includegraphics[width=6.5in]{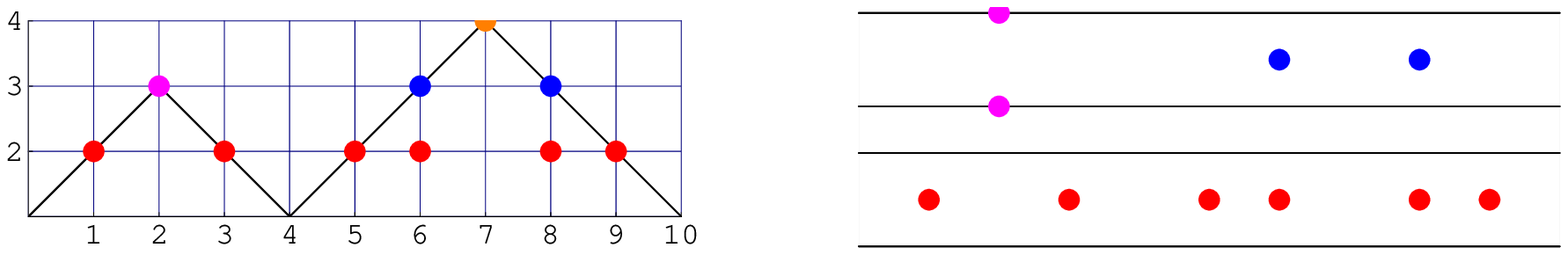}\\[-38pt]
\caption{A particle of type $2$ moving through a particle of type $3$. Like solitons, after the interaction, the particles maintain their integrity. On the right, the type $1$ and $2$ particle and dual-particle content is shown as a string pattern in two strips.}
\end{figure}

\subsection{Particle content and string patterns}

The content of particles and dual particles is conveniently encoded in a pattern of $1$- and $2$-strings in a number of {\it physical} strips as in Figures~5 and 6. The strips can be displayed horizontally or vertically as convenient. Anticipating the connection with lattice models, we call such patterns {\it string patterns} or {\it zero patterns}. A particle of type $a$ at position $j$ corresponds to a $2$-string in strip 
$a$ at position $j$, that is, a pair of zeros at the two edges of the strip with a common coordinate along the strip. A dual-particle of type $a$ at position $j$ corresponds to a $1$-string in the center of strip $a$ at position $j$. Although the positions $j$ are used to initially generate the string patterns, only the relative ordering of $1$- and $2$-strings in each separate strip is relevant. String patterns with the same relative orderings of $1$- and $2$-strings are regarded as equivalent.

We assert, without proof, that the particle decomposition provides a bijection between RSOS paths and distinct string patterns. The RSOS paths and string patterns are known to give the same finitized characters for arbitrary system size $N$. In addition, we have checked this conjecture explicitly for sizes out to $N=16$ by coding the decomposition algorithm in Mathematica~\cite{Wolfram}. We hope to give a proper proof of this conjectured bijection elsewhere.

\subsection{$q$-binomials and particle energies}

The distinct string patterns in strip $a$ with $n_a$ $2$-strings and $m_a$ $1$-strings are enumerated combinatorially by a $q$-binomial generating function $\sqbin{m_a+n_a}{m_a}q$ as illustrated in Figure~7. The strip configurations are assigned a weight $q^E$ and graded by the energy $E$. The lowest energy configuration, with all $1$-strings above all $2$-strings, is assigned the relative energy $E=0$. Each time a $1$-string is pushed down through a $2$-string, the energy increases by one unit. This {\em excitation} energy is associated with the 1-strings or dual-particles.

\begin{figure}[htbp]
\setlength{\unitlength}{.8pt}
\begin{center}
\begin{picture}(360,220)
\multiput(0,0)(40,0){2}{\line(0,1){100}}
\multiput(80,0)(40,0){2}{\line(0,1){100}}
\multiput(160,0)(40,0){2}{\line(0,1){100}}
\multiput(240,0)(40,0){2}{\line(0,1){100}}
\multiput(320,0)(40,0){2}{\line(0,1){100}}
\multiput(160,120)(40,0){2}{\line(0,1){100}}
\multiput(-5,0)(80,0){5}{\line(1,0){50}}
\put(155,120){\line(1,0){50}}
\multiput(60,50)(80,0){4}{\makebox(0,0)[c]{\large $\to$}}
\multiput(0,20)(40,0){2}{\color{green}\circle*{6}}
\multiput(0,40)(40,0){2}{\color{green}\circle*{6}}
\put(20,60){\color{red}\circle*{6}}
\put(20,80){\color{red}\circle*{6}}
\multiput(80,20)(40,0){2}{\color{green}\circle*{6}}
\multiput(80,60)(40,0){2}{\color{green}\circle*{6}}
\put(100,40){\color{red}\circle*{6}}
\put(100,80){\color{red}\circle*{6}}
\multiput(160,40)(40,0){2}{\color{green}\circle*{6}}
\multiput(160,60)(40,0){2}{\color{green}\circle*{6}}
\put(180,20){\color{red}\circle*{6}}
\put(180,80){\color{red}\circle*{6}}
\multiput(240,40)(40,0){2}{\color{green}\circle*{6}}
\multiput(240,80)(40,0){2}{\color{green}\circle*{6}}
\put(260,20){\color{red}\circle*{6}}
\put(260,60){\color{red}\circle*{6}}
\multiput(320,60)(40,0){2}{\color{green}\circle*{6}}
\multiput(320,80)(40,0){2}{\color{green}\circle*{6}}
\put(340,20){\color{red}\circle*{6}}
\put(340,40){\color{red}\circle*{6}}
\multiput(160,140)(40,0){2}{\color{green}\circle*{6}}
\multiput(160,200)(40,0){2}{\color{green}\circle*{6}}
\put(180,160){\color{red}\circle*{6}}
\put(180,180){\color{red}\circle*{6}}
\put(20,-20){\makebox(0,0)[c]{\large $1$}}
\put(100,-20){\makebox(0,0)[c]{\large $q$}}
\put(180,-20){\makebox(0,0)[c]{\large $2q^2$}}
\put(260,-20){\makebox(0,0)[c]{\large $q^3$}}
\put(340,-20){\makebox(0,0)[c]{\large $q^4$}}
\multiput(60,-20)(80,0){4}{\makebox(0,0)[c]{\large $+$}}
\thicklines
\put(100,110){\vector(1,1){50}}
\put(210,160){\vector(1,-1){50}}
\end{picture}
\end{center}
\smallskip
\caption{Enumeration, by the $q$-binomial $\sqbin 42q=1+q+2q^2+q^3+q^4$, of distinct string patterns in strip $a$ with $m_a=n_a=2$. For convenience, the strips are displayed vertically with position $j=1$ at the top.}
\end{figure}
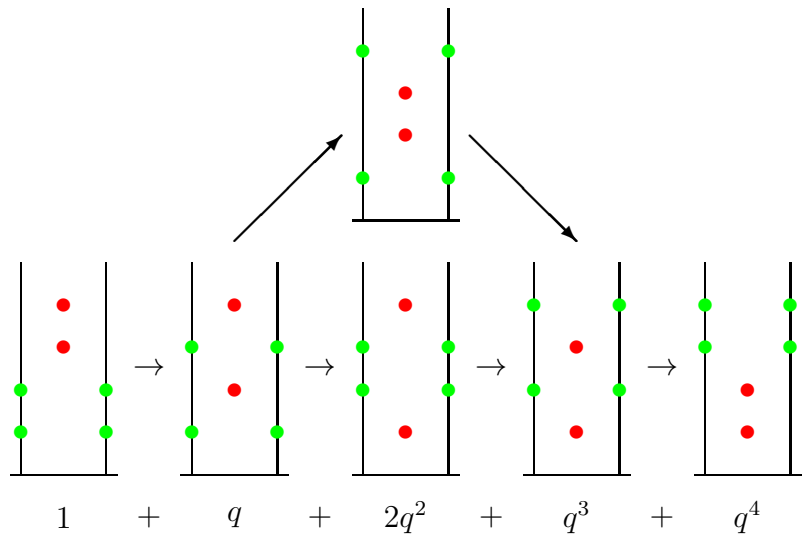

Integer string quantum numbers 
\bea
I^{(a)}=(I_1^{(a)},I_2^{(a)},\ldots,I_{m_a}^{(a)})
\eea
{\it uniquely} label the energy levels of dual-particles in each strip. These are defined by 
\bea\label{qnumb}
I_j^{(a)}&=&\left\{\mbox{number of
2-strings above the 1-string labelled $j$ in strip $a$}\right\}
\eea
Clearly,
\bea
n_a\ge I_1^{(a)}\ge I_2^{(a)}\ge\cdots\ge I_{m_a}^{(a)}\ge 0
\eea

In addition to the excitation energies described by $q$-binomials, there is an energy associated with the creation of the particle content, that is, particles or equivalently dual-particles. In terms of dual-particles, this energy is given by the Cartan matrix. The total energy assigned to a string pattern is 
\bea
E=-\frac{c}{24}+{1\over 4}\,\vec m^T C\vec m+\sum_a \sum_{j_a} I_{j_a}^{(a)}
\label{stringEnergy}
\eea
where the first term involves the central charge $c$ given by (\ref{mincentralcharge}) or $c=1$, 
the second term is the creation energy of the particles and the third term is the excitation energies arising from the various strips as in Figure~7. The central charge term is added to make contact with finitized characters given by the generating function or configurational sum
\bea
\chi_0^{(N)}(q)=\sum_{\rm config} q^E
\eea
where the configurations are enumerated by paths or string patterns. The total energies (\ref{stringEnergy}) including the central charge term are  correctly recovered in the TBA calculations of Sections~3 and 4 based on the string patterns.

\subsection{$A_L$ finitized fermionic characters}

In the $A_L$ case, the generating function for the spectrum of energy (\ref{stringEnergy}) in the vacuum sector is~\cite{Melzer94,Berkovich94}
\bea
\chi_{0,A}^{(N)}(q)=q^{-c/24} \sum_{(m,n)} q^{{1\over 4}\,\svec m^T C\svec m} \prod_{a=1}^{L-2} \qbin{m_a+n_a}{m_a}q
\eea
where $C$ is the $A_{L-2}$ Cartan matrix. This is precisely the finitized vacuum fermionic character of the unitary minimal model ${\cal M}(L,L+1)$ with central charge $c$ given by (\ref{mincentralcharge}). 
In the limit $N\to\infty$, with $N$ even, one obtains the vacuum fermionic character
\bea
\chi_{0,A}(q)=q^{-c/24}\sum_{m_a=0\;\text{(mod 2)}\atop m_1\ge m_2\ge \ldots\ge m_{L-2}\ge 0}
 {q^{{1\over 4}\,\svec m^T C\svec m}\over (q)_{m_1}} \prod_{a=2}^{L-2} \qbin{{1\over 2}(m_{a-1}+m_{a+1})}{m_a}q
\eea
where $m_{L-1}=0$ and the $q$-factorial is
\bea
(q)_m=\prod_{j=1}^m (1-q^j)
\eea

\subsection{$T_L$ finitized fermionic characters}

Combinatorially, in the $T_L$ case corresponding to the XXX model with 
central charge $c=1$, the finitized $T_L$ vacuum character 
admits several different forms related to spinons, Young tableaux and rigged configurations. We recall here the finitized~\cite{Melzer} and infinite $N$  forms~\cite{BernardEtAl,Bouwknegt} of this vacuum character.
%and give the combinatorial details in Appendix~A.

In the spinon formulation, the finitized $T_L$ vacuum character admits the fermionic form
\begin{equation*}
  \chi^{(N)}_{0,T}(q,z) 
 = q^{-1/24}\sum^{N/2}_{S_z=-N/2}q^{S_z^2}z^{2S_z}\qbin{N}{N/2-S_z}q 
\end{equation*}
Setting $q=z=1$ gives the counting of states
\begin{equation*}
  \chi^{(N)}_{0,T}(1,1) \;
 = \sum^{N/2}_{S_z=-N/2}{{N}\choose{N/2-S_z}} \;=\; 2^N 
\end{equation*}
In the XXX spin chain language, $N_+$, $N_-$ and $S_z$ are quantum numbers. $N_+$ is the number of up spins and $N_-$ is the number of down spins so that the number of spins is $N=N_++N_-$ and the $z$-component of spin is $S_z=\half(N_+-N_-)$.
Defining the quantum-dimension
\bea
  \chi_{n}(z) 
 = \frac{z^{n}-z^{-n}}{z-z^{-1}} ,\quad n \in \ZZ
\eea
%with $ \chi_{n}(1) = n $ and $ \chi_{-n}(z)=-\chi_{n}(z) $.
there is also a bosonic formula for this character~\cite{BernardEtAl}
\bea
  \chi^{(N)}_{0,T}(q,z) 
 = q^{-1/24}\sum^{\lfloor N/2\rfloor}_{N_+=0}\chi_{N-2N_++1}(z)
   \bigg( q^{(N/2-N_+)^2}\qbin{N}{N_+}q
          -q^{(N/2-N_++1)^2}\qbin{N}{N_+-1}q \bigg) 
\eea
As a consequence we have  
\bea
 2^N =  \chi^{(N)}_{0,T}(1,1) = \sum^{\lfloor N/2\rfloor}_{N_+=0}(N-2N_++1)  Z(N,N_+),\qquad
  Z(N,N_+) = {{N}\choose{N_+}} - {{N}\choose{N_+-1}}
\eea
and $ N-2N_++1=2S+1 $ is the degeneracy of the spin sector with total spin $S$.

An alternative fermionic representation of the $T_L$ vacuum character is given by~\cite{BernardEtAl}
\bea
  \chi^{(N)}_{0,T}(q,z)  &=& 
q^{-1/24} \sum_{(m,n)} \chi_{m_{\floor{N/2}}+1}(z)\, q^{{1\over 4}\,\svec m^T C\svec m} \prod_{a=1}^{\floor{N/2}} \qbin{m_a+n_a}{m_a}q
\eea
where $C$ is the $T_{L-1}'$ Cartan matrix. Let us consider the projection of this character onto the $z^0$ $s\ell(2)$ charge sector ($S_z=0$)
%We specialize the character by setting $z=1$ and ignore the degeneracies $\chi_{m_{\floor{N/2}}+1}(1)= m_{\floor{N/2}}+1$. This is equivalent to working in the sector $N_+=N/2$ which, from the lattice, contains all distinct eigenvalues. The trivial degeneracies are easily put back in.  We therefore work with
\bea
 \chi^{(N)}_{0,T}(q)\Big|_{z^0}  &=& 
q^{-1/24} \sum_{(m,n)} q^{{1\over 4}\,\svec m^T C\svec m} \prod_{a=1}^{\floor{N/2}} \qbin{m_a+n_a}{m_a}q\;=\;q^{-1/24}\qbin{N}{N/2}q\\
&=&q^{-1/24} \sum_{(m,n)} q^{{1\over 4}\,\svec m^T C\svec m} \prod_{a=1}^{K} \qbin{m_a+n_a}{m_a}q
\eea
%This is not a proper character but is convenient for our purposes. 
This is convenient since, from the lattice, this projected character contains all of the distinct eigenvalues (as Laurent polynomials) and removes the degeneracies. 
In the second formula we have taken an arbitrary integer $K>\floor{N/2}=L-1$ and extended the definitions of $\vec m$ and $\vec n$ by
\bea
n_a=0,\quad m_a=m_{L-1},\qquad a>L-1
\eea
with $C=T'_K$. 
Taking the limit $K\to\infty$ followed by $N\to\infty$, with $N$ even, we then obtain the projected vacuum fermionic character
\bea\label{xxxvac}
\chi_{0,T}(q)\Big|_{z^0}=q^{-1/24} \sum_{m_a=0\;\text{(mod 2)}\atop m_1\ge m_2\ge\ldots\ge 0}
 {q^{{1\over 4}\,\svec m^T C\svec m}\over (q)_{m_1}} \prod_{a=2}^{\infty}
  \qbin{{1\over 2}(m_{a-1}+m_{a+1})}{m_a}q
\eea
where $C$ is now the Cartan matrix of $T'_\infty=A_\infty$ and the sum is over asymptotically constant configurations $\vec m$ for which there exists  some finite $K>0$ such that $m_a=m_K$ whenever $a>K$.

\subsection{One-dimensional configurational sums}

Typically, finitized characters can be expressed as one-dimensional configurational sums, that is, via an energy statistic applied directly to RSOS paths. These one-dimensional configurational sums first appeared in the work of Baxter~\cite{BaxterBook82} but in the context of off-critical Corner Transfer Matrices (CTMs) where the elliptic nome $\hat{q}$ (\ref{cmx}) is related to the departure from criticality. By contrast, we work here only at criticality and $q$ is always the modular nome.

The one-dimensional configurational sums are defined by
\begin{equation}
X^{(N)}_{abc}(q)=\sum_{\{\sigma\}} q^{{1\over 2}\sum_{j=1}^Nj 
H(\sigma_{j-1},\sigma_j,\sigma_{j+1})},
\qquad \sigma_0=a,\ \  \sigma_N=b,\ \  \sigma_{N+1}=c
\end{equation}
where 
$H(\sigma_{j-1},\sigma_j,\sigma_{j+1})$
%=\half|\sigma_{j-1}-\sigma_{j+1}|=0,1$ 
is a local energy (energy density) and the sum is over all RSOS paths 
$\sigma=\{\sigma_0,\sigma_1,\ldots,\sigma_{N+1}\}$ on $A_L$ or $T_L$ with 
$\sigma_j\in\{1,2,\ldots,L\}$. 
These one-dimensional configurational sums satisfy mathematically 
powerful recursion relations in $N$ relating different boundary conditions  
specified by $a,b$ and $c$. In the vacuum sector with $N$ even, $a=b=1$ and $c=2$.

The form of the local energy function is not uniquely determined due to the possibility of local gauge transformations. 
Let us choose, however, the local energy function to be of the simple form
\bea
H(\sigma_{j-1},\sigma_j,\sigma_{j+1})=\begin{cases}
1,&(\sigma_{j-1},\sigma_j,\sigma_{j+1})=(1,1,1)\\
\half,&(\sigma_{j-1},\sigma_j,\sigma_{j+1})=(1,1,2)\ \mbox{or}\ (2,1,1)\\
1,&|\sigma_{j-1}-\sigma_{j+1}|=2\\
0,&\mbox{otherwise}
\end{cases}
\eea
It is then confirmed that, for both $A_L$ and $T_L$ paths, the energy statistic
\bea
E=-{c\over 24}+{1\over 2}\sum_{j=1}^Nj\, H(\sigma_{j-1},\sigma_j,\sigma_{j+1})\label{pathEnergy}
\eea
exactly reproduces the string energies (\ref{stringEnergy}) and hence
\bea
\chi_{0,A}^{(N)}(q)=q^{-c/24} X_{112}^{(N)}(q)
\eea
This relation has been checked directly in Mathematica~\cite{Wolfram} for all sizes up to to $N=26$ and confirms the equivalence of (\ref{stringEnergy}) and (\ref{pathEnergy}). This largest size involves 10.6 million paths.

\subsection{Combinatorial classification of states}

The combinatorial classification of states for the $A_5$ minimal model for paths of length $N=8$ is shown in Figures~8 and 9. In this case there are 14 such states with the finitized vacuum character
\bea
\chi_{0,A_5}^{(8)}(q)=q^{-1/30}(1+q^2+q^3+2q^4+q^5+2q^6+q^7+2q^8+q^9+q^{10}+q^{12})
\eea
The combinatorial classification of states for the $T_5$ XXX model for paths of length $N=6$ is shown in Figures~10 through 12. In this case there are 20 such states with the finitized vacuum character
\bea
\chi_{0,T}^{(6)}(q)=q^{-1/24}(1+q+2q^2+3q^3+3q^4+3q^5+3q^6+2q^7+q^8+q^9)
\eea
In both of these cases, there are 3 physical strips and we see that the patterns of 1-strings and 2-strings in these strips arising from the particle decompositions of the paths exactly match the patterns of eigenvalue zeros in the upper-half $u$-plane obtained by direct diagonalization of the commuting double row transfer matrices with $N$ faces along a row. 

The particular system sizes in these examples were chosen for convenience in illustrating the combinatorial classification of states. Many values for $L$ and $N$ have been checked, in particular considerably larger system sizes, and in all cases perfect agreement was found between the combinatorial picture in terms of paths and the numerical location of zeros of eigenvalues of the transfer matrices.

\begin{figure}[p]
\mbox{}\vspace{-.7in}\mbox{}
\bea
\mbox{}\hspace{-.5in}\mbox{}\raisebox{.5in}{1. $E=0$:\ }\quad\mbox{}\hspace{-1.in}\mbox{}\includegraphics[width=6.in]{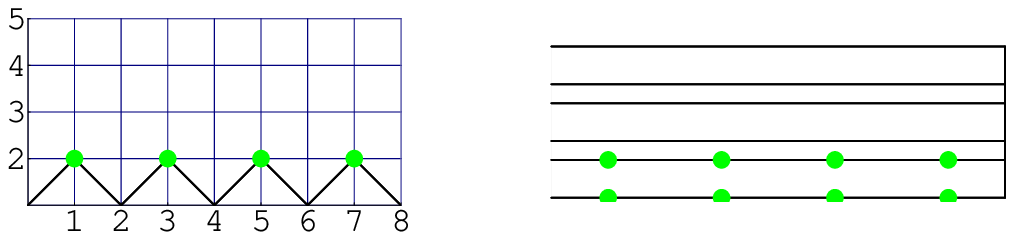}
\hspace{-.4in}
\includegraphics[width=1.6in]{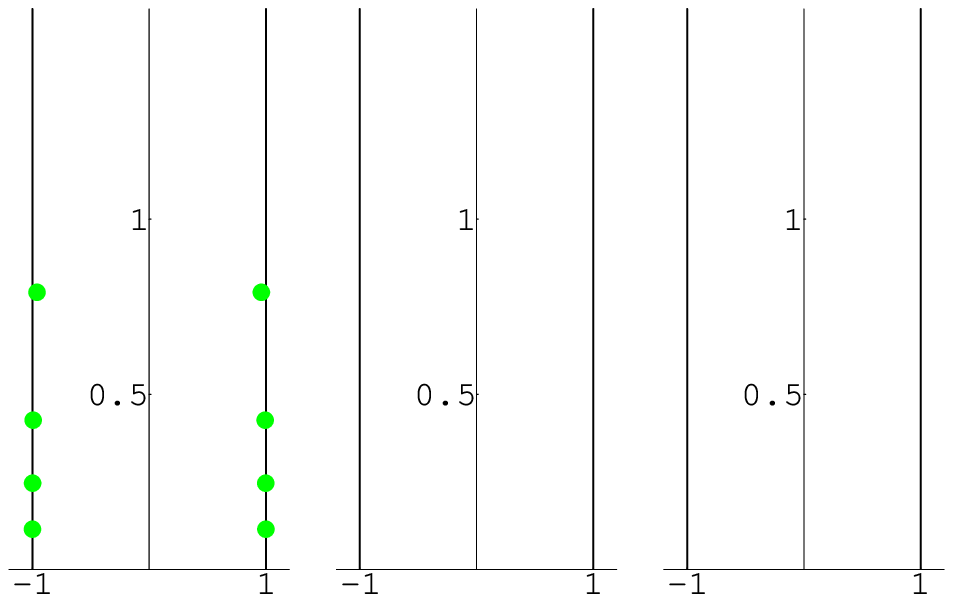}\nonumber\\[8pt]
\mbox{}\hspace{-.5in}\mbox{}\raisebox{.5in}{2. $E=2$:\ }\quad\mbox{}\hspace{-1.in}\mbox{}\includegraphics[width=6.in]{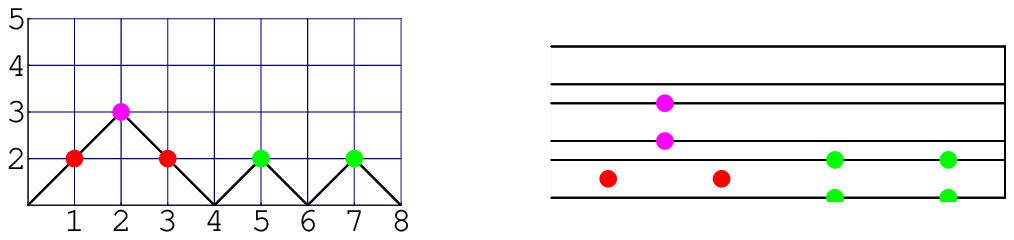}
\hspace{-.4in}
\includegraphics[width=1.6in]{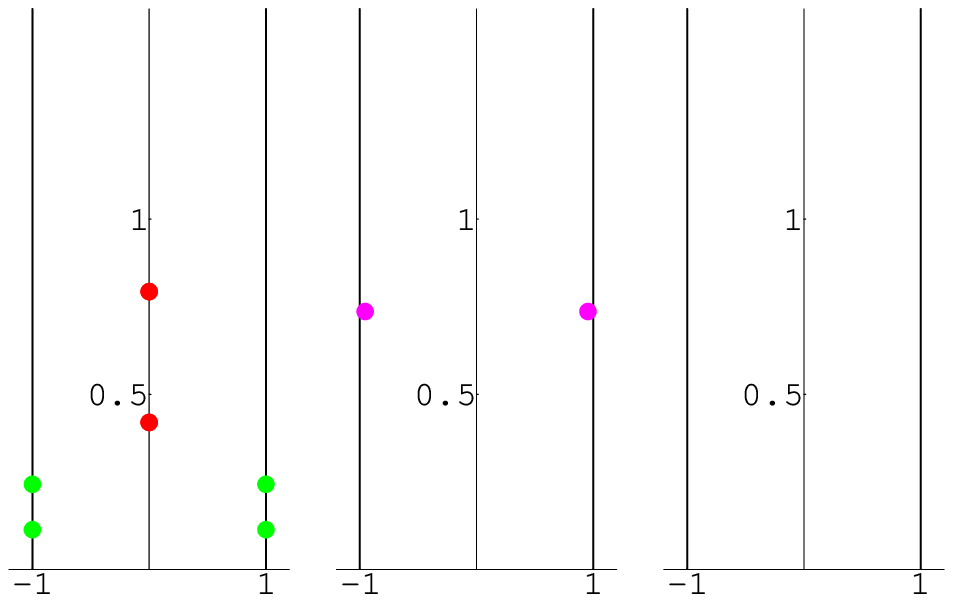}\nonumber\\[8pt]
\mbox{}\hspace{-.5in}\mbox{}\raisebox{.5in}{3. $E=3$:\ }\quad\mbox{}\hspace{-1.in}\mbox{}\includegraphics[width=6.in]{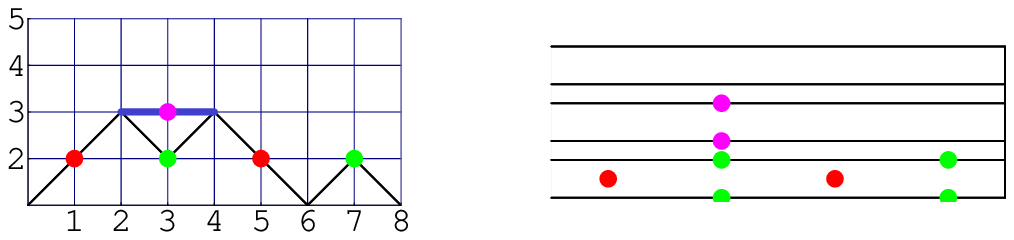}
\hspace{-.4in}
\includegraphics[width=1.6in]{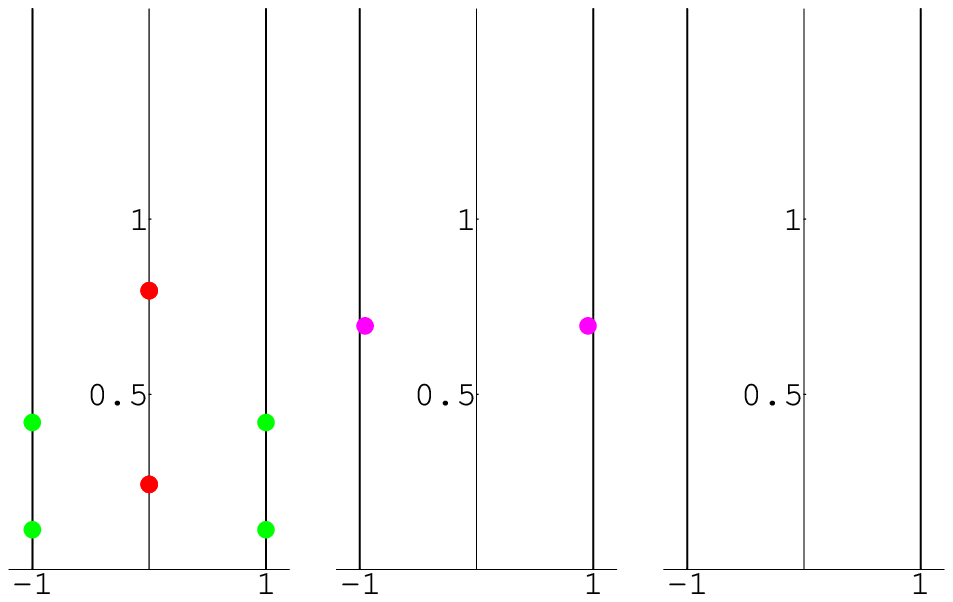}\nonumber\\[8pt]
\mbox{}\hspace{-.5in}\mbox{}\raisebox{.5in}{4. $E=4$:\ }\quad\mbox{}\hspace{-1.in}\mbox{}\includegraphics[width=6.in]{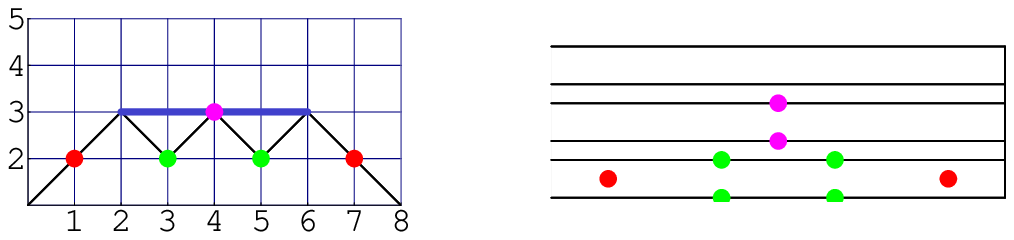}
\hspace{-.4in}
\includegraphics[width=1.6in]{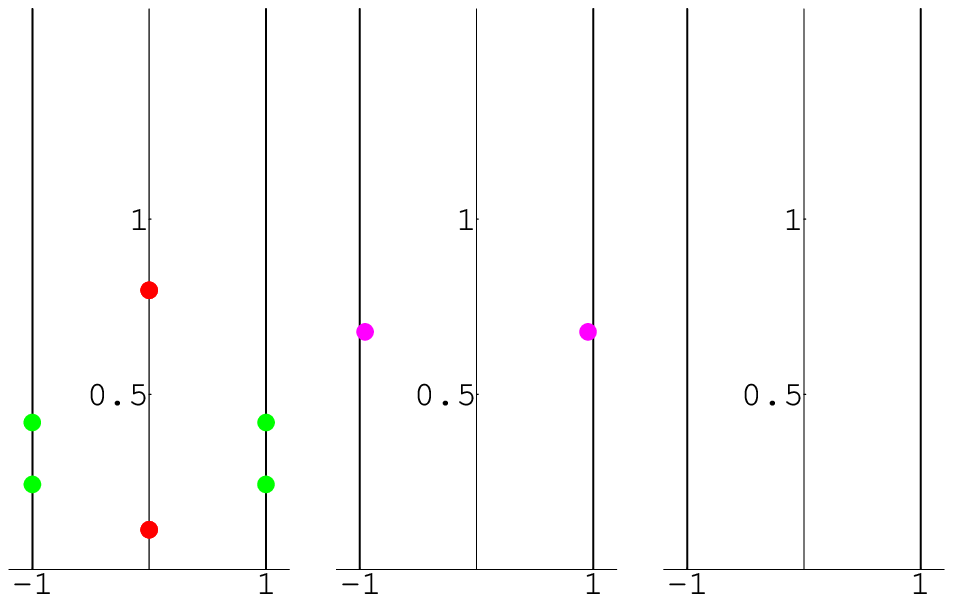}\nonumber\\[8pt]
\mbox{}\hspace{-.5in}\mbox{}\raisebox{.5in}{5. $E=4$:\ }\quad\mbox{}\hspace{-1.in}\mbox{}\includegraphics[width=6.in]{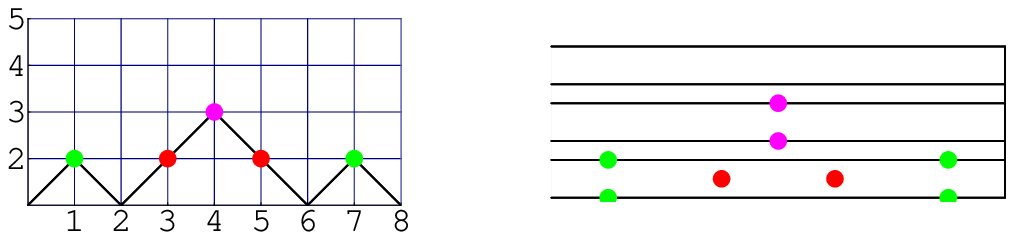}
\hspace{-.4in}
\includegraphics[width=1.6in]{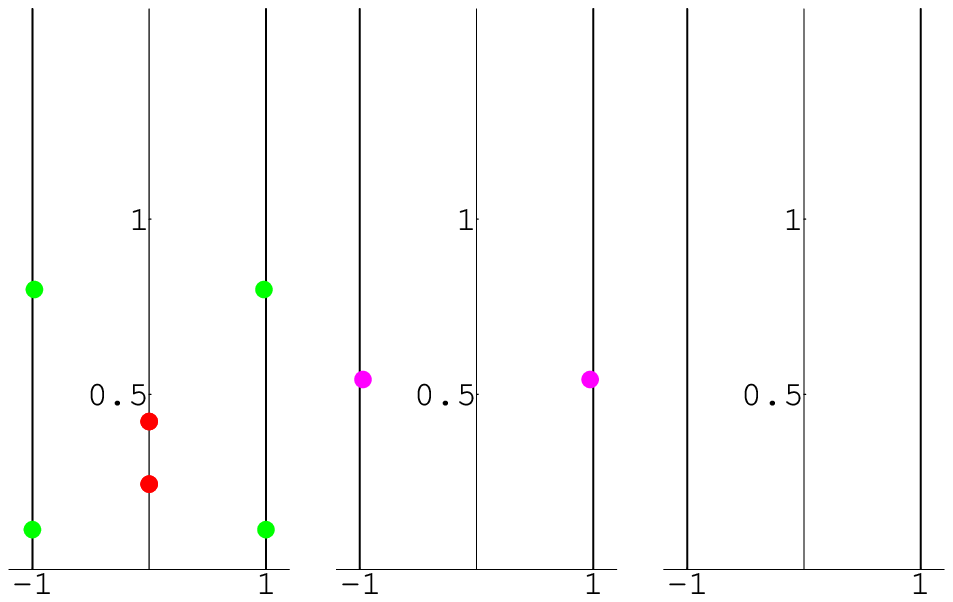}\nonumber\\[8pt]
\mbox{}\hspace{-.5in}\mbox{}\raisebox{.5in}{6. $E=5$:\ }\quad\mbox{}\hspace{-1.in}\mbox{}\includegraphics[width=6.in]{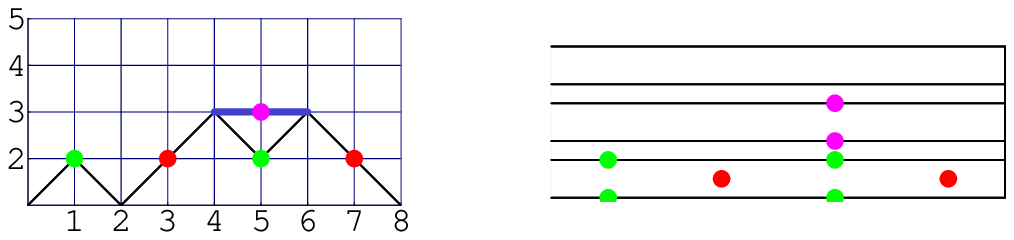}
\hspace{-.4in}
\includegraphics[width=1.6in]{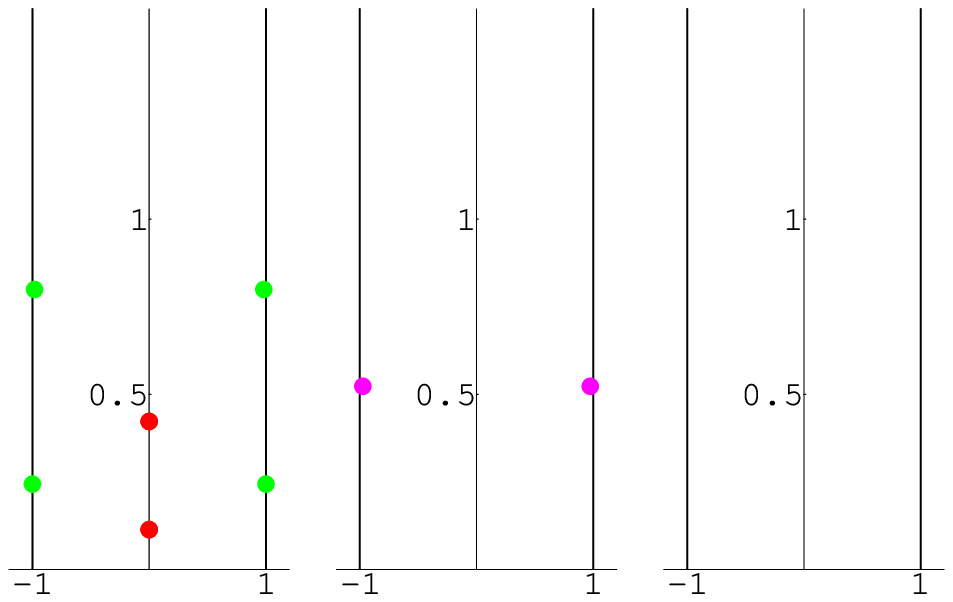}\nonumber\\[8pt]
\mbox{}\hspace{-.5in}\mbox{}\raisebox{.5in}{7. $E=6$:\ }\quad\mbox{}\hspace{-1.in}\mbox{}\includegraphics[width=6.in]{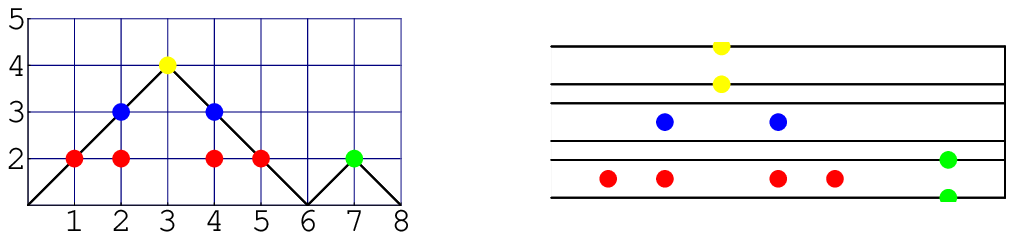}
\hspace{-.4in}
\includegraphics[width=1.6in]{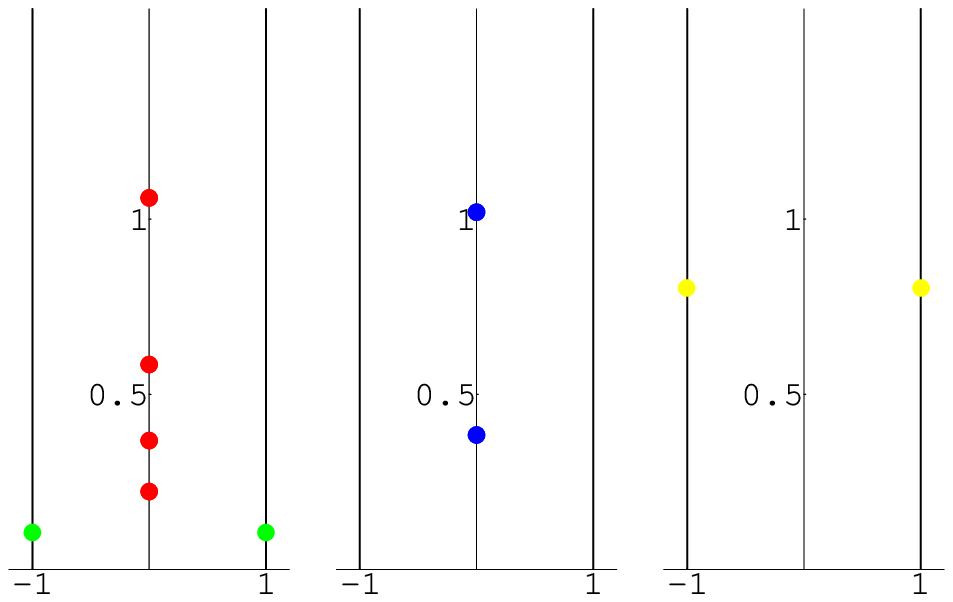}\nonumber
\eea
\caption{Eigenvalues 1 to 7 for $A_5$ with $N=8$ showing the energies (omitting the $-c/24$ term), paths, predicted patterns of zeros and actual patterns of zeros (rotated through 90 degrees) from numerical diagonalization of transfer matrices. Only the relative order of 1- and 2-strings in each strip is relevant. The relative order across strips is not relevant.}
\end{figure}
\begin{figure}[p]
\mbox{}\vspace{-.7in}\mbox{}
\bea
\mbox{}\hspace{-.5in}\mbox{}\raisebox{.5in}{8. $E=6$:\ }\quad\mbox{}\hspace{-1.in}\mbox{}\includegraphics[width=6.in]{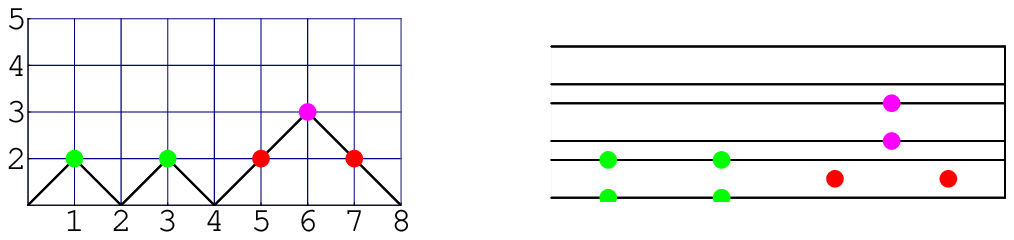}
\hspace{-.4in}
\includegraphics[width=1.6in]{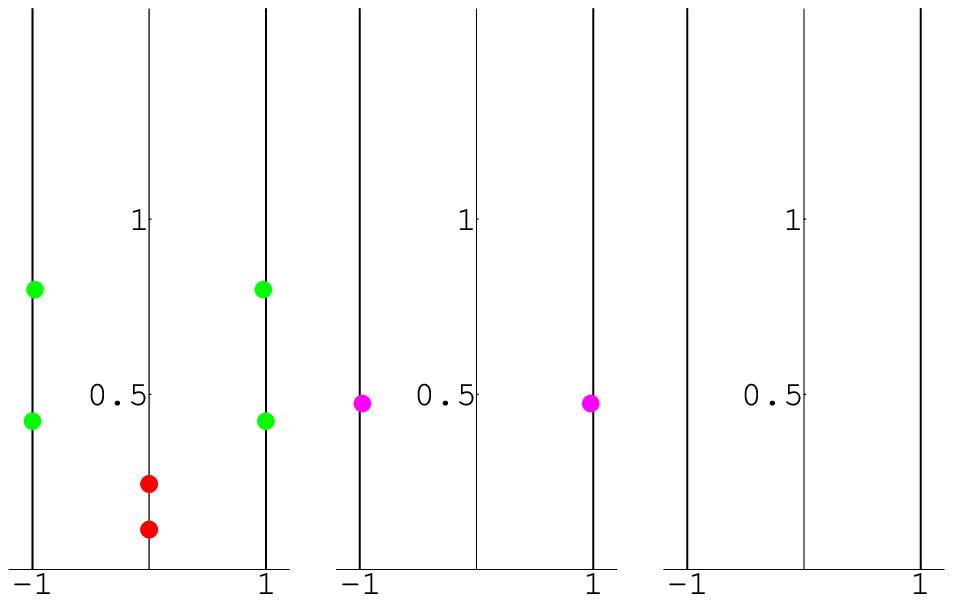}\nonumber\\[8pt]
\mbox{}\hspace{-.5in}\mbox{}\raisebox{.5in}{9. $E=7$:\ }\quad\mbox{}\hspace{-1.in}\mbox{}\includegraphics[width=6.in]{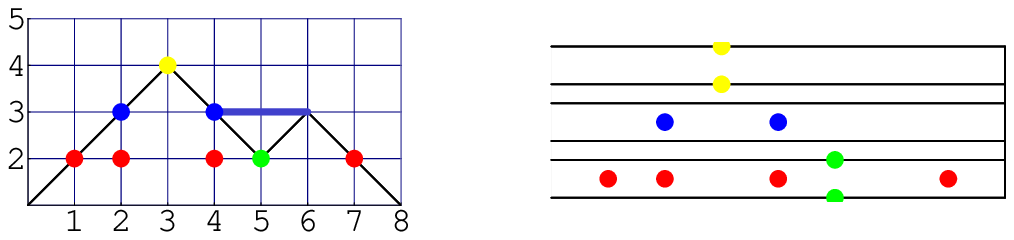}
\hspace{-.4in}
\includegraphics[width=1.6in]{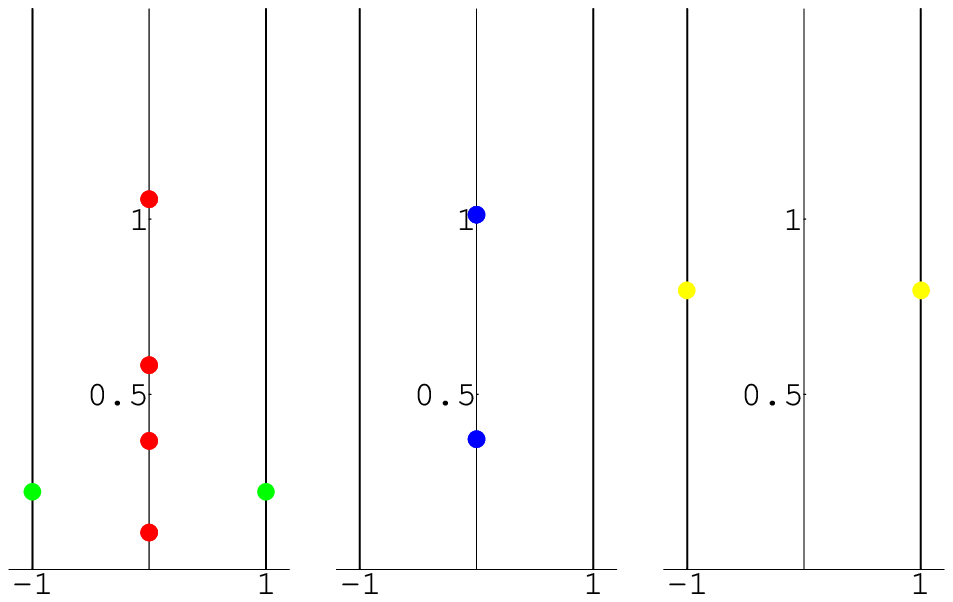}\nonumber\\[8pt]
\mbox{}\hspace{-.5in}\mbox{}\raisebox{.5in}{10. $E=8$:\ }\quad\mbox{}\hspace{-1.in}\mbox{}\includegraphics[width=6.in]{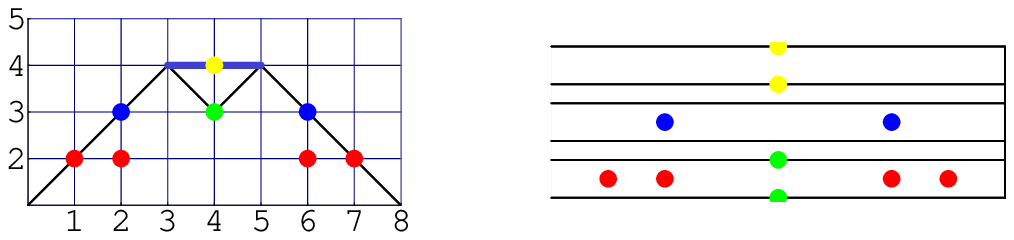}
\hspace{-.4in}
\includegraphics[width=1.6in]{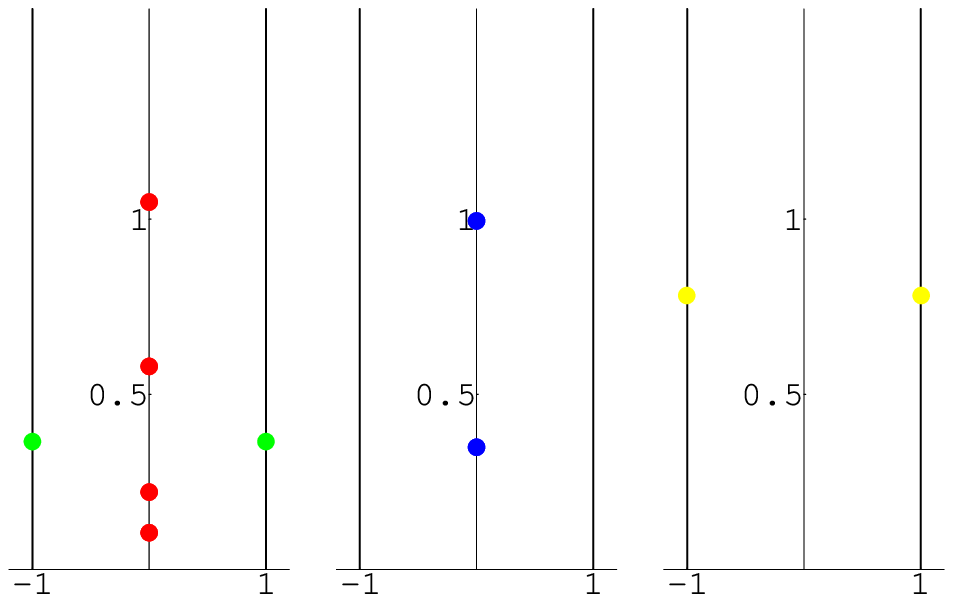}\nonumber\\[8pt]
\mbox{}\hspace{-.5in}\mbox{}\raisebox{.5in}{11. $E=8$:\ }\quad\mbox{}\hspace{-1.in}\mbox{}\includegraphics[width=6.in]{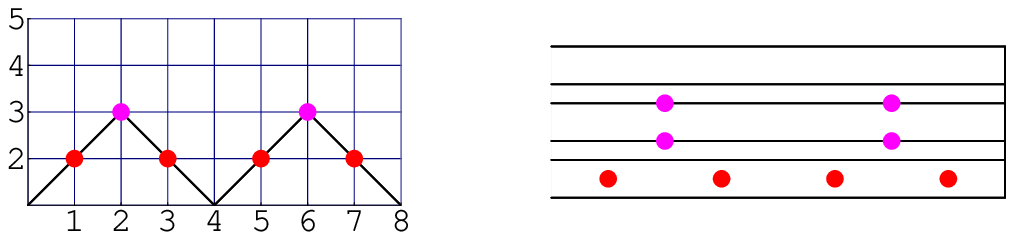}
\hspace{-.4in}
\includegraphics[width=1.6in]{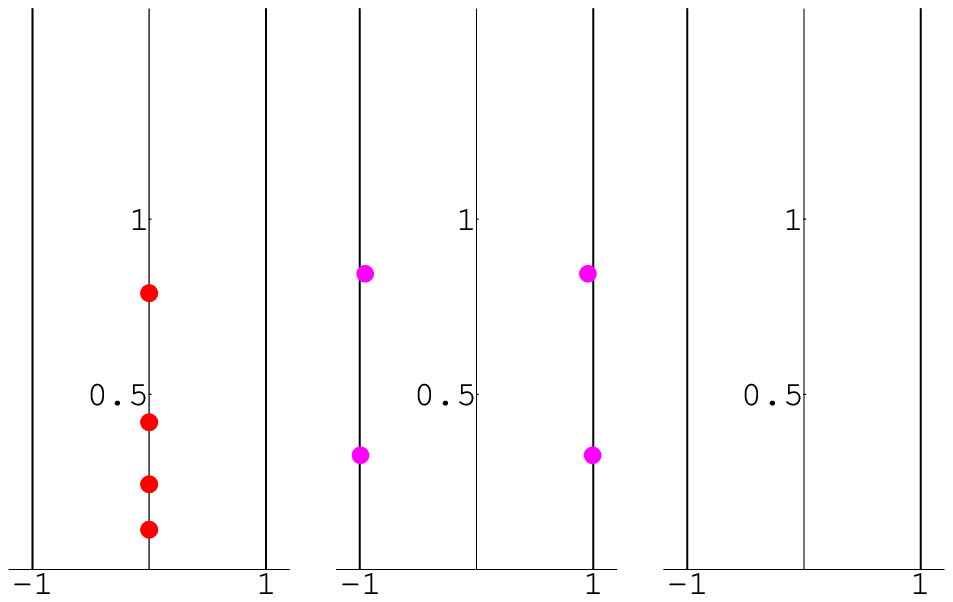}\nonumber\\[8pt]
\mbox{}\hspace{-.5in}\mbox{}\raisebox{.5in}{12. $E=9$:\ }\quad\mbox{}\hspace{-1.in}\mbox{}\includegraphics[width=6.in]{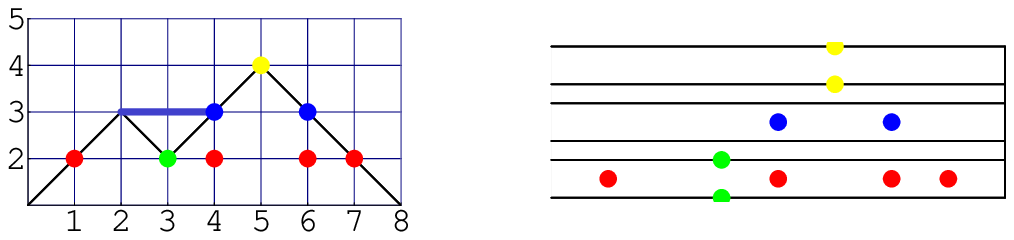}
\hspace{-.4in}
\includegraphics[width=1.6in]{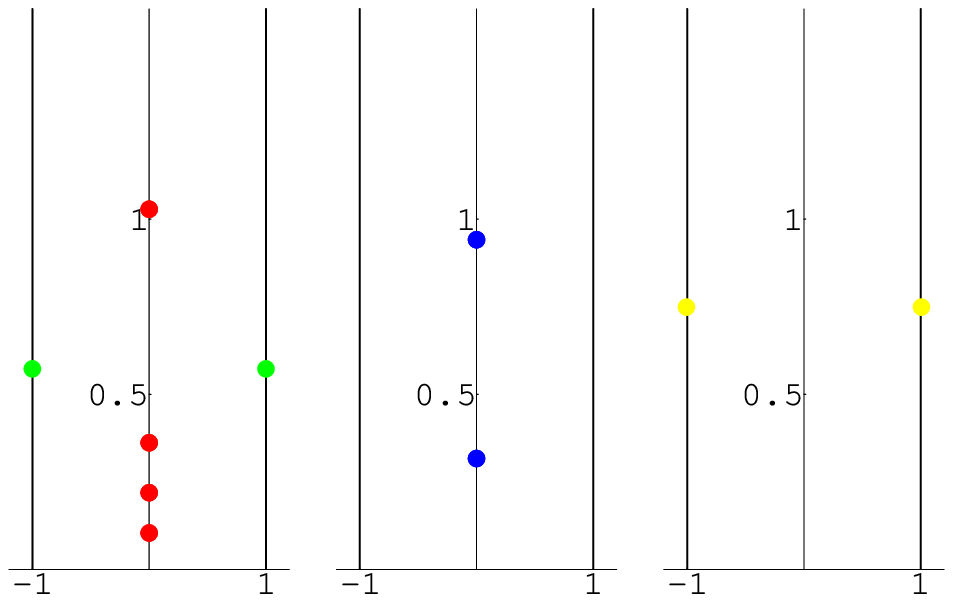}\nonumber\\[8pt]
\mbox{}\hspace{-.5in}\mbox{}\raisebox{.5in}{13. $E=10$:\ }\quad\mbox{}\hspace{-1.in}\mbox{}\includegraphics[width=6.in]{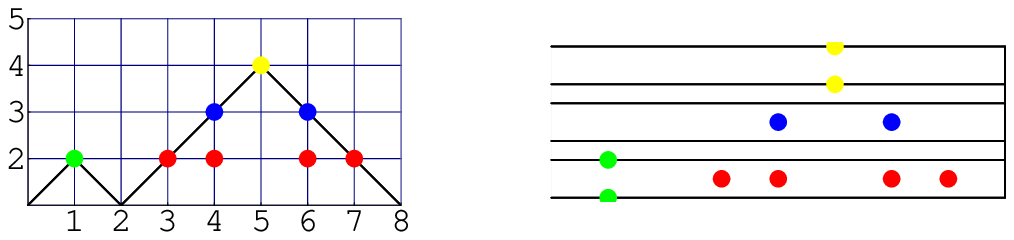}
\hspace{-.4in}
\includegraphics[width=1.6in]{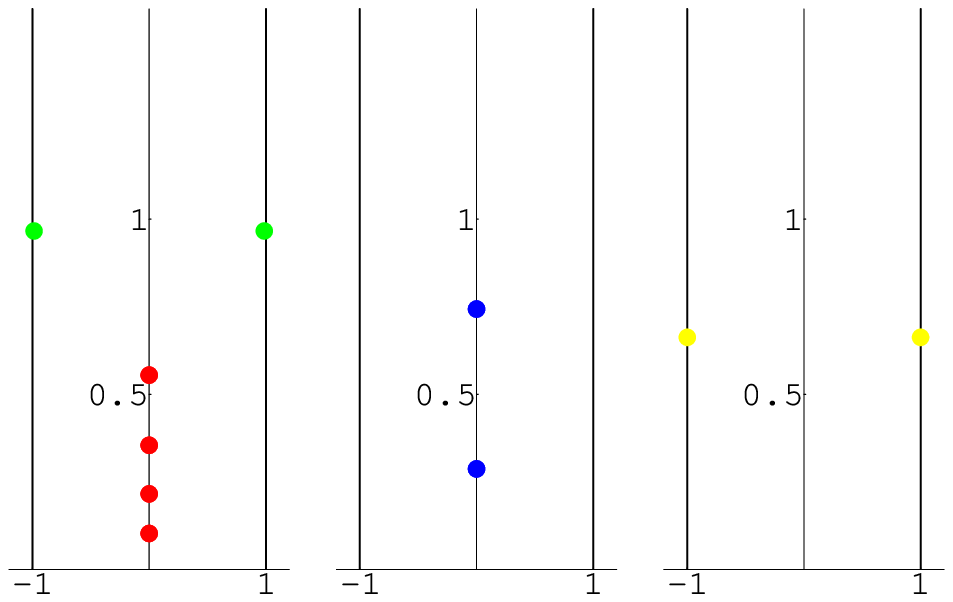}\nonumber\\[8pt]
\mbox{}\hspace{-.5in}\mbox{}\raisebox{.5in}{14. $E=12$:\ }\quad\mbox{}\hspace{-1.in}\mbox{}\includegraphics[width=6.in]{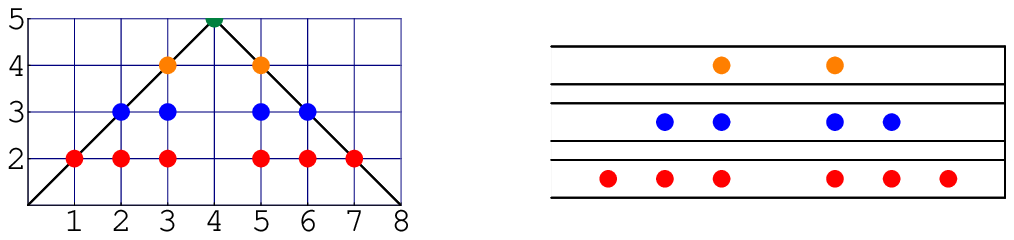}
\hspace{-.4in}
\includegraphics[width=1.6in]{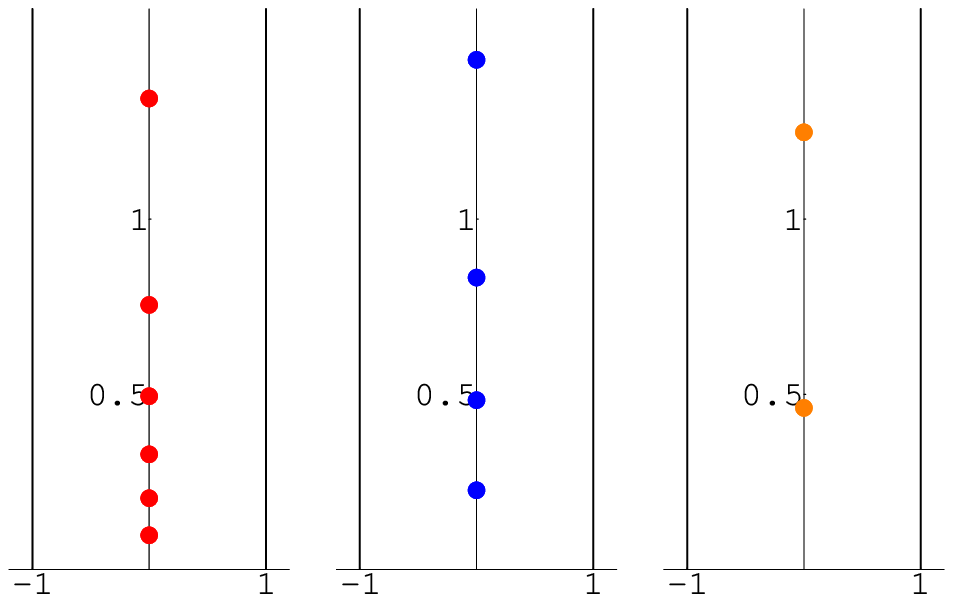}\nonumber
\eea
\caption{This is a continuation of Figure~8 showing eigenvalues 8 to 14 for $A_5$ with $N=8$.}
\end{figure}

\begin{figure}[p]
\mbox{}\vspace{-.4in}\mbox{}
\bea
\raisebox{.5in}{1. $E=0$\qquad}\raisebox{.1in}{\includegraphics[width=2in]{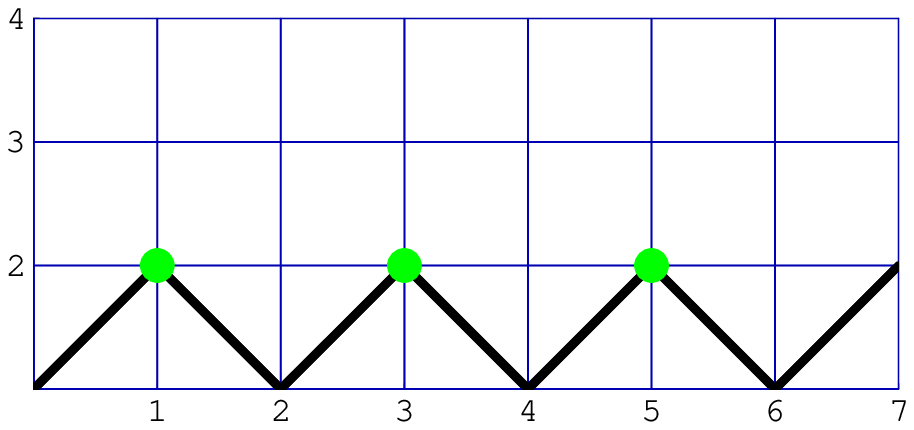}}\qquad\quad
\includegraphics[width=3in]{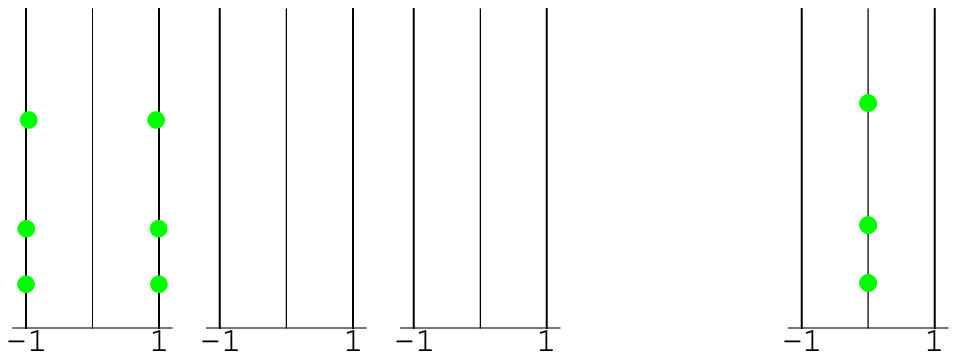}\nonumber\\[8pt]
\raisebox{.5in}{2. $E=1$\qquad}\raisebox{.1in}{\includegraphics[width=2in]{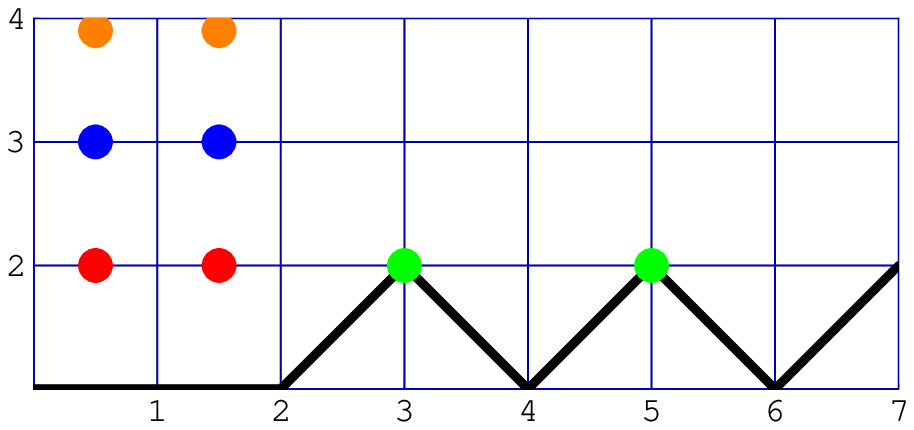}}\qquad\quad
\includegraphics[width=3in]{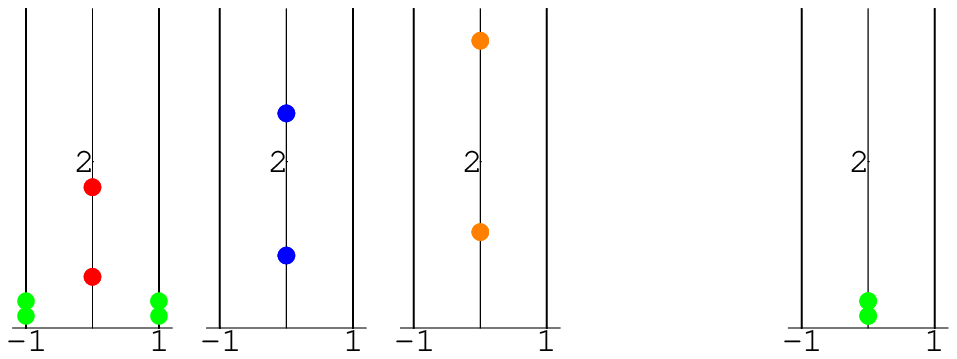}\nonumber\\[8pt]
\raisebox{.5in}{3. $E=2$\qquad}\raisebox{.1in}{\includegraphics[width=2in]{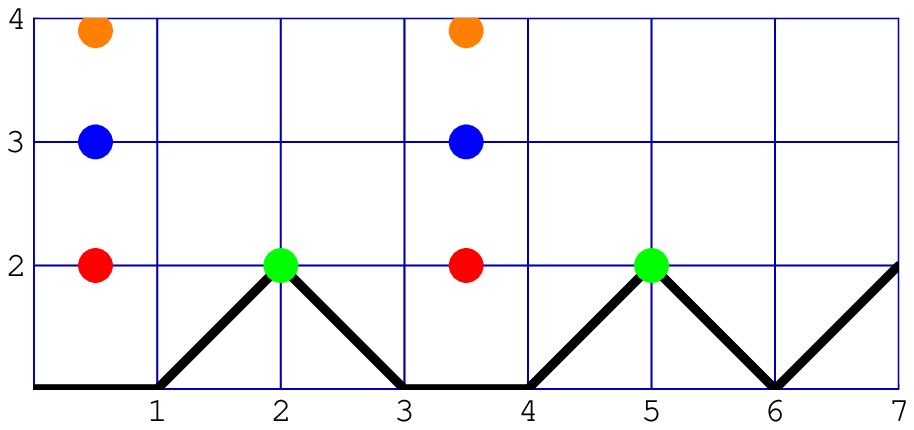}}\qquad\quad
\includegraphics[width=3in]{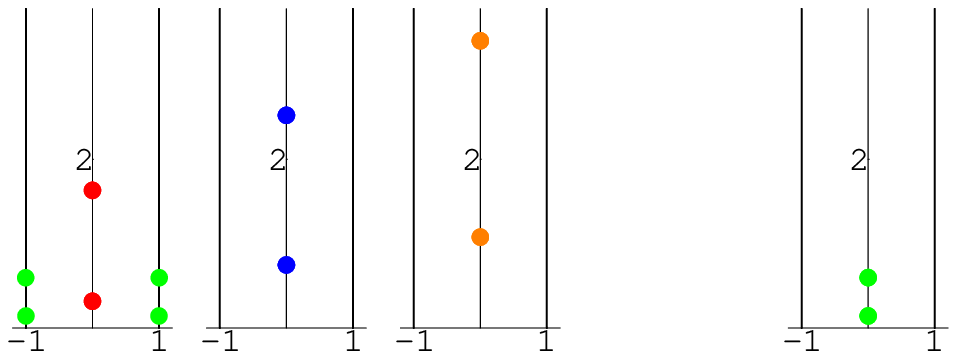}\nonumber\\[8pt]
\raisebox{.5in}{4. $E=2$\qquad}\raisebox{.1in}{\includegraphics[width=2in]{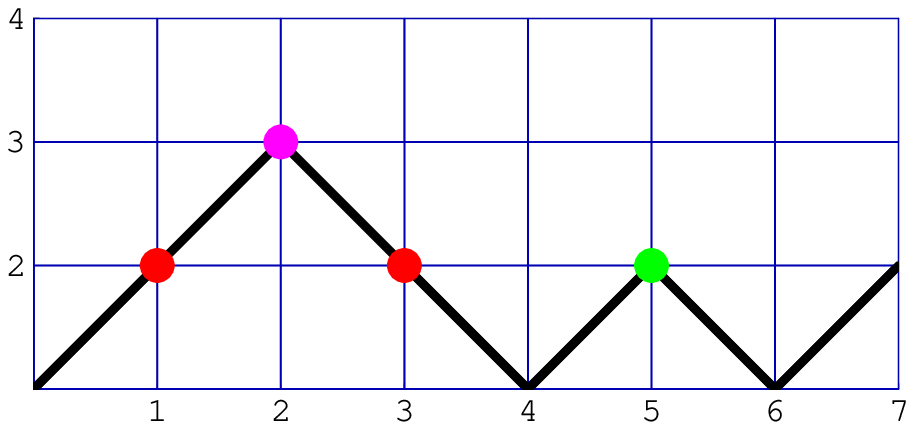}}\qquad\quad
\includegraphics[width=3in]{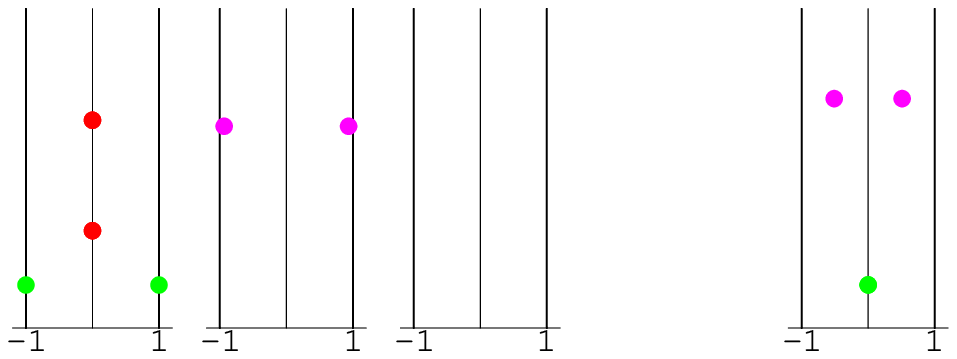}\nonumber\\[8pt]
\raisebox{.5in}{5. $E=3$\qquad}\raisebox{.1in}{\includegraphics[width=2in]{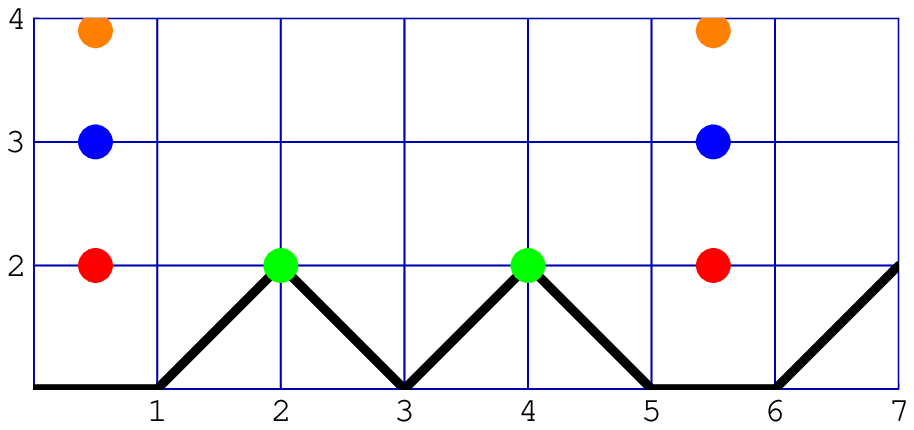}}\qquad\quad
\includegraphics[width=3in]{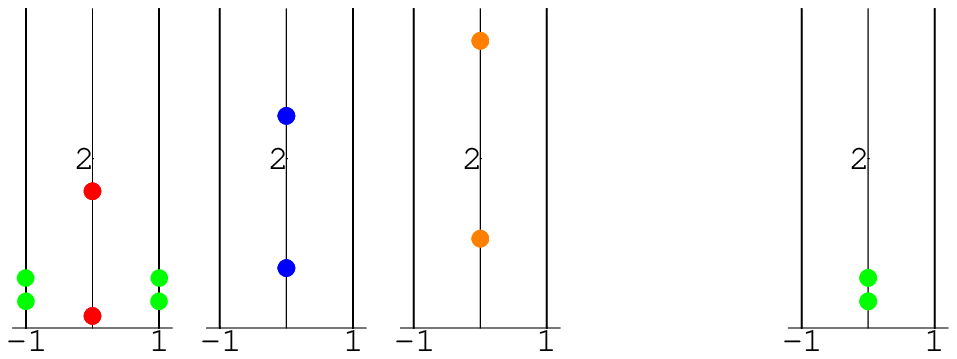}\nonumber\\[8pt]
\raisebox{.5in}{6. $E=3$\qquad}\raisebox{.1in}{\includegraphics[width=2in]{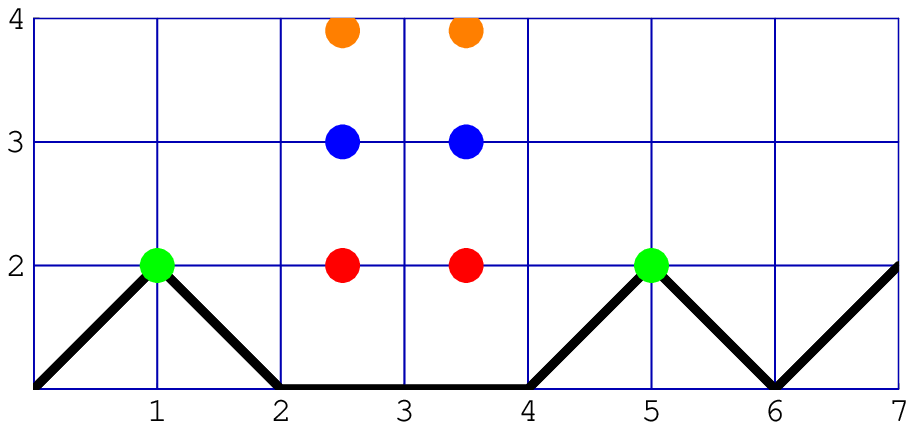}}\qquad\quad
\includegraphics[width=3in]{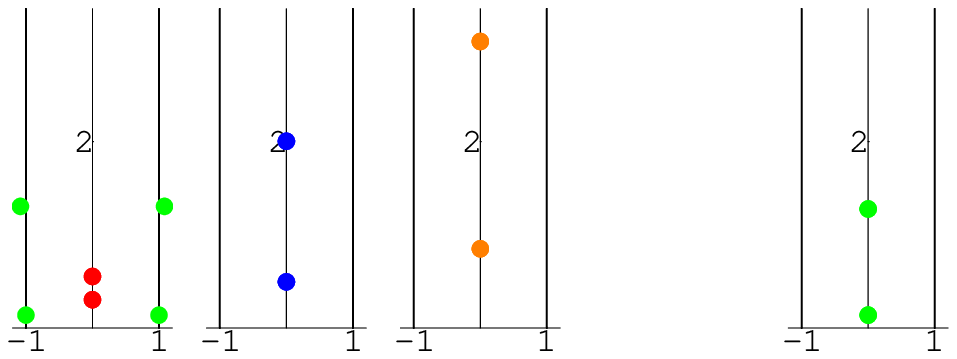}\nonumber\\[8pt]
\raisebox{.5in}{7. $E=3$\qquad}\raisebox{.1in}{\includegraphics[width=2in]{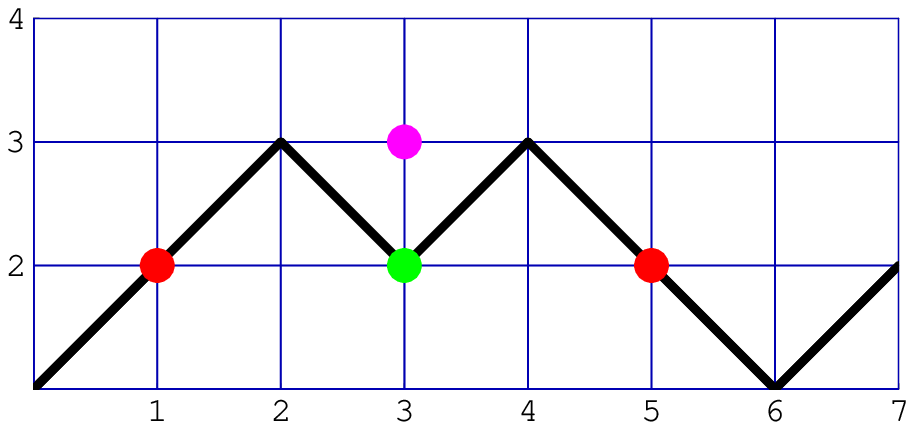}}\qquad\quad
\includegraphics[width=3in]{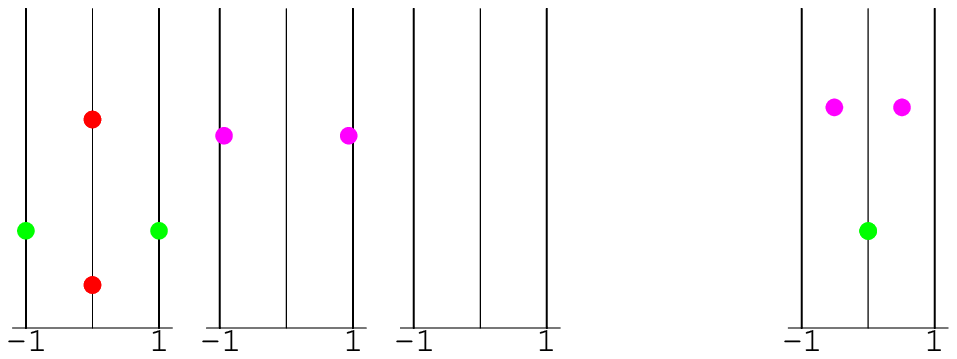}\nonumber
\eea
\caption{Eigenvalues 1 to 7 for the XXX model with $N=6$ showing the energies (omitting the $-c/24$ term), paths and actual patterns of zeros (rotated through 90 degrees) from numerical diagonalization of transfer matrices. Only the relative order of 1- and 2-strings in each strip is relevant. The last column shows the numerical patterns of zeros for the eigenvalues of the auxiliary matrix $Q$. The $Q$ eigenvalue has a $j$-string at the same relative position for each 2-string in strip $j$ of the transfer matrix eigenvalues. The path out to $j=7$ is needed to apply (2.33).\label{fig:wzw1}}
\end{figure}
\begin{figure}[p]
\mbox{}\vspace{-.4in}\mbox{}
\bea
\raisebox{.5in}{8. $E=4$\qquad}\raisebox{.1in}{\includegraphics[width=2in]{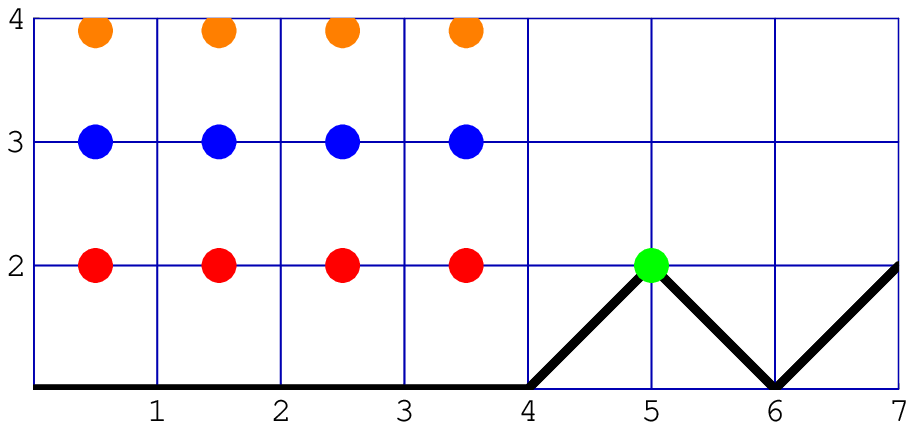}}\qquad\quad
\includegraphics[width=3in]{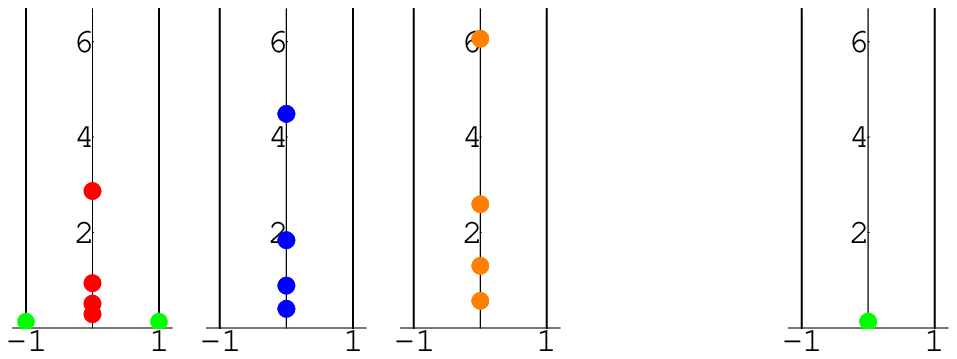}\nonumber\\[8pt]
\raisebox{.5in}{9. $E=4$\qquad}\raisebox{.1in}{\includegraphics[width=2in]{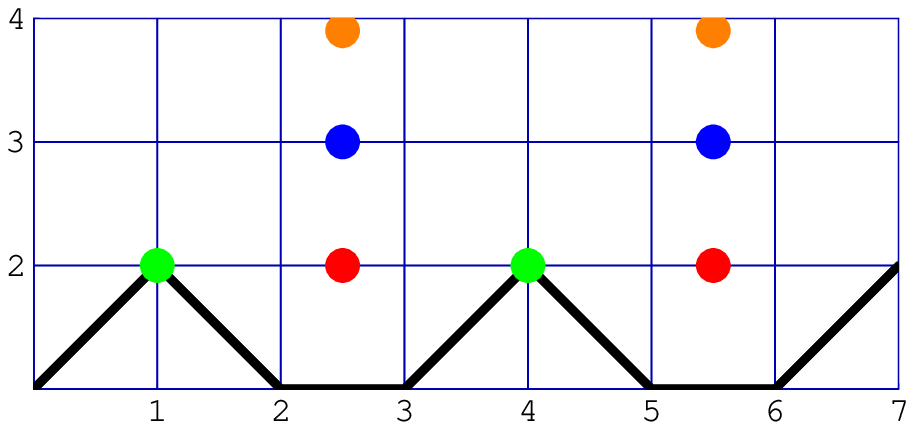}}\qquad\quad
\includegraphics[width=3in]{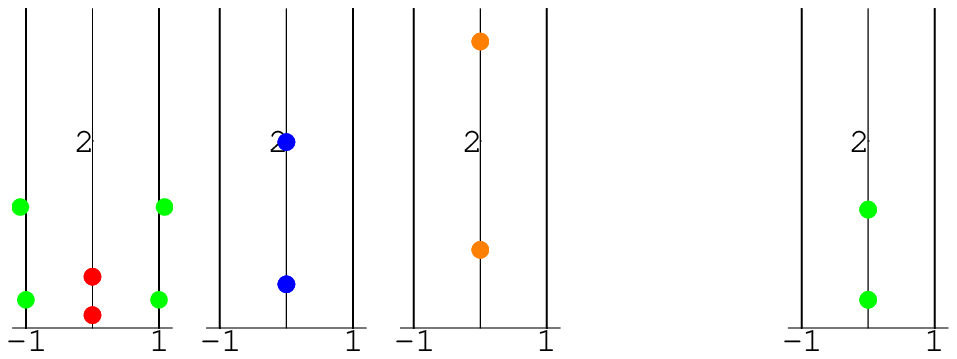}\nonumber\\[8pt]
\raisebox{.5in}{10. $E=4$\qquad}\raisebox{.1in}{\includegraphics[width=2in]{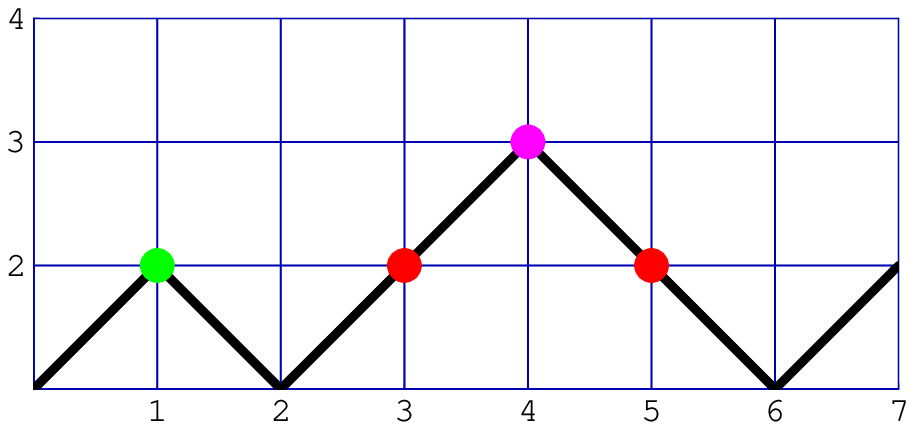}}\qquad\quad
\includegraphics[width=3in]{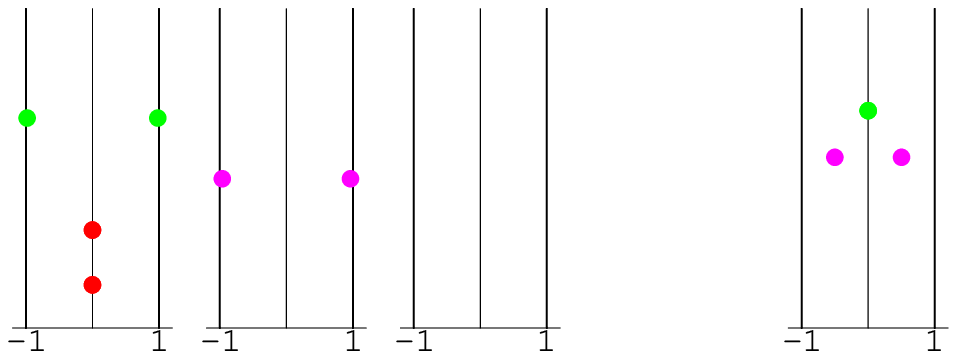}\nonumber\\[8pt]
\raisebox{.5in}{11. $E=5$\qquad}\raisebox{.1in}{\includegraphics[width=2in]{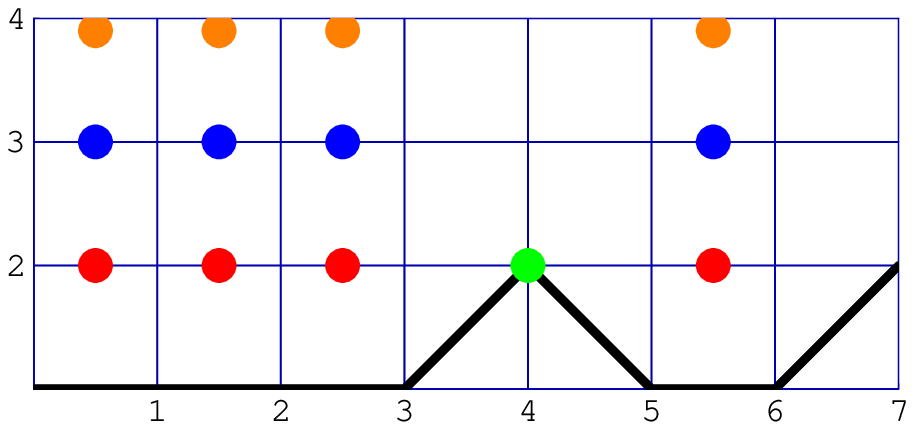}}\qquad\quad
\includegraphics[width=3in]{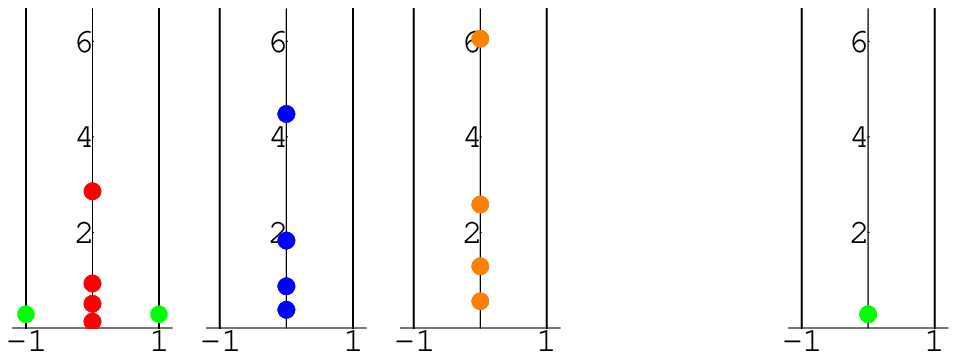}\nonumber\\[8pt]
\raisebox{.5in}{12. $E=5$\qquad}\raisebox{.1in}{\includegraphics[width=2in]{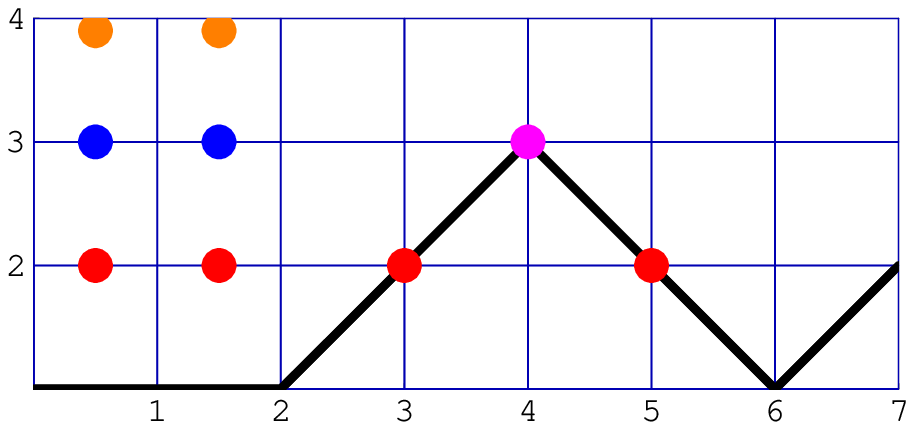}}\qquad\quad
\includegraphics[width=3in]{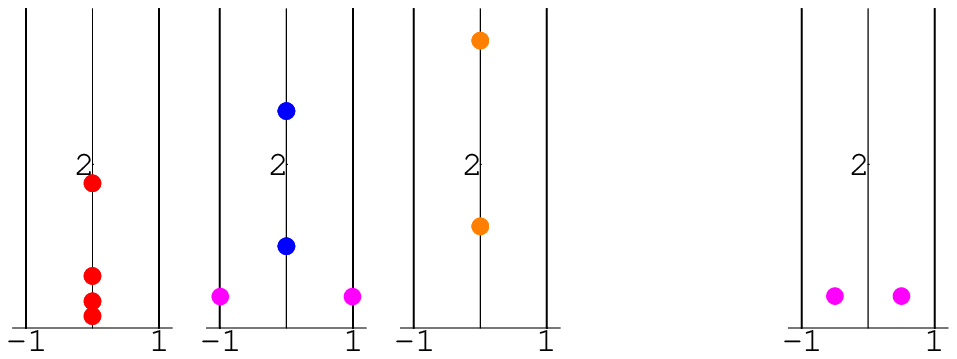}\nonumber\\[8pt]
\raisebox{.5in}{13. $E=5$\qquad}\raisebox{.1in}{\includegraphics[width=2in]{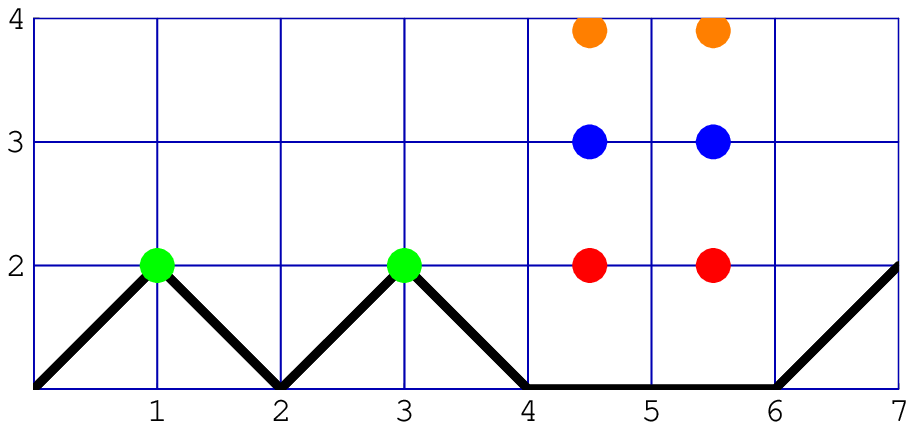}}\qquad\quad
\includegraphics[width=3in]{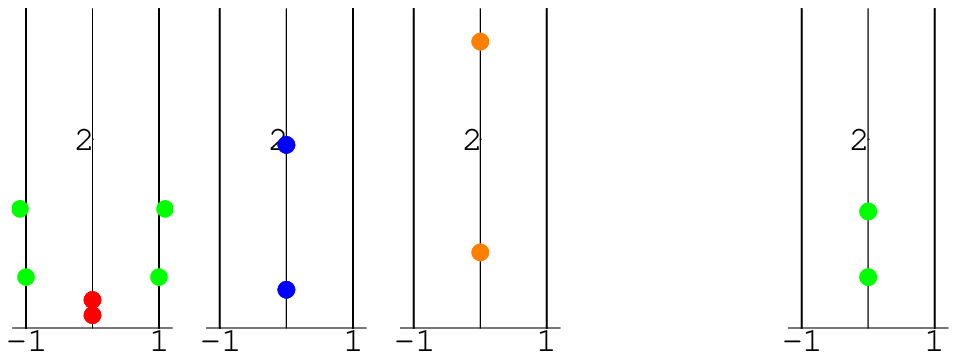}\nonumber\\[8pt]
\raisebox{.5in}{14. $E=6$\qquad}\raisebox{.1in}{\includegraphics[width=2in]{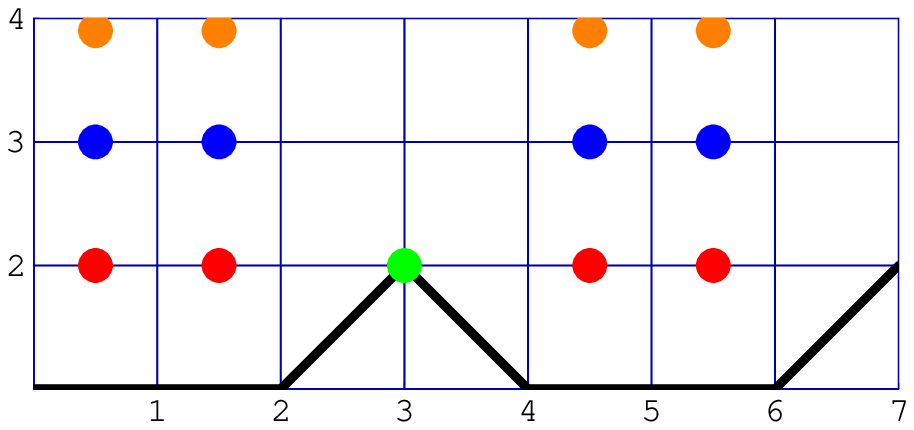}}\qquad\quad
\includegraphics[width=3in]{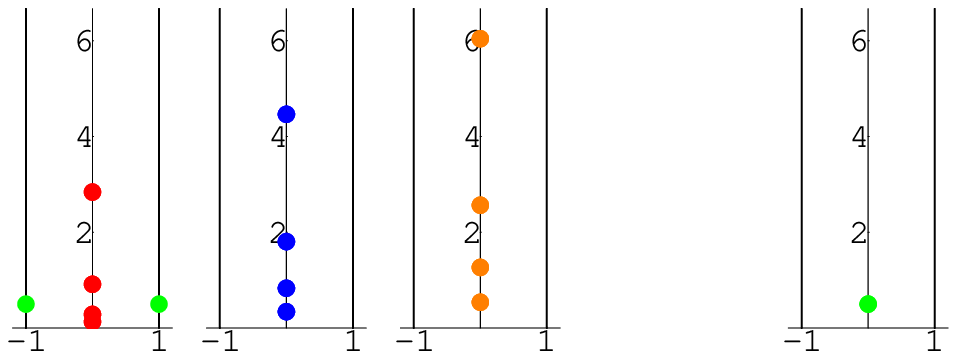}\nonumber
\eea
\caption{This is a continuation of Figure~10 showing eigenvalues 8 to 14 for the XXX model with $N=6$.\label{fig:wzw2}}
\end{figure}
\begin{figure}[p]
\mbox{}\vspace{-.4in}\mbox{}
\bea
\raisebox{.5in}{15. $E=6$\qquad}\raisebox{.1in}{\includegraphics[width=2in]{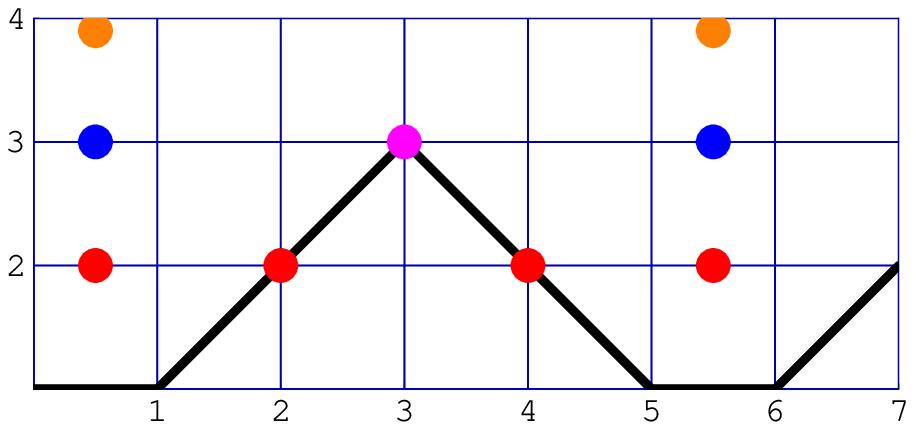}}\qquad\quad
\includegraphics[width=3in]{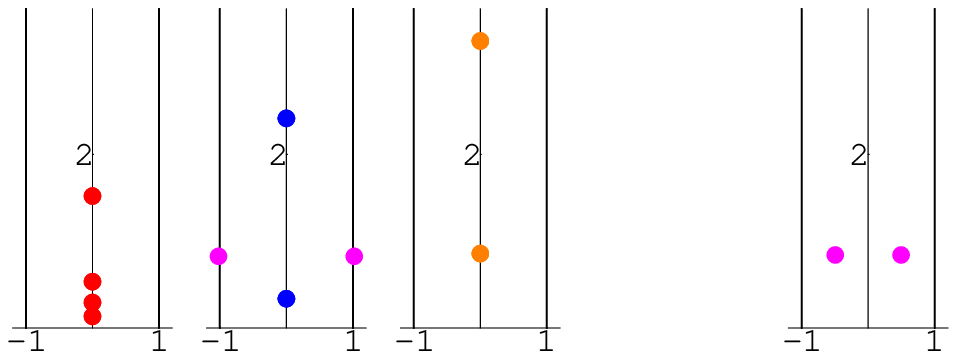}\nonumber\\[8pt]
\raisebox{.5in}{16. $E=6$\qquad}\raisebox{.1in}{\includegraphics[width=2in]{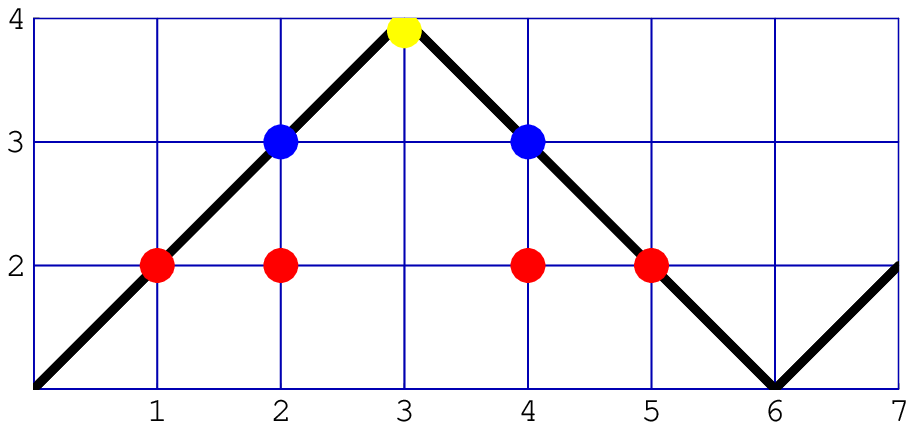}}\qquad\quad
\includegraphics[width=3in]{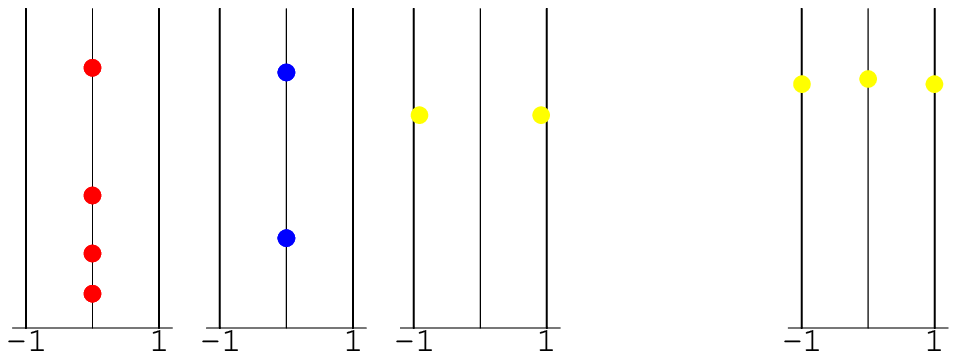}\nonumber\\[8pt]
\raisebox{.5in}{17. $E=7$\qquad}\raisebox{.1in}{\includegraphics[width=2in]{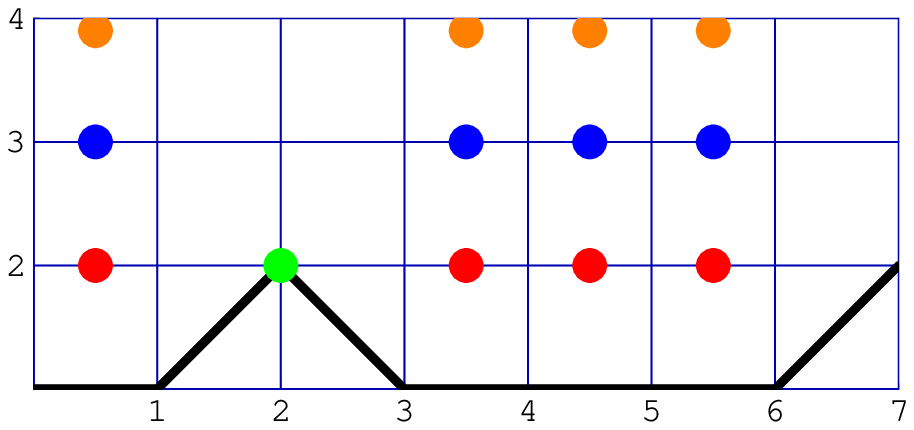}}\qquad\quad
\includegraphics[width=3in]{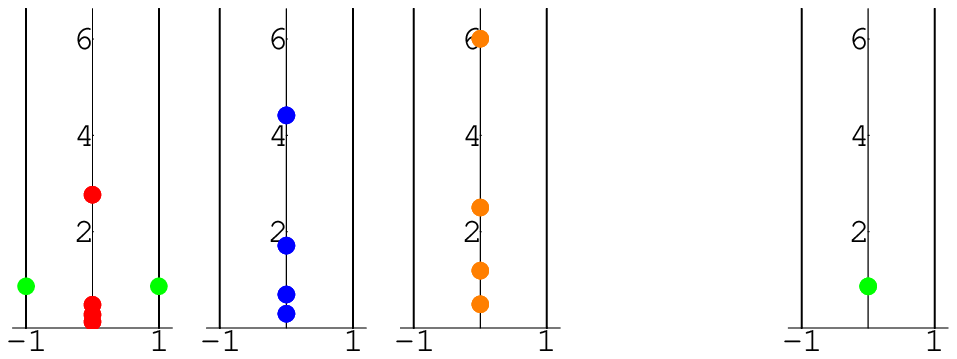}\nonumber\\[8pt]
\raisebox{.5in}{18. $E=7$\qquad}\raisebox{.1in}{\includegraphics[width=2in]{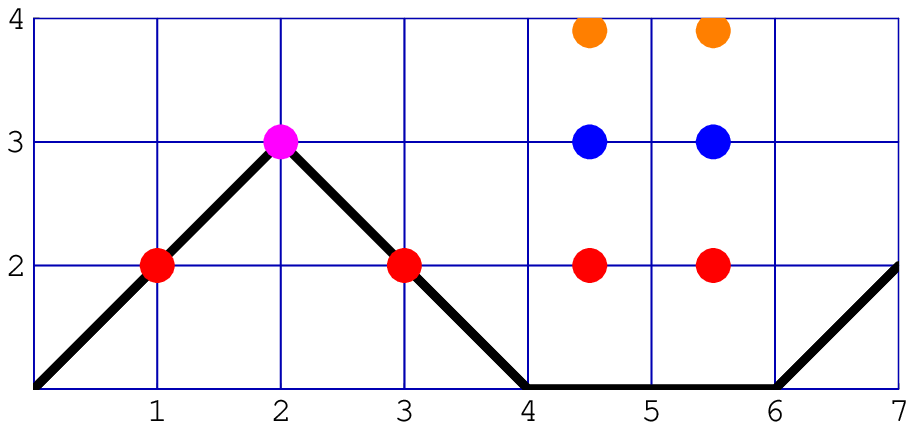}}\qquad\quad
\includegraphics[width=3in]{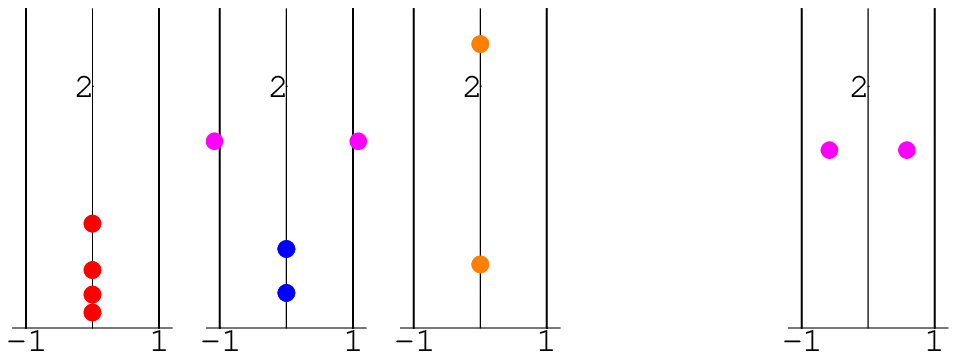}\nonumber\\[8pt]
\raisebox{.5in}{19. $E=8$\qquad}\raisebox{.1in}{\includegraphics[width=2in]{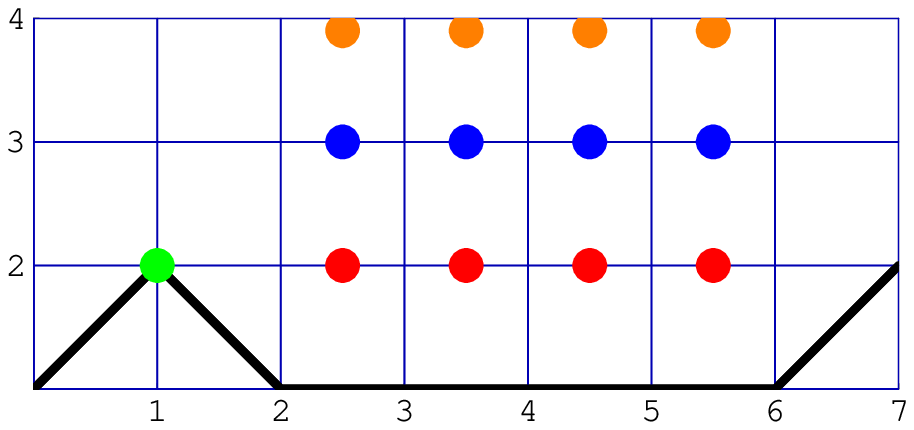}}\qquad\quad
\includegraphics[width=3in]{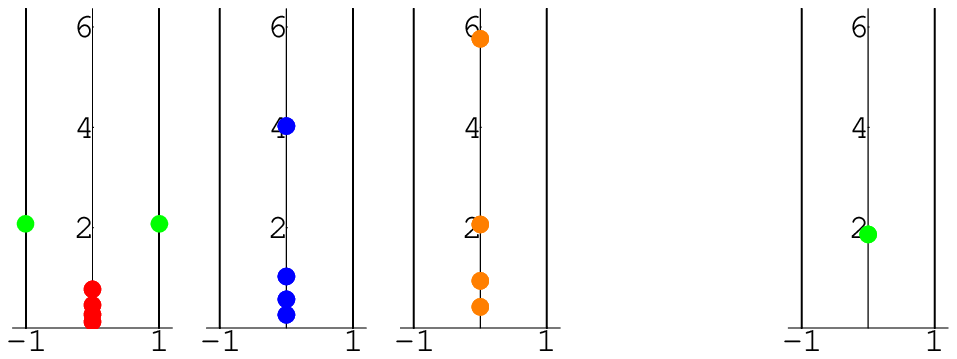}\nonumber\\[8pt]
\raisebox{.5in}{20. $E=9$\qquad}\raisebox{.1in}{\includegraphics[width=2in]{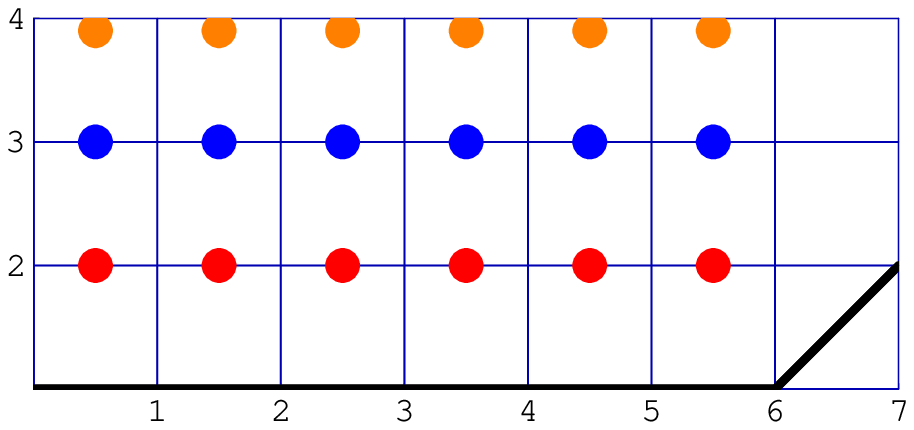}}\qquad\quad
\includegraphics[width=3in]{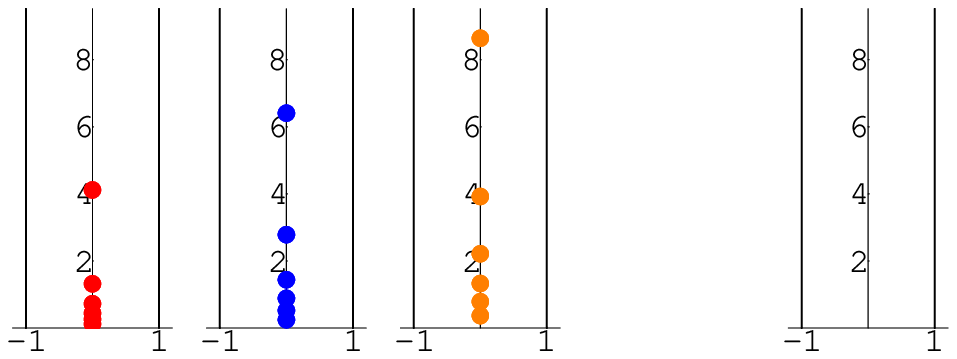}\nonumber
\eea
\caption{This is a continuation of Figures~10 and 11 showing eigenvalues 15 to 20 for the XXX model with $N=6$.\label{fig:wzw3}}
\end{figure}

\subsection{Bethe roots and transfer matrix 2-strings for XXX model}

In Figures 10 through 12, relevant to the XXX model, we have also shown the eigenvalue zeros of Baxter's auxiliary matrix 
$\vec Q(u)$ as in (\ref{tq}) and (\ref{Qop}). These zeros are precisely the Bethe ansatz roots. The transfer matrices $\vec D^q(u)$ are completely determined by knowledge of the matrix $\vec Q(u)$ through Baxter's $T$-$Q$ relation (\ref{tq}) and the fusion hierarchy.

It is readily confirmed, in this example, that an eigenvalue $Q(u)$ has a $j$-string at the same relative position as each 2-string in strip $j$ of the transfer matrix eigenvalues with $j=1,2,\ldots,N/2$. 
Identifying $S=n_0/2$ as the total spin, we see that there are $5, 9, 5, 1$ eigenvalues with total spin $S=0,1,2,3$ respectively in the $S_z=0$ sector.
% in agreement with the tabulated data in Appendix~A. 
The zero content of an eigenvalue $Q(u)$ thus determines the complete particle content $n_a$ with $a=0,1,\ldots,N/2-1$. The dual-particle content then follows from the $(m,n)$ system (\ref{Tmnsystem}). Again we have verified these observations for many different system sizes $N$. We further conjecture that these observations hold true for all $N$ and that in the limit $N\to\infty$ the locations of the $j$-strings of $Q(u)$ exactly coincide with the locations of the 2-strings of the transfer matrices in strip $j$.

%%%%%%%%%%%%%%%%%%%%%%
\section{Unitary Minimal RSOS Lattice Models}
\label{sec:minimal}

The bulk face and boundary triangle Boltzmann weights of the critical unitary minimal RSOS models~\cite{ABF84,BPO96} on $A_L$ in the vacuum sector are
\begin{equation}
\psset{unit=.2cm}
\begin{pspicture}[shift=-2.4](-1,-1)(6,6)
\pspolygon[linewidth=.25pt](0,0)(4,0)(4,4)(0,4)
\rput(-.5,-.7){\small $a$}
\rput(4.5,-.5){\small $b$}
\rput(4.5,4.5){\small $c$}
\rput(-.5,4.7){\small $d$}
\rput(2,2){\small $u$}
\psarc(0,0){.35}{0}{90}
\end{pspicture}\;=\;s(\lambda-u)\,\delta_{a,c}+s(u)\sqrt{S_a S_c\over S_b S_d}\,\delta_{b,d},\qquad\qquad
\begin{pspicture}[shift=-2.4](-1,-1)(4,6)
\pspolygon[linewidth=.25pt](0,-1)(0,5)(3,2)
\rput(3.75,2){\small $b$}
\rput(-.75,-.75){\small $a$}
\rput(-.75,4.75){\small $a$}
%\rput(2,2){\small $u$}
\end{pspicture}\;=\;
\begin{pspicture}[shift=-2.4](-1,-1)(4,6)
\pspolygon[linewidth=.25pt](3,-1)(3,5)(0,2)
\rput(-.75,2){\small $b$}
\rput(3.75,-.75){\small $a$}
\rput(3.75,4.75){\small $a$}
%\rput(2,2){\small $u$}
\end{pspicture}\;=\;\sqrt{S_2}\,\delta_{a,1}\delta_{b,2}
\end{equation}
where
\bea\label{parameters}
s(u)={\sin u\over\sin\lambda}={z-z^{-1}\over x-x^{-1}},\quad z=e^{iu},\quad x=e^{i\lambda},\quad \lambda={\pi\over L+1},\quad S_a=s(a\lambda)
\eea
Commuting double row transfer matrices can now be defined following \cite{BPO96}.

\subsection{Bethe ansatz}
The double row transfer matrices of the unitary minimal models in the vacuum sector satisfy the $T$-$Q$ functional equation~\cite{BaxterTQ} modified to account for the boundary
\bea
s(2u)\,\DD(u+{\lambda\over 2})\vec Q(u)=s(2u+\lambda)f(u+{\lambda\over 2})
\vec Q(u-\lambda)+s(2u-\lambda)f(u-{\lambda\over 2})\vec Q(u+\lambda)\label{BaxTQ}
\eea
where $f(u)=s(u)^{2N}$ and $\vec Q(u)$ is an auxiliary family of matrices 
satisfying $[\vec Q(u),\vec Q(v)]=[\vec Q(u),\DD(v)]=\vec 0$. The same 
functional equation is satisfied by the eigenvalues $D(u)$ and $Q(u)$ and 
so the Bethe ansatz equations result by setting
\bea
Q(u)=\prod_{j=1}^n s(u-u_j)s(u+u_j)\label{BaxQ}
\eea
where $u_j$ are the $r=2n$ Bethe roots. The physical analyticity strip of $D(u)$ is $-\lambda/2<\Re u<3\lambda/2$ and the physical analyticity strip of $\tilde{D}(u)=D(u+{\lambda\over 2})$ is $-\lambda<\Re u<\lambda$.

Let $\DD(u)=\DD_0^1$ and
\bea
\DD_k^{q}=\DD^{q}(u+k\lambda),\quad \vec Q_k=\vec Q(u+k\lambda),\quad 
s_k(u)={s(2u+k\lambda)},\quad
f_k(u)=(-1)^{N}{s(u+k\lambda)}^{2N}
\eea
then the $T$-$Q$ relation implies that the eigenvalues $D(u)$ are determined 
by the eigenvalues $Q(u)$ in the compact form
\bea
\tD_0={s_1f_{1/2}Q_{-1}+s_{-1}f_{-1/2}Q_1\over s_0 Q_0}
\eea

\subsection{Functional equations}
\label{sec:functional}

The fused transfer matrices $\DD^q$ satisfy the fusion hierarchy of functional equations~\cite{BazhResh89,BPO96}
\bea
s_{q-2}s_{2q-1}\DD_0^{q}\DD_q^{1}=s_{q-3}s_{2q}f_q\DD_0^{q-1}
		+s_{q-1}s_{2q-2}f_{q-1}\DD_0^{q+1},\qquad q=1,2,\ldots,L-1\label{EqnFusionHier}
\eea
subject to the closure conditions
\bea
\DD_0^{-1}=0,\qquad \DD_0^{0}=f_{-1}\vec I,
\qquad \DD_0^{L}=0\label{closure}
\eea

Starting with the fusion hierarchy, induction can be used to derive the
$T$-system of functional equations~\cite{KP92,BPO96}
\bea\label{funct}
s_{q-2}s_{q}\DD_0^{q}\DD_1^{q}=
s_{-2}s_{2q}f_{-1}f_q\vec I+s_{q-1}^2\DD_0^{q+1}\DD_1^{q-1},\qquad q=1,2,\ldots,L-1
\eea 
For $q=1$, this is just the usual inversion identity given by the fusion 
hierarchy
with $q=1$. 

If we further define
\bea\label{deft}
\dd_0^{q}={s_{q-1}^2\DD_1^{q-1}\DD_0^{q+1}\over
s_{-2}s_{2q}f_{-1}f_q},\qquad q=1,2,\ldots, L-2
\eea
then the inversion identity hierarchy can be recast in the form of the 
$Y$-system~\cite{KP92,BPO96}
\bea
\dd_0^{q}\dd_1^{q}&=&{s_{q-1}^2s_{q+1}^2
(\DD_0^{q+1}\DD_1^{q+1})(\DD_1^{q-1}\DD_2^{q-1})
			\over  s_{-2}s_0s_{2q}s_{2q+2}
f_{-1}f_0f_qf_{q+1}}\nonumber\\ 
&=&\Big(\vec I+{s_{q}^2\DD_0^{q+2}\DD_1^{q}\over
s_{-2}s_{2q+2}f_{-1}f_{q+1}}\Big)\!\!
   \Big(\vec I+{s_{q}^2\DD_1^{q}\DD_2^{q-2}\over
s_{0}s_{2q}f_0f_q}\Big)
%\nonumber\\
%&=&
\;=\;\big(\vec I+\dd_0^{q+1}\big)\big(\vec I+\dd_1^{q-1}\big)\label{Ysystem}
\eea 
with closure
\bea
\dd_0^{0}=\dd_0^{L-1}=0
\eea

We wish to solve the $Y$-system (\ref{Ysystem}) 
subject to the constraints of periodicity, conjugation and crossing symmetries, analyticity and asymptotic limits. The periodicity, conjugation and crossing symmetries are
\bea
\DD^{q}(u)=\DD^{q}(u+\pi),\qquad
\DD^{q}(u)=\DD^{q}(\lambda-u)
\label{periodicity}
\eea
\bea
\overline{D^q(u)}=D^{q}(u)=D^{q}\big((2-q)\lambda-u\big),\qquad
\overline{d^q(u)}=d^{q}(u)=d^{q}\big((1-q)\lambda-u\big)
\eea
To discuss analyticity, we consider the eigenvalues of the transfer matrices 
$\DD^{q}(u)$ and $\dd^{q}(u)$ at
each fusion level in their respective analyticity strips
\bea
-{q\over 2}\,\lambda<\Re u<{4-q\over 2}\,\lambda,\qquad
-{q+1\over 2}\,\lambda<\Re u<{3-q\over 2}\,\lambda
\label{analyticitystrips}
\eea
Defining the shifted transfer matrices
\be
\tilde{\DD}^{q}(u)=\DD^{q}\Big(u+{2-q\over 2}\,\lambda\Big),\qquad
\tilde{\dd}^{q}(u)=\dd^{q}\Big(u+{1-q\over 2}\,\lambda\Big)
\ee
it follows that these transfer matrices have the common analyticity strip
\bea\label{phystrip}
-\lambda< \Re(u)<\lambda
\eea
and satisfy the same periodicity (\ref{periodicity})
with crossing symmetries
\be
\tD^{q}(u)=\tD^{q}(-u),\qquad \td^{q}(u)=\td^{q}(-u)
\ee
In terms of shifted transfer matrices, the TBA functional 
equations take the form
\bea \label{tbafunc}
\tilde{\dd}^q\big(u-\frac{\lambda}2\big)\ 
\tilde{\dd}^q\big(u+\frac{\lambda}2\big)=
\big(1+\tilde{\dd}^{q-1}(u)\big)\big(1+\tilde{\dd}^{q+1}(u)\big)
\eea
Lastly, the asymptotic values $d^j(+ i\infty)$ were computed in \cite{KP92}
\bea\label{asym1}
%\td^j(-i\infty)&=&\frac{\sin[(j-1)\tau]\:\sin[(j+1)\tau]}{\sin^2\tau}
%\;=\;\frac{\sin^2 j\tau}{\sin^2\tau}-1,
%\qquad \tau=\frac{r\pi}{L}\\[8pt]
%\label{asym2}
\td^j(+i\infty)&=&\frac{\sin [j\theta]\:\sin[(j+2)\theta]}{\sin^2\theta}
\;=\;\frac{\sin^2 (j+1)\theta}{\sin^2\theta}-1,
\qquad \theta=\frac{s\pi}{L+1}
\eea
Here, %$r$ and 
$s$ is a Kac label and plays the role of selecting the eigenstates of 
the transfer matrix that, in the 
scaling limit, will be in the $(r,s)$ sector of the conformal field theory. 
More precisely, $s$ is a good quantum number for the finite system corresponding to the braid 
limit $u\rightarrow i\infty$. In contrast, $r$ is only a good quantum number in the limit 
$N\to\infty$ and so is not accessible directly on a finite lattice (see after (\ref{asymp2})).
In the vacuum sector of interest here, $(r,s)=(1,1)$.

%%%%%%%%%%%%%%%%%%%%%%
\subsection{Analyticity and solution of TBA}

We use analyticity properties of the
transfer matrices to transform the $Y$-system (\ref{Ysystem}) into
the form of Thermodynamic Bethe Ansatz (TBA) equations. 
For the $A_L$ lattice model, these TBA equations take the form of coupled integral equations 
whose structure is governed by the Dynkin diagram $G^*=A_{L-2}$. 
The analyticity is encoded in the relative locations of the 1- and 2-strings in the analyticity strips (\ref{analyticitystrips}). 
The TBA equations can be completely solved in the scaling limit for the locations of the 1-strings as well as the excitation energies. Following \cite{OPW97}, the strategy is as follows:
\begin{enumerate}
\item The special case $q=1$ of (\ref{funct}) provides the
expression for the energies or, equivalently, the eigenvalues of the 
transfer matrix. This expression contains an unknown function 
(pseudoenergy in the first strip) and a number of unknown real numbers (location of 1-strings in the first strip). 
\item The Y-system (\ref{Ysystem}) can be transformed into the set of 
TBA equations. 
\item The Y-system (\ref{Ysystem}) allows us to access the location of 
the 1-strings by a set of coupled auxiliary equations, that perform the 
job of quantisation conditions.
\item The simultaneous solution of TBA plus auxiliary equations completely 
fixes the pseudoenergies, the locations of the 1-strings and the energy expression.
\end{enumerate}

\subsubsection{Analyticity: the energy}
We aim at finding the transfer matrix eigenvalues in the physical strip $\frac{\lambda}2<\re u <\frac{3\lambda}2$. We start from 
the case $q=1$ of (\ref{funct}), written for the eigenvalues of the 
corresponding matrices  
\be\label{energ}
\left(\frac{\sin 2\lambda}{\sin \lambda}\right)^2
\frac{s_{-1}s_{1}}{s_{-2}s_{2}\ f_{-1}f_1}~
D^{1}(u)\ D^{1}(u+\lambda)=1+ d^{1}(u)
\ee
Ignoring the term $d^1(u)$ which is exponentially small for large $N$, this functional equation factorizes into bulk and boundary
equations with $D^1(u)=\kappa_{\text{bulk}}(u) \kappa_0(u)$
\be
[\kappa_{\text{bulk}}(u)\ \kappa_{\text{bulk}}(u+\lambda)]^{2N}=f_{-1}f_1=\left[\frac{\sin(\lambda-u)\sin(\lambda+u)}{\sin^2\lambda}\right]^{2N},
\qquad \re u\in (-\frac{\lambda}2,\frac{\lambda}2)
\ee
\be
\kappa_0(u)\ \kappa_0(u+\lambda)
=\frac{s_{-2}s_{2}}{s_{-1}s_{1}}\left(\frac{\sin \lambda}{\sin 2\lambda}\right)^2
=\frac{\sin(\lambda-2u)\sin(\lambda+2u)}
{\sin(2\lambda-2u)\sin(2\lambda+2u)}
\left(\frac{\sin \lambda}{\sin 2\lambda}\right)^2,
\quad \re u\in (-\frac{\lambda}2,\frac{\lambda}2)
\ee
The expressions for these factors are given in \cite{NepomechiePearce} and immediately 
extend to the whole physical strip $\re u\in (-\frac{\lambda}2,\frac32 \lambda)$. 
It follows that the ``finite'' part of the transfer matrix is defined 
\be\label{deffinite}
D_{\text{finite}}^1(u)=\frac{D^1(u)}{[\kappa_{\text{bulk}}(u)]^{2N}\ \kappa_0(u)}=
\tD_{\text{finite}}^1(u-\frac{\lambda}2)\,,
\qquad \re u\in (-\frac{\lambda}2,\frac32 \lambda)\,.
\ee
Here we do not make use of these 
factors we only need to know that, from \cite{NepomechiePearce}, they do not introduce new zeros or poles to 
$D_{\text{finite}}^1(u)$ in the physical strip. 
It follows that the finite transfer matrix has precisely the same zeros as 
$D^1(u)$ and no poles.   
The functional equation (\ref{energ}) now takes the form
\be\label{energ1}
\tD^{1}_{\text{finite}}(u-\frac{\lambda}2)\ 
\tD_{\text{finite}}^{1}(u+\frac{\lambda}2)=
1+ d^{1}(u) \,, \qquad \re u\in (-\frac{\lambda}2,\frac{\lambda}2)\,.
\ee

Let $y_k^{(1)}$ denote the location of the zeros of $\tD^{1}(u)$ such that
\be
\tD_{\text{finite}}^{1}(\pm\frac{i\lambda}{\pi}y_k^{(1)})=
\tD^{1}(\pm\frac{i\lambda}{\pi}y_k^{(1)})=0\,, 
\qquad y_k^{(1)}>0\,.
\ee
To remove these zeros from $\tD^{1}(u)$, we construct the function
\be\label{zeros}
Z(u)=\prod_{k=1}^{m_1}\left[\tanh\frac{y_k^{(1)}+i(L+1) u}{2}\ 
\tanh\frac{y_k^{(1)}-i(L+1) u}{2}\right]\,,\qquad \re u\in (-\lambda,\lambda)
\ee
In the indicated region this function has the same zeros as $\tD^{1}(u)$ 
and no other zeros or poles. Moreover, it satisfies the inversion identity
\be\label{inversione}
Z(u-\frac{\lambda}2)Z(u+\frac{\lambda}2)=1\,.
\ee
Let us define the function
\be\label{ANZ}
\A(x)=\frac{\tD^{1}_{\text{finite}}(u)}
{Z(u)}\Big|_{u=\frac{i x}{L+1}}\,,\qquad \re u\in (-\lambda,\lambda)
\ee
which is free of zeros and poles in the analyticity strip $\im x\in (-\pi,\pi)$. 
%Observe that the shift
%in the interval with respect to (\ref{deffinite}) is consistent with the
%use of the tilde version of the transfer matrix.
Dividing (\ref{energ1}) by (\ref{inversione}) we then obtain 
\be\label{energ2}
\A(x+\frac{i\pi}2)\,\A(x-\frac{i\pi}2)=
1+{\tilde\varepsilon}^{1}(x) \,, \qquad \im\,x\in (-\frac{\pi}2,\frac{\pi}2)
\ee
where
\be\label{ttd}
{\tilde\varepsilon}^{q}(x)=\td^{q}(u)\big|_{u=\frac{i x}{L+1}}\,.
\ee
The left side of (\ref{energ2}) is free of zeros and poles in the 
indicated region so the right side never vanishes. 
The function $\td^1(u)$ vanishes at $u=0$ by (\ref{deft}) and (\ref{closure}). We conclude that
\be\label{posit}
1+{\tilde\varepsilon}^1(x)>0 \quad \mbox{for} \quad x\in \mathbb{R}
\ee
so that we can take the logarithmic derivative
\be\label{logder}
\frac{d}{dx}\log\A(x+\frac{i\pi}2)+\frac{d}{dx}\log\A(x-\frac{i\pi}2)=
\frac{d}{dx}\log(1+{\tilde\varepsilon}^{1}(x)) 
\ee
Taking Fourier transforms
\bea
\F(f,k)=\frac{1}{2\pi}\int_{-\infty}^{\infty}dx\,f(x)\,e^{-ikx},\qquad 
f(x)=\int_{-\infty}^{\infty}dk\,\F(f,k)\,e^{ikx}
\eea
we obtain 
\be
\F\Big(\frac{d}{dx}\log\A(x),k\Big)\ 2\cosh (k\frac{\pi}2)=
\F\Big(\frac{d}{dx}\log(1+{\tilde\varepsilon}^{1}(x)),k\Big)
\ee
Transforming back, we obtain a convolution with a kernel and
an integration constant
\be\label{aconv}
\log\A(x)=\int_{-\infty}^{\infty}dy\ \frac{\log(1+{\tilde\varepsilon}^{1}(y))}
{2\pi\cosh (x-y)}+\mbox{const} = \big(K*\log(1+{\tilde\varepsilon}^1)\big)(x)+\mbox{const}
\ee
where
\bea
(f*g)(x)=\frac{1}{2\pi}\int_{-\infty}^{\infty}dy\: f(x-y)\,g(y),\qquad
K(x)=\frac{1}{\cosh x}\,.\label{kernel}
\eea

Using (\ref{deffinite}), (\ref{ANZ}) and the needed functions we can 
reconstruct the initial double row transfer matrix eigenvalues   
$\log D^1(\frac{\lambda}2+\frac{i x}{L+1})$.
Actually, here we are concerned with the finite size corrections to it namely 
those terms that characterise the scaling limit of the theory. 
These terms behave as $1/N$ in the log of the transfer matrix.
The bulk term in (\ref{deffinite}) is of order $N$; the boundary one and the 
integration constant are of order 1 and will be ignored.
What remains expresses the scaling part of the free energy (in (\ref{aconv})
we ignored the constant)
\begin{gather}\label{free}
f_{\text{scal.}}(x)=
-\log D^1_{\text{finite}}(\frac{\lambda}2+\frac{i x}{L+1})=
-\log \A(x) -Z(\frac{i x}{L+1})= \\[2mm]
-\big(K*\log(1+{\tilde\varepsilon}^1)\big)(x)-\sum_{k=1}^{m_1}
\left[\log\tanh\frac{y_k^{(1)}-x}2+\log\tanh\frac{y_k^{(1)}+x}{2}\right]
\nonumber 
\end{gather}
We can fix $x=0$ namely choose the isotropic point of the system and get
\begin{gather}\label{free2}
f_{\text{scal.}}=
-\log D^1_{\text{finite}}(\frac{\lambda}2)=
-\int_{-\infty}^{\infty}dy\ \frac{\log(1+{\tilde\varepsilon}^{1}(y))}{2\pi\cosh (y)}
-2\sum_{k=1}^{m_1}\log\tanh\frac{y_k^{(1)}}2\\
=\frac{2\pi}{N}E+\text{higher order corrections in } \frac{1}{N}\nonumber
\end{gather}
and $E$ is the energy in the language of conformal field theory, corresponding 
to (\ref{stringEnergy})
\be
E=-\frac{c}{24}+\text{conformal dimension}
\ee
So far, the zeros' positions and the function ${\tilde\varepsilon}^1$ are unknown: 
we will evaluate them in the next sections.  

To summarise, starting from (\ref{energ}), we have discussed the technique, 
which we will largely use later, 
consisting in the removal of all zeros and poles from the analyticity strip, 
then using the Fourier transforms on the non-singular part of the functional equation
to extract the unknown function $D^1(u)$.

\subsubsection{Analyticity: the TBA}
We aim at solving the Y-system (\ref{tbafunc}), for $q$ generic. 
We start examining zeros and poles of $\td^q(u)$ in the analyticity strip 
(\ref{phystrip}), using the definition (\ref{deft}) 
converted to the tilde form 
\be\label{dtilde}
\td^q(u)=\frac{\sin^2 (2u)}{\sin(2u-(q+1)\lambda)\ \sin(2u+(q+1)\lambda)}
\cdot  \frac{\tD^{q-1}(u)\tD^{q+1}(u)}
{\left[\frac{\sin(u-\frac{q+1}2\lambda)\sin(u+\frac{q+1}2\lambda)}
{\sin^2\lambda}\right]^{2N}}
\ee
The bulk term in square brackets vanishes if 
\be\label{bulkp}
u=\pm \frac{q+1}{2}\lambda
\ee
These points are outside or on the border of the analyticity strip for all 
$q$ so they will play no role later. On the contrary, for the special case
$q=1$, the transfer matrix of level zero has a zero at $u=0$
$$
\tD^{0}(u)=(-1)^N\left[\frac{\sin u}{\sin \lambda}\right]^{2N}
$$
that implies the same type of zero for $\td^1(u)$
so we need to remove it from the case $q=1$ in (\ref{tbafunc}); to that purpose we
need a function with a zero of order $2N$ in $u=0$, with no other poles 
or zeros in the analyticity strip and satisfying 
\be\label{fbulk}
[f(u-\frac{\lambda}2)\,f(u+\frac{\lambda}2)]^{2N}=1
\ee
We notice that $N$ is even, here.
Clearly we have (an overall sign is irrelevant)
\be\label{fbulk1}
f(u)=i\tan \Big(\frac{L+1}2 u\Big) 
\ee
for $q=1$ only.  
The first factor in (\ref{dtilde}) has poles at the points indicated in (\ref{bulkp}), 
that lie outside the 
analyticity strip; it has also a double zero in $u=0$ that we will remove 
as in (\ref{fbulk}), (\ref{fbulk1}) with the same  
function\footnote{With more general boundary conditions, these two functions 
need not to coincide and $g$ must contain the boundary parameter $\xi$.}
\bea
&g(u)=[f(u)]^2=-\tan^2 \Big(\frac{L+1}2 u\Big)&\\
&g(u-\frac{\lambda}2)\,g(u+\frac{\lambda}2)=1& \label{gfunc}
\eea

In addition, we have to remove the zeros of the strips $q\pm 1$. These zeros 
are in the center of the analyticity strip so, remembering (\ref{zeros})
and (\ref{inversione}), we introduce
\begin{gather}\label{zerosq}
Z^{q}(u)=
\prod_{j=1}^{L-2}\prod_{k=1}^{m_{j}}\left[\tanh\frac{-i(L+1)u-y_k^{(j)}}{2}\ 
\tanh\frac{-i(L+1)u+y_k^{(j)}}{2}\right]^{A_{j,q}} \\
\label{inversioneq}
Z^q(u-\frac{\lambda}2)Z^q(u+\frac{\lambda}2)=1
\end{gather}
where the adjacency matrix (\ref{adjacency}) selects the appropriate strips. 
As indicated, this function satisfies an inversion identity.
Now we divide the functional equation (\ref{tbafunc}) by 
(\ref{fbulk}), (\ref{gfunc}), (\ref{inversioneq}) and obtain a newer equation 
for the analytic and non-zero function $\A^{q}(x)$
\bea\label{ANZq}
\A^{q}(x)&=&\left.\frac{\td^{q}(u)}{[f(u)]^{2N\,\delta_{1,q}}g(u)Z^{q}(u)}
\right|_{u=\frac{i x}{L+1}}\qquad \re\,u\in (-\lambda,\lambda) \\
\A^{q}(x+\frac{i\pi}2)\,\A^{q}(x-\frac{i\pi}2)&=&
(1+{\tilde\varepsilon}^{q-1}(x))(1+{\tilde\varepsilon}^{q+1}(x)) \,, 
\qquad \im\,x\in (-\frac{\pi}2,\frac{\pi}2)\,.\quad
\eea
As in (\ref{energ2} and \ref{aconv}), we solve for $\A^{q}(x)$ and 
obtain 
\bea\label{aqconv}
\log\A^{q}(x)&=&\sum_{j=1}^{L-2} A_{q,j}\int_{-\infty}^{\infty}dy\ 
\frac{\log(1+{\tilde\varepsilon}^{j}(y))}{2\pi\cosh (x-y)}+C^{(q)}
\\ &=& \sum_{j=1}^{L-2} A_{q,j}
\big(K*\log(1+{\tilde\varepsilon}^{j})\big)(x)+C^{(q)}\,.\nonumber
\eea
We now replace the definition of $\A^{q}(x)$ and get the TBA equations
\bea\label{tbaeq}
\log {\tilde\varepsilon}^{q}(x)&=&\delta_{1,q}\,N \log \big[f(\frac{i x}{L+1})\big]^2+
\log g(\frac{i x}{L+1})+C^{(q)}\\ 
&+&\log \prod_{j=1}^{L-2}\prod_{k=1}^{m_j}\Big[\tanh\frac{x-y_k^{(j)}}{2} 
\tanh\frac{x+y_k^{(j)}}{2}\Big]^{A_{j,q}} +
\sum_{j=1}^{L-2} A_{q,j}\,\big(K*\log(1+{\tilde\varepsilon}^{j})\big)(x)
\nonumber
\eea
The $+i\infty$ asymptotic values (\ref{asym1}) fix the integration constants;
we make use of the integral
\be\label{integr}
\int_{-\infty}^{\infty}dy\, \frac{1}{\cosh y}=\pi
\ee
and also observe that (\ref{ANZq}) being a non-zero function,
the expression $1+{\tilde\varepsilon}^{(q)}$ cannot vanish on the real axis
and its sign can be evaluated from its asymptotic value (\ref{asym1})
\be
1+{\tilde\varepsilon}^{q}(+\infty)=\frac{\sin^2(q+1)\lambda}{\sin^2\lambda}>0
\ee
therefore on the whole real axis we have
\be
1+{\tilde\varepsilon}^{q}(x)> 0 \qquad x \in \mathbb{R}\,,\qquad q=1,\ldots,L-2 
\label{asym3}
\ee
and its logarithm does not introduce imaginary terms.

We can now evaluate the limit in (\ref{tbaeq})  
\be
\log {\tilde\varepsilon}^{q}(+\infty)=\frac12 \sum_{j=1}^{L-2} A_{q,j}
\log (1+{\tilde\varepsilon}^{j}(+\infty))+C^{(q)}
\ee
where logarithms are taken in the fundamental branch. 
Using (\ref{asym1}), (\ref{asym3}) we conclude that 
\be \label{intcon}
C^{(q)}=0 \qquad q=1,\ldots,L-2\,.
\ee
Here we restrict ourselves to the vacuum sector but in other cases the 
integration constants could be non-zero, usually multiples of $i\,\pi$. 
The $L-2$ TBA equations are now
\bea\label{tbaeq2}
\log {\tilde\varepsilon}^{q}(x)&=&\delta_{1,q}\, N\log \tanh^2 \frac{x}{2}+
\log \tanh^2 \frac{x}{2}+ \\ 
&+&\log\prod_{j=1}^{L-2}\prod_{k=1}^{m_j}\Big[\tanh\frac{x-y_k^{(j)}}{2}
\tanh\frac{x+y_k^{(j)}}{2}\Big]^{A_{q,j}}+
\sum_{j=1}^{L-2} A_{q,j}\,\big(K*\log(1+{\tilde\varepsilon}^{j})\big)(x)
\nonumber
\eea

\subsubsection{Analyticity: the zeros}
The zero's positions are still undetermined and, again, the functional 
equations (\ref{tbafunc}) will
help us. We shift them of $\pm\frac{\lambda}2$ so we can have one of the 
two forms 
\bea\label{vanishp}
\td^{(q)}(u)\td^{q}(u+\lambda)&=&\Big(1+\td^{q-1}\big(u+\frac{\lambda}2\big)\Big)
\Big(1+\td^{q+1}\big(u+\frac{\lambda}2\big)\Big)\\\label{vanishm}
\td^{q}(u-\lambda)\td^{q}(u)&=&\Big(1+\td^{q-1}\big(u-\frac{\lambda}2\big)\Big)
\Big(1+\td^{q+1}\big(u-\frac{\lambda}2\big)\Big)
\eea
and if we are on a zero of $\td^{q}(u)$ both expressions vanish. 
The zeros of $\td^{q}(u)$ are in both the strips $q\pm 1$ (\ref{dtilde}). 
Let's
take a zero of $q+1$, $u_0=\frac{i\,y_k^{(q+1)}}{L+1}$; in that case 
also $\td^{q+2}(u_0)$ will vanish but not $\td^{q+4}(u_0)$ so
\bea
\td^{q}(u_0)=0 &\Rightarrow& \Big(1+\td^{q-1}\big(u_0+\frac{\lambda}2\big)\Big)
\Big(1+\td^{q+1}\big(u_0+\frac{\lambda}2\big)\Big)=0\nonumber\\
\td^{q+2}(u_0)=0 &\Rightarrow& \Big(1+\td^{q+1}\big(u_0+\frac{\lambda}2\big)\Big)
\Big(1+\td^{q+3}\big(u_0+\frac{\lambda}2\big)\Big)=0 \label{vanish2} \\
\td^{q+4}(u_0)\neq 0 &\Rightarrow& \Big(1+\td^{q+3}\big(u_0+\frac{\lambda}2\big)\Big)
\Big(1+\td^{q+5}\big(u_0+\frac{\lambda}2\big)\Big)\neq 0\,.\nonumber
\eea
The non-vanishing of $1+\td^{q+3}(u_0+\frac{\lambda}2)$ forces 
$1+\td^{q+1}(u_0+\frac{\lambda}2)$ to vanish on the zeros of strip $q+1$ 
and to be different from zero on the zeros of the remaining strips. The same
is true for the `minus' form in (\ref{vanishm}) so we state that
\be
0=1+\td^{q}\Big(\frac{i\,y_k^{(q)}}{L+1}\pm\frac{\lambda}2\Big)=
1+{\tilde\varepsilon}^{q}\Big(y_k^{(q)}\mp i\frac{\pi}2\Big)
\ee
Actually, by (\ref{dtilde}), $\td^{q}(u)$ also vanishes if $u$ is a 
2-string of $q+1$ (and also of $q-1$). The 2-strings are located in
$u_0=\pm\lambda+\frac{i\,w_h^{(q+1)}}{L+1}$ with real $w_h^{(q+1)}$. 
Taking the ``$-$'' case, we use (\ref{vanishm}) and repeat precisely 
the same construction as in 
(\ref{vanish2}); this yields ($q+1\rightarrow q$)
\be
0=1+\td^{q}\Big(u_0+\frac{\lambda}2\Big)=1+\td^{q}\Big(-\lambda+
\frac{i\,w_k^{(q)}}{L+1}+\frac{\lambda}2\Big)=
1+{\tilde\varepsilon}^{q}\Big(w_k^{(q)}+i\frac{\pi}2\Big)
\ee
This calculation holds true for the other sign so we conclude that
a unique equation, equivalently in the `minus' or the `plus' form,
\be
0=1+{\tilde\varepsilon}^{q}\Big(v\mp i\frac{\pi}2\Big)\,,\qquad 
v\in\{y_k^{(q)},\,k=1,\ldots,m_q;~ w_h^{(q)},\,h=1,\ldots,n_q\}
\ee
fixes both the one- and the 2-strings of the given $q$ strip.
We don't need the 2-strings positions but we need their order with respect to
the 1-strings. Focusing on the ``minus'' form, we introduce the function
\bea\label{psi}
\Psi^{q}(x)&=& i\log {\tilde\varepsilon}^{q}\Big(x-i\frac{\pi}2\Big)=
i\delta_{1,q}\, N\log \tanh^2 \Big(\frac{x}{2}-i\frac{\pi}4\Big)+
i\log \tanh^2 \Big(\frac{x}{2}-i\frac{\pi}4\Big)+ \\ &+& 
i\sum_{j=1}^{L-2}{A_{q,j}}\sum_{k=1}^{m_j}
\Big[\log\tanh\Big(\frac{x-y_k^{(j)}}{2}-i\frac{\pi}4\Big)+
\log\tanh\Big(\frac{x+y_k^{(j)}}{2}-i\frac{\pi}4\Big)\Big]+\nonumber\\
&-&\sum_{j=1}^{L-2} A_{q,j} \pv dy\,
\frac{\log(1+{\tilde\varepsilon}^{j}(y))}{2\pi\sinh(x-y)}
\nonumber 
\eea
that is real on the real axis and becomes an odd multiple of $\pi$ on 
one- and 2-string zeros. Differently from (\ref{tbaeq2}), here the sum on 
different zeros is done after taking the logarithm so we can get an extremely 
useful information by keeping track of the winding.
On the zeros' positions the principal value is not 
required as both the numerator and the denominator vanish while it is 
required for generic values of $x$.
We introduce the 1-string quantum numbers by
\be\label{quantum}
\Psi^{q}(y_k^{(q)})=\pi\,n_k^{(q)}\, \qquad n_k^{(q)}=1 \mbox{ mod } 2\,.
\ee
The $y_k^{(q)}$ are always single zeros and also each quantum number 
uniquely fixes one zero. 
These features have been largely observed in our numerical analysis
and also in previous cases as in \cite{FPR03};
this is also typical of models solved by Bethe Ansatz. For these reasons, these
quantum numbers were called ``non-degenerate'' in \cite{FPR03}.
In the framework of 
ABF models, exceptions are known \cite{FPR03} and are related to bulk 
or boundary renormalisation flows and do not play any role here.
Using (\ref{psi}) and also (\ref{asym1}) we can derive
the asymptotic values of the function $\Psi^q$. At $+\infty$ 
$1+d^j$ is non-zero for all $j$ so we can safely take the limit of the 
integral part and get zero
\be\label{psiasym}
\Psi^{q}(+\infty)=0\,\qquad q=1,\ldots,L-2
\ee
From (\ref{asym1}) we find that $1+d^j(-\infty)$ is non-zero for all $j>1$
but is exactly zero for $j=1$ so the evaluation of the integral in (\ref{psi})
requires special care. Actually, the case $j=1$ occurs just once for $q=2$
so in all the other cases we have
\be
\Psi^{q}(-\infty)=2\pi(\delta_{1,q}N+1+\sum_{j=1}^{L-2} A_{q,j}m_j)>0 \,,
\qquad q\neq 2
\ee
We believe that the divergence of $\log(1+d^j(-\infty))$ does not affect 
the result so we write
\be
\Psi^{2}(-\infty)=2\pi (1+\sum_{j=1}^{L-2} A_{2,j}m_j)>0
\ee
These asymptotic behaviours say that the function is globally decreasing. 
Moreover, the real function
\be
i\log \tanh \Big(\frac{x}{2}-i\frac{\pi}4\Big)
\ee
is monotonically decreasing. We cannot prove the behaviour of the integral part
but it is usually subdominant so we conclude that the function 
$\Psi^{q}(x)$ is monotonically decreasing. This has been widely confirmed 
numerically; again, minor exceptions are known \cite{FPR03} and are 
related to bulk or boundary renormalisation flows and do not play any role here.

The $+\infty$ asymptotic fixes the smallest quantum number in (\ref{quantum}); 
indeed, the largest zero is $y_{m_q}^{(q)}$ so 
\be\label{quanta}
\pi\,n_{m_q}^{(q)}=\Psi^{q}(y_{m_q}^{(q)}) > \Psi^{q}(+\infty) =0
\ee
so
\be\label{quantb}
n_{1}^{(q)}>\ldots>n_{k}^{(q)}>n_{k+1}^{(q)}>\ldots>n_{m_q}^{(q)}\geqslant 1
\ee
Given a zero $y_{k}^{(q)}$ with quantum number $n_{k}^{(q)}$, if a lower
2-string moves up and exchanges its position with the 1-string, 
the position $y_{k}^{(q)}$ gets smaller and the decreasing monotonicity of 
$\Psi^{(q)}(x)$ implies that the quantum number has to grow of two. It cannot 
grow of one because it must be odd. At the same time, the 
so called non-negative quantum number $I_{k}^{(q)}$ grows of one by definition
(\ref{qnumb}) so the two quantum numbers must be proportional
\be\label{quantc}
n_{k}^{(q)}=2I_k^{(q)}+\text{integer}
\ee
Imposing (\ref{quanta} and \ref{quantb}) we get
\be\label{quantd}
n_{k}^{(q)}=1+2(I_k^{(q)}+m_q-k)
\ee

Both $+\infty$ and $-\infty$ asymptotics are even multiples of $\pi$; 
as one- and 2-strings are quantised with odd multiples of $\pi$, the number 
of odd positions from $\Psi^{q}(-\infty)$ to $\Psi^{q}(+\infty)$ 
must contain the number of one- and 2-strings $2(m_q+n_q)$, the $2$ comings 
from the upper and lower half planes, so
\be\label{countmn}
\frac{\Psi^{q}(-\infty)-\Psi^{q}(+\infty)}{2\pi}=
\delta_{1,q}N+1+\sum_{j=1}^{L-2} A_{q,j}m_j\geqslant 2(m_q+n_q)
\ee
The $1$ is the spurious zero $u=0$ of (\ref{dtilde}).

The next step will be to evaluate the scaling limit of these equations.

\subsubsection{The scaling limit}
\label{sec:scal}

As we are interested in the conformal energy (\ref{free2}), we need to evaluate
\begin{gather}\label{confE}
E=\lim_{N\rightarrow\infty}\left\{\frac{N}{2\pi}\left[
-\int_{-\infty}^{\infty}dy\ \frac{\log(1+{\tilde\varepsilon}^{1}(y))}{2\pi\cosh (y)}
-2\sum_{k=1}^{m_1}\log\tanh\frac{y_k^{(1)}}2 \right]\right\}
\end{gather}
so we are pushing the number of sites to infinity.
In doing this, we need to pay attention to the movement of the one- and 
2-strings in consequence of the growth of $N$. They depend by $N$ with the law 
\be\label{shift}
y_k^{(q)}=y_k^{(q)}(N)=\hy_k^{(q)}+\log N>0
\ee
where the scaled position $\hy_k^{(q)}$ is not constrained to be positive, 
as it was for its lattice partner.
We also need to consider that the reduced transfer matrix eigenvalues
${\tilde\varepsilon}^{1}(y)$ also depend by $N$ both by the TBA equations 
(\ref{tbaeq2}) and by the zero positions.

The ``zeros term'' in (\ref{confE}) can be easily worked out using
the scaling (\ref{shift})
\be\label{limzero}
\lim_{N\rightarrow\infty}\left[-\frac{N}{2\pi}\log\tanh\frac{\hy_k^{(1)}+\log N}2
\right]=\frac{1}{\pi}e^{-\hy_k^{(1)}}
\ee
The integral term has to be split and treated by changing the 
integration variables
\begin{gather}
I=\lim_{N\rightarrow\infty}\:-\frac{N}{2\pi}\,\int_{-\infty}^{\infty}dy\ 
\frac{\log(1+{\tilde\varepsilon}^{1}(y))}{2\pi\cosh (y)}=\\  \nonumber
=\lim_{N\rightarrow\infty}\:-\frac{N}{2\pi^2}\left[
\int_{0}^{\infty}dy\, \frac{\log(1+{\tilde\varepsilon}^{1}(y))}{2\cosh (y)}+
\int_{-\infty}^{0}dy\, \frac{\log(1+{\tilde\varepsilon}^{1}(y))}{2\cosh (y)}\right]=\\
\nonumber
=\lim_{N\rightarrow\infty}\:-\frac{N}{2\pi^2}\left[
\int_{-\log N}^{\infty}dy'\, \frac{\log(1+{\tilde\varepsilon}^{1}(y'+\log N))}
{2\cosh (y'+\log N)}+
\int_{-\infty}^{\log N}dy''\, \frac{\log(1+{\tilde\varepsilon}^{1}(y''-\log N))}
{2\cosh (y''-\log N)}\right]\\
=\lim_{N\rightarrow\infty}\:-\frac{1}{2\pi^2}\left[
\int_{-\log N}^{\infty}dy'\, \frac{\log(1+{\tilde\varepsilon}^{1}(y'+\log N))}
{e^{y'}+N^{-2}e^{-y'}}+ 
\int_{-\infty}^{\log N}dy''\, \frac{\log(1+{\tilde\varepsilon}^{1}(y''-\log N))}
{N^{-2}e^{y''}+e^{-y''}}\right] \nonumber
\end{gather}
In the limit, both the integrations go over the whole real axis;
the subleading $N^{-2}$ term disappears and we are lead to define
the following scaled eigenvalues
\bea\label{scald}
\hd^{q}(x)&=&\lim_{N\rightarrow\infty}{\tilde\varepsilon}^{q}(x+\log N)\\
\hd^{q}_{\text{lower}}(x)&=&\lim_{N\rightarrow\infty}{\tilde\varepsilon}^{q}(-x-\log N)\nonumber
\eea
The first one comes from the original upper half plane the second one
from the lower one. For finite $N$, here we have perfect mirror symmetry
between the two half-planes so we don't need to consider the lower one. 
Because of this mirror symmetry, it is convenient to put $-x$ in the definition
of the lower component.
We will show later that scaling the eigenvalues as in (\ref{scald}) is
perfectly consistent with the TBA and auxiliary equations.

We finalise the  energy expression ($y''\rightarrow -y''$) using
the mirror symmetry
\bea
I&=&-\frac{1}{2\pi^2}\left[
\int_{-\infty}^{\infty}dy\,e^{-y}\log(1+\hd^{1}(y))+
\int_{-\infty}^{\infty}dy\,e^{-y}\log(1+\hd^{1}(y))\right]=\\
&=&-\frac{1}{\pi^2}\int_{-\infty}^{\infty}dy\,e^{-y}\log(1+\hd^{1}(y))\nonumber
\eea
and finally the conformal energy is given by
\be\label{sc_energ}
E=-\frac{1}{\pi^2}\int_{-\infty}^{\infty}dy\,e^{-y}\log(1+\hd^{1}(y))+
2\sum_{k=1}^{m_1}\frac{1}{\pi}e^{-\hy_k^{(1)}}
\ee
The scaling eigenvalues in (\ref{scald}) can be determined by performing 
the same limit on the TBA equations (\ref{tbaeq2}). We initially 
treat the special term occurring for $q=1$
\be\label{leading}
\lim_{N\rightarrow\infty}\:\delta_{1,q}\, N\log \tanh^2 \frac{x+\log N}{2}=
-4\delta_{1,q} e^{-x}
\ee
Then, the limit on the boundary term (second term on the right hand side of 
(\ref{tbaeq2})) trivially vanishes. 
Among the terms containing the zeros, we look at those where there is a 
difference 
\be\label{differ}
\lim_{N\rightarrow\infty}\:\log\tanh\frac{x+\log N-y_k^{(q)}}2=
\log\tanh\frac{x-\hy_k^{(q)}}{2}
\ee 
while those where there is a sum disappear
\be
\lim_{N\rightarrow\infty}\:\log\tanh\frac{x+\log N+y_k^{(q)}}2=
\lim_{N\rightarrow\infty}\:\log\tanh\frac{x+\hy_k^{(q)}+2\log N}2=0
\ee 
Clearly, these terms were produced by zeros in the lower half plane 
so their contribution vanishes in the upper half plane.
The integral term is evaluated using the properties of the convolution
\be\label{integral}
\big(K*\log(1+{\tilde\varepsilon}^{j})\big)(x+\log N)=
\big(K*\log(1+\hd^{j})\big)(x)
\ee
We are ready to recompose the TBA equations
\bea
\log \hd^{q}(x)&=&-4\ \delta_{1,q}\ e^{-x}+ 
\log\prod_{j=1}^{L-2}\prod_{k=1}^{m_j} \Big[\tanh\frac{x-\hy_k^{(j)}}{2}\Big]^{A_{q,j}}
\nonumber\\
&+&\sum_{j=1}^{L-2} A_{q,j}\,\big(K*\log(1+\hd^{j})\big)(x)\,,
~\qquad q=1,\ldots,L-2 \label{sc_tba}
\eea
Observe that the positivity condition 
\be\label{positiv}
1+\hd^{q}(x)> 0
\ee
remains true and the 
asymptotic value (\ref{asym1}) is not changed by the scaling limit (\ref{scald}); 
that asymptotic value was used to fix the integration constants (\ref{intcon}).
We now use the TBA equations to evaluate the opposite asymptotic $\hd^{(q)}(-\infty)$.
First, we observe that the zeros contribution is 
\be
\lim_{x\rightarrow -\infty}\log\prod_{j=1}^{L-2}\prod_{k=1}^{m_j} \Big[\tanh
\frac{x-\hy_k^{(j)}}{2}\Big]^{A_{q,j}}=0
\ee
because the zeros are in even number, from section~\ref{sec:geo}.
If we exponentiate (\ref{sc_tba}), this contributes a factor one.
The integral term is evaluated as in (\ref{integr}) and following equations,
the positivity condition (\ref{positiv}) ensuring that no imaginary parts are introduced
so we have 
\be
\lim_{x\rightarrow -\infty}\hd^{q}(x)=\lim_{x\rightarrow -\infty}\exp(-4\ \delta_{1,q}\ e^{-x})
\ \prod_{j=1}^{L-2}\, \Big[1+\hd^{j}(-\infty)\Big]^{\frac12 A_{q,j}}
\ee
If $q=1$, the first factor forces a transcendentally fast vanishing
\be\label{asymp1}
\hd^{q}(-\infty)=0
\ee
that is an indirect manifestation of the presence 
of a sea of 2-strings in the first strip, namely pyramids of hight one.
For the remaining values of $q$ we write 
\be\label{minf}
\hd^{q}(-\infty)=\prod_{j=2}^{L-2}\, \Big[1+\hd^{j}(-\infty)\Big]^{\frac12 A_{q,j}}\qquad
q=2,\ldots,L-2
\ee
where the range of the product starts from $j=2$ because of (\ref{asymp1}). 
This equation can be solved with the technique of \cite{KP92}: simple trigonometry
shows that 
\be\label{trigo}
1+\hd^{q}(-\infty)=\frac{\sin^2 (q+a)\tau}{\sin^2 \tau}
\ee
is the solution of (\ref{minf}) for generic $q$, with two arbitrary parameters
$a,\tau$ to be fixed ((\ref{minf}) is like a second order finite difference equation in
the integer $q$, if the logarithm is taken).
The case $q=2$ fixes $a=0$ and the case $q=L-2$ fixes $\tau=\pi/L$ so we conclude with
\be\label{asymp2}
\hd^{q}(-\infty)=\frac{\sin^2 q\tau}{\sin^2 \tau}-1\,,\qquad q=1,\ldots,L-2\,,
\qquad \tau=\frac{\pi}{L}
\ee
Actually, the $q=1$ case is also contained here.
For generic parities of $m_q$, the value of $\tau$ would have been $\tau=r\pi/L$
with an integer parameter $r$ in the role of the Kac index of $(r,s)$. The 
parameter $s$ was introduced on the finite lattice in (\ref{asym1});
the parameter $r$ first appears here after the scaling limit; in the
present case $(r,s)=(1,1)$ namely we investigate the vacuum sector.

In literature, the following pseudoenergies are often used 
\be\label{pseudo}
\epsilon_q(x)=-\log \hd^{q}(x), \qquad q=1,\ldots,L-2.
\ee 
A standard notation is also the function  
\be \label{elj}
L_q(x)=\log(1+\hd^{q}(x))=\log(1+ e^{-\epsilon_q(x)}), 
\qquad L_q(x)\in\mathbb{R} \quad \text{for } x\in\mathbb{R} 
\ee 
From (\ref{dtilde}), this function vanishes on the zeros of the adjacent strips
\be\label{Lvanish}
L_q(\hy_k^{(q\pm 1)})=0 \qquad \forall \quad k.
\ee 
The quantisation condition (\ref{quantum}), according to the shift of the 
zeros (\ref{shift}), becomes
\be\label{sc_quant}
\lim_{N\rightarrow\infty}\Psi^{q}(y_k^{(q\pm 1)})=
\Psi^{q}(\hy_k^{(q)}+\log N)=\pi\,n_k^{(q)}\, \qquad n_k^{(q)}=1 
\mbox{ mod } 2
\ee
so we are lead to scale the function as we did in (\ref{scald}). Actually, we 
recycle all our previous steps (\ref{leading}), (\ref{differ}), (\ref{integral})
and get
\bea\label{sc_psi}
\hat{\Psi}^{q}(x)&=&\lim_{N\rightarrow\infty}\Psi^{q}(x+\log N)\\
&=&4\delta_{1,q}\, e^{-x}+i\sum_{r=1}^{L-2}{A_{q,r}}\sum_{k=1}^{m_r} 
\log\tanh(\frac{x-\hy_k^{(r)}}{2}-i\frac{\pi}{4}) \nonumber \\
&-&\sum_{r=1}^{L-2} A_{q,r} \pv dy\,
\frac{\log(1+\hd^{r}(y))}{2\pi\sinh(x-y)}\nonumber\\
\hat{\Psi}^{q}(\hy_k^{(q)})&=&\pi\,n_k^{(q)}
\eea
Observe that $\Psi^{1}$ keeps track of the
infinite number of 2-strings in the first strip that form the Dirac sea.
In a familiar terminology, it is the coloured node of the $A_{L-2}$ 
TBA diagram.
Equations (\ref{sc_tba}) and (\ref{sc_psi}) are fully consistent with the 
scaling limit taken in the energy expression (\ref{sc_energ}). 

We can use asymptotics to count the number of one- and 2-strings. 
Because of (\ref{sc_psi}), the spurious zero $u=0$ of (\ref{countmn}) now 
has left the analyticity strips moving toward $-\infty$ therefore 
all the odd multiples of $\pi$ from $\hat{\Psi}^{q}(-\infty)$ to 
$\hat{\Psi}^{q}(+\infty)$ are exclusively occupied by one or two strings; 
their amount is $m_q+n_q$ as the lower half plane is not visible here.
The number of odd positions from $\Psi^{q}(-\infty)$ to $\Psi^{q}(+\infty)$
is evaluated from (\ref{sc_psi}) and must be equated to the number of 
objects that can be allocated in the strip, namely $m_q+n_q$
\be\label{tba_mn}
\frac{\hat{\Psi}^{q}(-\infty)-\hat{\Psi}^{q}(+\infty)}{2\pi}=
\frac12 \sum_{j=1}^{L-2} A_{q,j}m_j=m_q+n_q \qquad q=2,\ldots,L-2\,.
\ee
The case $q=1$ simply gives $\infty=m_1+n_1$, consistently with the
notion that in the first strip there is a Dirac sea of $n$-family
particles of high 1, $n_1=\infty$, and $m_1$ is not upper bounded. 
In summary, we feed in the functional equations (\ref{tbafunc}) the 1-string 
positions and from the equivalent TBA and auxiliary equations we get out 
the part of the $(m,n)$-system that survives after taking the scaling 
limit in (\ref{Amnsystem}).

\subsubsection{Energy from excited TBA}
\label{sec:exact}

Extending the calculations introduced in \cite{OPW97}, the expression 
for the energy (\ref{sc_energ}) can be cast into a closed form.
We rewrite the quantization condition in the first strip, as given 
in (\ref{quantd}) with $q=1$, in the following form:
\bea
\frac2{\pi} e^{-\hy_h^{(1)}} &= & \frac12 n_h^{(1)}-\frac1{2\pi} 
\Psi^{1}(\hy_h^{(1)}) + \frac2{\pi} e^{-\hy_h^{(1)}}
\eea
We now sum on all the first strip zeros to obtain
\bea
\frac2{\pi} \sum_{h=1}^{m_1} e^{-\hy_h^{(1)}}&=&\frac12\sum_{h=1}^{m_1} n_h^{(1)}
-\frac{i}{2\pi}\sum_{h=1}^{m_1} \sum_{k=1}^{m_2} \log \tanh 
(\frac{\hy_h^{(1)}-\hy_k^{(2)}}{2}-i\frac{\pi}{4}) \\
&+&  \int_{-\infty}^{\infty}
\frac{dy}{4\pi^2} L_2(y)\sum_{h=1}^{m_1} \frac1{\sinh(\hy_h^{(1)}-y)} \nonumber
\eea
being the integral evaluated on a zero we do not need to prescribe the
principal value, according to (\ref{Lvanish}).
A similar manipulation in the other strips leads to
\bea
0 &= & \sum_{h=1}^{m_q} \Big(\frac12 n_h^{(q)}-\frac1{2\pi} 
\Psi^{q}(\hy_h^{(q)}) \Big)\\
&=&\frac12\sum_{h=1}^{m_q} n_h^{(q)}
-\frac{i}{2\pi}\sum_{h=1}^{m_q} \sum_{r=1}^{L-2}A_{q,r}\sum_{k=1}^{m_r} 
\log \tanh (\frac{\hy_h^{(q)}-\hy_k^{(r)}}{2}-i\frac{\pi}{4})\nonumber \\
&+&  \int_{-\infty}^{\infty}
\frac{dy}{4\pi^2}\sum_{r=1}^{L-2}A_{q,r} L_r(y)\sum_{h=1}^{m_q} 
\frac1{\sinh(\hy_h^{(q)}-y)}\quad \qquad q=2,\ldots,L-2 \nonumber
\eea
We add all these expressions for different strips ~$q=1,\ldots,L-2$ 
\bea
\frac2{\pi} \sum_{h=1}^{m_1} e^{-\hy_h^{(1)}} &=& \frac12\sum_{q=1}^{L-2}
\sum_{h=1}^{m_q} n_h^{(q)} - \frac{i}{2\pi}\sum_{q,r=1}^{L-2} A_{q,r}
\sum_{h=1}^{m_q}
\sum_{k=1}^{m_r}\log \tanh (\frac{\hy_h^{(q)}-\hy_k^{(r)}}{2}-i\frac{\pi}{4}) \\
&+ &\ \int_{-\infty}^{\infty} \frac{dy}{4\pi^2} \left(\sum_{q,r=1}^{L-2} 
A_{q,r} L_{r}(y)\ 
\sum_{h=1}^{m_q} \frac1{\sinh(\hy_h^{(q)}-y)}\right) \nonumber
\eea
We can use the following identity to evaluate the second term on the right 
hand side 
\be\label{logid}
\log \tanh (x-i\frac{\pi}4)+\log \tanh (-x-i\frac{\pi}4)=-i\pi
\ee
This identity is true in the fundamental determination of the logarithm and
yields
\bea  \label{summed}
\frac2{\pi} \sum_{h=1}^{m_1} e^{-\hy_h^{(1)}} &=& \frac12\sum_{q=1}^{L-2}
\sum_{h=1}^{m_q} n_h^{(q)} - \frac{1}{2}\sum_{q=1}^{L-3} m_q\,m_{q+1}\\
&+ &\ \int_{-\infty}^{\infty} \frac{dy}{4\pi^2} \left(\sum_{q,r=1}^{L-2} 
A_{q,r} L_{r}(y)\ 
\sum_{h=1}^{m_q} \frac1{\sinh(\hy_h^{(q)}-y)}\right) \nonumber
\eea
In the expression for the energy (\ref{sc_energ}) we recognise 
the sum of exponentials computed in (\ref{summed}). We substitute it and
we obtain an algebraic term and an integration term:
\bea \label{energy1}
E&=& \frac12 \sum_{q=1}^{L-2}\sum_{h=1}^{m_q} n_h^{(q)} 
- \frac12 \sum_{q=1}^{L-3} m_q m_{q+1} -\frac{S}{8\pi^2}.
\eea
The integral term is defined as
\be \label{esse} S=
2 \int_{-\infty}^{+\infty} dy \left\{ 4 e^{-y} L_1(y)
-\left( \sum_{q,r=1}^{L-2}
A_{q,r} L_{r}(y)
\sum_{h=1}^{m_q} \frac1{\sinh(\hy_h^{(q)}-y)} 
\right) \right\}  
\ee
and it will be evaluated later with Rogers dilogarithms. 
The first and second of the algebraic terms can be easily worked out 
using the relation (\ref{quantd}) between the two families of quantum 
numbers: 
\be \label{sum}
\frac12 \sum_{q=1}^{L-2}\sum_{h=1}^{m_q} n_{h}^{(q)}=
\sum_{q=1}^{L-2}\sum_{h=1}^{m_q} I_{h}^{(q)}
+\frac12 \sum_{q=1}^{L-2}m_q^2\,.
\ee
Using the Cartan matrix (\ref{adjacency}) the following identity is easily 
proved:
\be\label{cartid}
\frac12 \sum_{q=1}^{L-2} m_q^2 - \frac12 \sum_{q=1}^{L-3} m_q m_{q+1}
=\frac14 \m^{T}  C \m
\ee
where we use the same notation as in section~\ref{sec:geo}.
Using (\ref{sum}) and (\ref{cartid}) in (\ref{energy1}) we obtain
the following expression where the quantum numbers and the 
content of zeros enter explicitly
\be \label{energy2}
E= \frac14 \m^{T}  \C \m+ \sum_{q=1}^{L-2}\sum_{h=1}^{m_q} I_{h}^{(q)}
-\frac{S}{8\pi^2}.
\ee
To evalutate the integral contribution $S$ in (\ref{esse}), 
it is convenient to 
compute the following expression in two alternative ways
\be
S_q = \int_{-\infty}^{\infty}dy\: \Big[ \frac{d\,\log |\hd^q(y)|}{dy}\ 
\log(1+\hd^q (y))-\log|\hd^q(y)|\ \frac{d\,\log(1+\hd^q(y))}{dy} \Big]
\label{esseq}
\ee
The absolute value here is convenient because it avoids us keeping track 
of the imaginary part that could appear when $\hd^q(y)$ is negative; 
this imaginary part does not contribute to the final value so we safely 
remove it from now on. Moreover, the full integrand is always real because 
of the reality condition (\ref{elj}). 
The quantity here defined will be evaluated in one way as sums and 
differencies of dilogarithms and in a second way by using the TBA equations.

\subsubsection{Evaluation with TBA equations}
We evaluate the whole sum 
\be \label{summation}
\sum_{q=1}^{L-2}   S_q
\ee
and we will see that many terms will cancel. 
We need to insert in (\ref{esseq}) the function 
$\log |\hd^q (y)|$ and its derivative as obtained from the TBA equations:
\be
\frac{d\,\log |\hd^q(y)|}{dy}  = 4e^{-y} \delta_{1,q} 
+\sum_{r=1}^{L-2} A_{q,r} \left( \sum_{k=1}^{m_{r}} 
\frac1{\sinh (y-\hy_k^{(r)})}
+K* L'_{r}(y)\right).
\ee
One can easily realize that, within the whole sum (\ref{summation}),
all the terms that contain a convolution mutually cancels. Indeed, 
each term of type 
\be\label{derlog}
\frac{d\,\log |\hd^r(y)|}{dy}\ L_r(y) 
\ee
contains a product $K*(L'_{r-1}+L'_{r+1}) L_r$ 
and each term of type 
\be\label{loghd}
-\log| \hd^{r+1}(y)|\ L'_{r+1}(y)
\ee
contains $-K*(L_{r}+L_{r+2}) L'_{r+1}$.
In these triple products the convolution and the ordinary 
product can be exchanged and this produces the cancellation of all the 
convolution terms. 

From the terms like (\ref{loghd}) we also have contributions of the form
\be
\int_{-\infty}^{\infty}dy\:(-1)\log\Big|\tanh\frac{y-\hy_k^{(r\pm1)}}2\Big|
L'_r(y)
\ee
where the absolute value enters because of (\ref{esseq}).
Integrating by parts we obtain
\be\label{contrib}\begin{array}{c}\displaystyle
\left[(-1)\log\Big|\tanh\frac{y-\hy_k^{(r\pm1)}}2\Big| 
L_r\right]_{-\infty}^{+\infty}
-\int_{-\infty}^{\infty}dy\:(-1)\frac{L_r(y)}{\sinh(y-\hy_k^{(r\pm1)})}\\[5mm]
\displaystyle
=-\int_{-\infty}^{\infty}dy\:\frac{L_r(y)}{\sinh(\hy_k^{(r\pm1)}-y)}
\end{array}\ee
where the first contribution vanishes. 
The remaining term in (\ref{contrib}) sums with a analogous contribution
from (\ref{derlog}) and we are left with the following expression
\be
\sum_{q=1}^{L-2}   S_q=
2\int_{-\infty}^{\infty}dy \left[4e^{-y}\ L_1 -
\left( \sum_{q,r=1}^{L-2} A_{q,r}\ L_{r}(y)
\sum_{h=1}^{m_q} \frac1{\sinh(\hy_h^{(q)}-y)}\right)\right] =S
\ee
that precisely matches the quantity (\ref{esse}) that enters the energy.

\subsubsection{Evaluation with dilogarithms}
\label{sec:dilogt}

We start the evaluation of the first line in (\ref{esseq}) 
by performing a change of the integration variable 
\be t=\hd^q(y)> -1 , \qquad dt=dy\ [\hd^q(y)]' \ee
so that $S_q$ is uniquely fixed by the asymptotic values of $\hd^q$
\be \label{sjasym}
S_q=\int_{\hd^q(-\infty)}^{\hd^q(+\infty)}dt\: \Big( \frac{\log(1+t)}{t}
-\frac{\log|t|}{1+t} \Big).
\ee
The integration interval can be modified passing through the point zero
\be \label{point0}
S_q=\int_{0}^{\hd^q(+\infty)}dt\: \Big( \frac{\log(1+t)}{t}
-\frac{\log|t|}{1+t} \Big)
-\int_0^{\hd^q(-\infty)}dt\: \Big( \frac{\log(1+t)}{t}
-\frac{\log|t|}{1+t} \Big)
\ee
so we need to evaluate single constituent blocks given by the integration 
from zero to a certain asymptotic value $t_a>-1$. We introduce the 
function $L_{+}$ to label the main block
\be \label{Lplus}\begin{array}{l}\displaystyle
L_{+}(t_a)\stackrel{\text{def}}{=}\frac12 \int_{0}^{t_a}dt\: 
\Big( \frac{\log(1+t)}{t}-\frac{\log|t|}{1+t} \Big)\\[5mm]
=\left\{ \begin{array}{l@{\qquad}l}
\displaystyle -\frac12\int_{0}^{\frac{t_a}{1+t_a}}dy\: 
\Big(\frac{\log(1-y)}{y}+\frac{\log y}{1-y}\Big)=
\mathcal{L}\Big(\frac{t_a}{1+t_a}\Big), & \text{if} 
\quad t_a\geqslant0,\\[5mm]
\displaystyle \frac12\int_{0}^{-t_a}dy\: 
\Big(\frac{\log(1-y)}{y}+\frac{\log y}{1-y}\Big)=
-\mathcal{L}\big(-t_a\big), & \text{if} 
\quad t_a\leqslant0
\end{array} \right. \end{array}
\ee
where we performed simple changes of the integration variable
and used the dilogarithm function $\calL$ defined in Appendix~A. 
Notice that, so far, we only used the dilogarithm function in the 
fundamental interval $[0,1]$. 

The integral under investigation (\ref{esseq}) can finally be expressed
as the difference of the constituent blocks in (\ref{point0}) through the 
function $L_+$
\be\label{sdil}
S_q=2 \Big[ L_+(\hd^q(+\infty))-L_+(\hd^q(-\infty))\Big].
\ee

\subsubsection{Excitation energies}
Our purpose is to make explicit the expression (\ref{sdil}).
Using the known identity (\ref{identity}) in the case 
$t\geqslant 0$ we obtain 
\be
L_{+}(t)=\calL\Big(\frac{t}{1+t}\Big)=\frac{\pi^2}6-\calL\Big(\frac1{1+t}\Big),
\qquad t\geqslant 0.
\ee
Similarly, using the continuation of the function (\ref{dilp}) we have
\be
L_{+}(t)=-\calL(-t)=\frac{\pi^2}6-\calL\Big(\frac1{1+t}\Big),
\qquad t\leqslant 0
\ee
so we can summarize these two cases into a single expression 
\be \label{block} 
\frac12 \int_{0}^{t_a}dt\: \Big( \frac{\log(1+t)}{t}-\frac{\log|t|}{1+t} \Big)=
L_{+}(t_a)=\frac{\pi^2}6-\calL\Big(\frac1{1+t}\Big).
\ee
Using now the asymptotic values (\ref{asym1}), (\ref{trigo}),
the argument of the dilogarithm in (\ref{block}) takes 
the quadratic form
\be
\frac1{1+t}=\frac{\sin^2 \tau}{\sin^2q\tau} \quad \text{or} \quad
\frac{\sin^2 \theta}{\sin^2(q+1)\theta}
\ee
respectively for the $\hd^q(-\infty)$ or $\hd^q(+\infty)$ case.
Finally, using (\ref{block}), the sum of expressions (\ref{sdil})
can be rewritten as
\be\label{dil1}
\frac3{\pi^2}S=\frac6{\pi^2}\sum_{q=1}^{L-2} 
\Big( L_+(\hd^q(+\infty))-L_+(\hd^q(-\infty))\Big)=
\frac6{\pi^2} \sum_{q=1}^{L-2}\Big[ 
\calL\Big(\frac{\sin^2 \tau}{\sin^2q\tau}\Big)
-\calL\Big(\frac{\sin^2 \theta}{\sin^2(q+1)\theta}\Big) \Big]
\ee
where we introduced the factor $3/\pi^2$ to make a comparison with 
the dilogarithm identity stated in (\ref{coroll}).
Indeed, the $\varphi$ in the first term specializes to
$\varphi=\tau=\frac{\pi}{L}$ so that
\be\label{idr}
s(0,2,L-2)=\frac6{\pi^2} \sum_{q=1}^{L-2} 
\calL\Big(\frac{\sin^2 \tau}{\sin^2(q+1)\tau}\Big)=
\frac6{\pi^2} \sum_{q=1}^{L-2} 
\calL\Big(\frac{\sin^2 \tau}{\sin^2q\tau}\Big)
\ee
and the last equality is justified because
\be 
\calL(1)=\calL\Big(\frac{\sin^2 \tau}{\sin^2(L-1)\tau}\Big)=\frac{\pi^2}6.
\ee
In the third term of (\ref{coroll}) we have $\varphi=\theta=\frac{\pi}{L+1}$
so that 
\be\label{ids}
s(0,2,L-1)=\frac6{\pi^2} \sum_{q=1}^{L-1} 
\calL\Big(\frac{\sin^2 \theta}{\sin^2(q+1)\theta}\Big)=
\frac6{\pi^2} \sum_{q=1}^{L-2} 
\calL\Big(\frac{\sin^2 \theta}{\sin^2(q+1)\theta}\Big)+1
\ee
where we used 
\be 
\calL(1)=\calL\Big(\frac{\sin^2 \theta}{\sin^2 L\theta}\Big)=\frac{\pi^2}6.
\ee
With the results (\ref{idr}), (\ref{ids}), the expression (\ref{dil1}) becomes
\be\label{dil2}
\frac3{\pi^2}S=s(0,2,L-2)+1-s(0,2,L-1)=c_L=1-\frac6{L(L+1)}
\ee
where the identities (\ref{coroll}), (\ref{coroll1}) were used and the 
vacuum boundary conditions were specified $r=s=1$.
We can insert this result in the expression for the energy (\ref{energy2})
to obtain
\be\label{energy3}
E=-\frac{c_L}{24}+ \frac14 \m^{T}  C \m 
+\sum_{q=1}^{L-2}\sum_{k=1}^{m_q} I_{k}^{(q)}
\ee
and we see the appearance of the central charge of unitary minimal models.
We have obtained a completely explicit expression for the energy of the 
states of the theory. 

The identities (\ref{coroll}), (\ref{coroll1}) remind us the analogous 
expressions for the conformal dimensions of fields in unitary minimal models.
Indeed, the full TBA calculation of this sector can be immediately
extended to all cases. 
More difficult is the task of deriving the \mn-system that applies 
to an arbitrary sector $(r,s)$. We do not discuss this issue in the 
present paper.

%%%%%%%%%%%%%%%%%%%%%%
\section{$s\ell(2)_1$ WZW Lattice Models and XXX Hamiltonian}
\label{sec:wzw}

\subsection{Definitions}

The bulk face and boundary triangle Boltzmann weights of the critical level-1 WZW vertex models in the vacuum sector are
\bea
\psset{unit=.275cm}
\begin{pspicture}[shift=-1.8](0,0)(4,4)
\pspolygon[linewidth=.25pt](0,0)(4,0)(4,4)(0,4)
\psline[linewidth=.75pt,arrowsize=6pt]{->}(0,2)(2,2)
\psline[linewidth=.75pt,arrowsize=6pt]{->}(2,0)(2,2)
\psline[linewidth=.75pt,arrowsize=6pt]{->}(2,2)(2,4)
\psline[linewidth=.75pt,arrowsize=6pt]{->}(2,2)(4,2)
\end{pspicture}\;=\;
\begin{pspicture}[shift=-1.8](0,0)(4,4)
\pspolygon[linewidth=.25pt](0,0)(4,0)(4,4)(0,4)
\psline[linewidth=.75pt,arrowsize=6pt]{->}(4,2)(2,2)
\psline[linewidth=.75pt,arrowsize=6pt]{->}(2,4)(2,2)
\psline[linewidth=.75pt,arrowsize=6pt]{->}(2,2)(0,2)
\psline[linewidth=.75pt,arrowsize=6pt]{->}(2,2)(2,0)
\end{pspicture}\;=\;1-z,\qquad
\begin{pspicture}[shift=-1.8](0,0)(4,4)
\pspolygon[linewidth=.25pt](0,0)(4,0)(4,4)(0,4)
\psline[linewidth=.75pt,arrowsize=6pt]{->}(4,2)(2,2)
\psline[linewidth=.75pt,arrowsize=6pt]{->}(2,0)(2,2)
\psline[linewidth=.75pt,arrowsize=6pt]{->}(2,2)(0,2)
\psline[linewidth=.75pt,arrowsize=6pt]{->}(2,2)(2,4)
\end{pspicture}\;=\;
\begin{pspicture}[shift=-1.8](0,0)(4,4)
\pspolygon[linewidth=.25pt](0,0)(4,0)(4,4)(0,4)
\psline[linewidth=.75pt,arrowsize=6pt]{->}(0,2)(2,2)
\psline[linewidth=.75pt,arrowsize=6pt]{->}(2,4)(2,2)
\psline[linewidth=.75pt,arrowsize=6pt]{->}(2,2)(2,0)
\psline[linewidth=.75pt,arrowsize=6pt]{->}(2,2)(4,2)
\end{pspicture}\;=\;z,\qquad
\begin{pspicture}[shift=-1.8](0,0)(4,4)
\pspolygon[linewidth=.25pt](0,0)(4,0)(4,4)(0,4)
\psline[linewidth=.75pt,arrowsize=6pt]{->}(2,4)(2,2)
\psline[linewidth=.75pt,arrowsize=6pt]{->}(2,0)(2,2)
\psline[linewidth=.75pt,arrowsize=6pt]{->}(2,2)(0,2)
\psline[linewidth=.75pt,arrowsize=6pt]{->}(2,2)(4,2)
\end{pspicture}\;=\;
\begin{pspicture}[shift=-1.8](0,0)(4,4)
\pspolygon[linewidth=.25pt](0,0)(4,0)(4,4)(0,4)
\psline[linewidth=.75pt,arrowsize=6pt]{->}(0,2)(2,2)
\psline[linewidth=.75pt,arrowsize=6pt]{->}(4,2)(2,2)
\psline[linewidth=.75pt,arrowsize=6pt]{->}(2,2)(2,0)
\psline[linewidth=.75pt,arrowsize=6pt]{->}(2,2)(2,4)
\end{pspicture}\;=\;1
\eea
\bea
\psset{unit=.2cm}
\begin{pspicture}[shift=-2.8](0,-1)(4,5)
\pspolygon[linewidth=.25pt](0,-1)(0,5)(3,2)
\psline[linewidth=.75pt,arrowsize=6pt]{->}(3,0)(1,2)
\psline[linewidth=.75pt,arrowsize=6pt]{->}(1,2)(3,4)
\end{pspicture}\;=\;
\begin{pspicture}[shift=-2.8](0,-1)(4,5)
\pspolygon[linewidth=.25pt](0,-1)(0,5)(3,2)
\psline[linewidth=.75pt,arrowsize=6pt]{->}(1,2)(3,0)
\psline[linewidth=.75pt,arrowsize=6pt]{->}(3,4)(1,2)
\end{pspicture}\;=\;
%1,\qquad\qquad
\begin{pspicture}[shift=-2.8](0,-1)(4,5)
\pspolygon[linewidth=.25pt](3,-1)(3,5)(0,2)
\psline[linewidth=.75pt,arrowsize=6pt]{->}(0,0)(2,2)
\psline[linewidth=.75pt,arrowsize=6pt]{->}(2,2)(0,4)
\end{pspicture}\;=\;
\begin{pspicture}[shift=-2.8](0,-1)(4,5)
\pspolygon[linewidth=.25pt](3,-1)(3,5)(0,2)
\psline[linewidth=.75pt,arrowsize=6pt]{->}(2,2)(0,0)
\psline[linewidth=.75pt,arrowsize=6pt]{->}(0,4)(2,2)
\end{pspicture}\;=\;1
\eea
where $z=e^{iu}$ is now an multiplicative spectral parameter.
Commuting double row transfer matrices can now be defined following \cite{BPO96}. The Hamiltonian limit of this model gives the XXX quantum spin chain. The quasiparticle description of this model, including the $(m,n)$ system, tower particles, string patterns and Cartan matrices, is given in Section~2. In the rest of this section we consider the functional equations and solution of the resulting TBA equations.

\subsection{Bethe ansatz}
The Bethe ansatz equations satisfied by the double row transfer matrices of the XXX model are given by
\bea\label{tq}
z\,D(z)Q(z)&=&(z+\half)^{2N+1} Q(z-1)+(z-\half)^{2N+1}Q(z+1)
\eea
where 
\bea\label{Qop}
Q(z)=\prod_{j=1}^n (z-z_j)
\eea
and the physical analyticity strip of $T(z)$ is $-1/2<\Re z< 3/2$.

\subsection{Functional equations}
The functional equations given in section~\ref{sec:functional} hold also 
in the present case, with the difference that here there is no 
truncation so the fusion level $q$ takes all integer values up to $\infty$.
Moreover, the parameter $\lambda$ is now an arbitrary real number. 
%(anisotropy of the spin chain).
We are interested in the limit $\lambda\rightarrow 0$ so it is useful to 
rescale the spectral parameter and take the limit 
\bea\label{rescale}
u&\mapsto& \frac{u\lambda}{\pi}\\
\lambda&\rightarrow &0
\eea
This rescaling and limit transform all trigonometric functions into  
polynomial functions of the rescaled variable $u$. We introduce
$\DD(u)=\DD_0^1$ and
\bea
\DD_k^{q}=\DD^{q}(u+k\pi),\quad \vec Q_k=\vec Q(u+k\pi),\quad 
s_k(u)=2\frac{u}{\pi}+k,\quad
f_k=(-1)^{N}(\frac{u}{\pi}+k)^{2N}
\eea
The definition (\ref{deft}) becomes 
\bea\label{defd}
\dd_0^{q}={s_{q-1}^2\DD_1^{q-1}\DD_0^{q+1}\over
s_{-2}s_{2q}f_{-1}f_q},\qquad 1\le q\,,
\eea
and the $Y$-system has a lower closure only 
\bea
\dd_0^{q}\dd_1^{q}
&=&\big(\vec I+\dd_0^{q+1}\big)\big(\vec I+\dd_1^{q-1}\big)\label{Ysyst} \\
\dd_0^{0}&=&0
\eea 
We also rewrite (\ref{funct}) for $q=1$ as
\bea\label{funcd}
\frac{s_{-1}s_{1}}{s_{-2}s_{2q}f_{-1}f_q}\DD_0^{1}\DD_1^{1}=
\vec I+\dd^{1}_0
\eea 
We remark that the periodicity of $\pi$ that was a feature of all the 
transfer matrices (\ref{periodicity}) is now lost after the rescaling 
of the spectral parameter.
The new analyticity strips for $\DD^{q}(u)$ and $\dd^{q}(u)$ at
each fusion level in their respective analyticity strips are
\bea
-{q\over 2}\,\pi<\Re u<{4-q\over 2}\,\pi,\qquad
-{q+1\over 2}\,\pi<\Re u<{3-q\over 2}\,\pi
\eea
Defining the shifted transfer matrices
\be
\tilde{\DD}^{q}(u)=\DD^{q}\Big(u+{2-q\over 2}\,\pi\Big),\qquad
\tilde{\dd}^{q}(u)=\dd^{q}\Big(u+{1-q\over 2}\,\pi\Big)
\ee
it follows that these transfer matrices have the common analyticity strip
\bea\label{phystr}
-\pi< \Re(u)<\pi
\eea
and satisfy the same crossing symmetries
\be
\tD^{q}(u)=\tD^{q}(-u),\qquad \td^{q}(u)=\td^{q}(-u)
\ee
In terms of shifted transfer matrices, the TBA functional 
equations take the form
\bea \label{tbafu}
\tilde{\dd}^q\big(u-\frac{\pi}2\big)\ 
\tilde{\dd}^q\big(u+\frac{\pi}2\big)=
\big(1+\tilde{\dd}^{q-1}(u)\big)\big(1+\tilde{\dd}^{q+1}(u)\big)
\eea
We evaluate the asymptotics taking the limit $\lambda\rightarrow 0$ of
$(\ref{asym1})$ and we get
\be
\td^q(+i\infty)=q(q+2)\label{asy2}
\ee
Observe that now the asymptotics are the same for all sectors.

\subsection{Analyticity and solution of TBA}
Most parts of the derivation of the TBA equations for the truncated case 
carry on here, without modifications, so we limit ourselves to the results
of the calculation. 

Starting with the energy calculations, the bulk and boundary factors 
survive in the limit (\ref{rescale}) so we can remove order $N$ and order 1
zeros from $D^1$ as in (\ref{deffinite})
\be\label{deffini}
D_{\text{finite}}^1(u)=\frac{D^1(u)}{[\kappa_{\text{bulk}}(u)]^{2N}\ \kappa_0(u)}=
\tD_{\text{finite}}^1(u-\frac{\pi}2)\,,
\qquad u\in (-\frac{\pi}2,\frac32 \pi)\,.
\ee
We easily get 
\be\label{ener1}
\tD^{1}_{\text{finite}}(u-\frac{\pi}2)\ 
\tD_{\text{finite}}^{1}(u+\frac{\pi}2)=
1+ \td^{1}(u) \,, \qquad u\in (-\frac{\pi}2,\frac{\pi}2)\,.
\ee
We now remove the zeros of the level=1 transfer matrix, that we denote as
\be
\tD_{\text{finite}}^{1}(\pm i y_k^{(1)})=
\tD^{1}(\pm i y_k^{(1)})=0\,, 
\qquad y_k^{(1)}>0\,.
\ee
To that purpose, we construct the following function
\be\label{zeros_xxx}
Z(u)=\prod_{k=1}^{m_1}\left[\tanh\frac{y_k^{(1)}+i u}{2}\ 
\tanh\frac{y_k^{(1)}-i u}{2}\right]\,,\qquad u\in (-\pi,\pi)
\ee
that guarantees the complete removal of all zeros from the analyticity strip.
Again, we rotate the variable 
\be
{\tilde\varepsilon}^q(x)=\td^{q}(ix)
\ee
and use the analytic and non-zero function
\be
\A(x)=\left. \frac{\tD_{\text{finite}}(u)}{Z(u)} \right|_{u=ix}
\ee
to write the functional equation
\be\label{ener2}
\A(x+i\frac{\pi}2)\ \A(x-i\frac{\pi}2)=
1+{\tilde\varepsilon}^{1}(x) \,, \qquad u\in (-\frac{\pi}2,\frac{\pi}2)\,.
\ee
We solve it by Fourier transform, as in (\ref{aconv}), and, at the isotropic 
point, the same equation obtained in (\ref{free2}).
\begin{gather}\label{freeen2}
f_{\text{scal.}}=
-\log D^1_{\text{finite}}(\frac{\pi}2)=
-\int_{-\infty}^{\infty}dy\ \frac{\log(1+{\tilde\varepsilon}^{1}(y))}{2\pi\cosh (y)}
-2\sum_{k=1}^{m_1}\log\tanh\frac{y_k^{(1)}}2\\
=\frac{2\pi}{N}E+\text{higher order corrections in } \frac{1}{N}\nonumber
\end{gather}
It is now clear that we can directly rescale and take the limit 
(\ref{rescale}) in the equations for the truncated case and we will get 
the equations for the untruncated one. 
From (\ref{tbaeq}) we have 
\bea\label{tbaequ}
\log {\tilde\varepsilon}^{q}(x)&=&\delta_{1,q}\,N\log \tanh^2 \frac{x}{2}+
\log \tanh^2 \frac{x}{2}+C^{(q)}\\ 
&+&\log\prod_{j=1}^{\infty}\prod_{k=1}^{m_j}\big[\tanh\frac{x-y_k^{(j)}}{2}
\tanh\frac{x+y_k^{(j)}}{2}\big]^{A_{q,j}}+
\sum_{j=1}^{\infty} A_{q,j}\,\big(K*\log(1+{\tilde\varepsilon}^{j})\big)(x)
\nonumber
\eea
with the same kernel as in (\ref{kernel}). 
We evaluate the $+\infty$ limit of the equation to fix the integration 
constant. Using (\ref{integr}) we get
\be
\log {\tilde\varepsilon}^{(q)}(+\infty)=\frac12 \sum_{j=1}^{\infty} A_{q,j}
\log (1+{\tilde\varepsilon}^{(j)}(+\infty))+C^{(q)}
\ee
and the asymptotic values (\ref{asy2}) force the vanishing of all 
integration constants
\be
C^{(q)}=0\,.
\ee
The zeros are fixed by the untruncated family of $\Psi$ functions
\bea
\Psi^{q}(x)&=& i\log {\tilde\varepsilon}^{q}(x-i\frac{\pi}2)=
i\delta_{1,q}\, N\log \tanh^2 (\frac{x}{2}-i\frac{\pi}4)+
i\log \tanh^2 (\frac{x}{2}-i\frac{\pi}4)+ \nonumber \\ &+& 
i\sum_{j=1}^{\infty}{A_{q,j}}\sum_{k=1}^{m_j}
\Big[\log\tanh(\frac{x-y_k^{(j)}}{2}-i\frac{\pi}4)+
\log\tanh(\frac{x+y_k^{(j)}}{2}-i\frac{\pi}4)\Big]-\nonumber\\
&-&\sum_{j=1}^{\infty} A_{q,j} \pv dy\,
\frac{\log(1+{\tilde\varepsilon}^{j}(y))}{2\pi\sinh(x-y)}\label{psi_xxx}
\eea
with the quantum numbers
\be\label{quantum_xxx}
\Psi^{q}(y_k^{(q)})=\pi\,n_k^{(q)}\, \qquad n_k^{(q)}=1+2(I_k^{(q)}+m_q-k)\,.
\ee
The scaling limit is computed precisely as in section~\ref{sec:scal}. 
The result for the energy is
\be\label{sc_energ_xxx}
E=-\frac{1}{\pi^2}\int_{-\infty}^{\infty}dy\,e^{-y}\log(1+\hd^{1}(y))+
2\sum_{k=1}^{m_1}\frac{1}{\pi}e^{-\hy_k^{(1)}}
\ee
with the following set of TBA equations 
\bea
\log \hd^{q}(x)&=&-4\ \delta_{1,q}\ e^{-x}+ 
\log\prod_{j=1}^{\infty}\prod_{k=1}^{m_j} \Big[\tanh\frac{x-\hy_k^{(j)}}{2}\Big]^{A_{q,j}}+
\nonumber\\
&+&\sum_{j=1}^{\infty} A_{q,j}\,\big(K*\log(1+\hd^{j})\big)(x)\,,
\qquad q=1,\ldots \label{sc_tba_xxx}
\eea
and the following quantisation conditions and auxiliary equations to fix 
the zeros
\bea\label{sc_psi_xxx}
\hat{\Psi}^{q}(x)&=&
4\delta_{1,q}\, e^{-x}+i\sum_{r=1}^{\infty}{A_{q,r}}\sum_{k=1}^{m_r} 
\log\tanh(\frac{x-\hy_k^{(r)}}{2}-i\frac{\pi}{4}) \nonumber \\
&-&\sum_{r=1}^{\infty} A_{q,r} \pv dy\,
\frac{\log(1+\hd^{r}(y))}{2\pi\sinh(x-y)}\nonumber\\
\hat{\Psi}^{q}(\hy_k^{(q)})&=&\pi\,n_k^{(q)}=\pi\Big[1+2(I_k^{(q)}+m_q-k)\Big]
\eea
along with the constraint on particle composition given by
\bea
m_1\ge m_2\ge \cdots \ge m_{k}\ge \cdots \ge 0
\eea
After, the $-\infty$ limit can be evaluated from the TBA equations or simply by taking 
the limit of (\ref{asymp2}) for $\lambda\rightarrow 0$ 
\be
\hd^q(-\infty)=q^2-1\label{asy1}
\ee 
as in (\ref{asy2}) we see that the asymptotics do not distinguish the sector.

\subsection{Exact energy from excited TBA}
\label{exact_xxx}
Our goal is to get an explicit expression for energy eigenvalues. The method 
is very similar to the one introduced in section~\ref{sec:exact} but we need to 
introduce slight modifications to avoid intermediate divergent sums as those 
appearing in (\ref{summed}). Moreover, the non-cancellation of a number of 
terms compared to the truncated case requires to introduce a very compact 
notation to keep expressions within a reasonable length. 
We define
\newcommand{\B}{\mathcal{B}}
\be\label{bi}
\B_{qh,rk}=
- \frac{i}{2\pi}\log \tanh (\frac{\hy_h^{(q)}-\hy_k^{(r)}}{2}-i\frac{\pi}{4})
\ee
and observe that the identity (\ref{logid}) takes now the form
\be\label{bid}
\B_{qh,rk}+\B_{rk,qh}=-\frac12
\ee
Also, we define
\be\label{gint}
G_{r,qh}=\int_{-\infty}^{\infty} dy\ \frac{L_r(y)}{4\pi^2\sinh(y_h^{(q)}-y)}
\ee
here the integrand is always finite because the numerator vanishes on the
adjacent strip zeros as in (\ref{Lvanish}) therefore the principal value is not 
necessary. 
We rewrite the quantization condition (\ref{sc_psi_xxx}) for a zero in a generic strip 
$y_h^{(q)}$ as 
\bea\label{quanz}
\delta_{1,q}\frac2{\pi} e^{-\hy_h^{(1)}}=\frac12 n_h^{(q)}+
\sum_{r,k} A_{q,r}\B_{qh,rk} 
+\sum_r A_{q,r} G_{r,qh}
\eea
For both $\B$ and $G$ notations, the comma separates variables that refere 
to different objects, for example in $G_{r,qh}$ the label $r$ referes to the numerator 
while $qh$ to the denominator. In the sums, an index as in $\sum_r$
means to sum on all possible values of $r$; which are its possible values 
it is clear from the context, for example the following sum
\be
\sum_{r,k}\B_{qh,rk}=\sum_{r=1}^{\infty}\sum_{k=1}^{m_r}\B_{qh,rk}
\ee 
has no other possible interpretations because $r$ ranges on positive integers
and $k$, according to (\ref{bi}), labels a zero so it can take values in the
corresponding strip $r$ only. Also, no sum on repeated indices will be used.

In (\ref{quanz}) we sum on all zeros of all strips up to a given $Q>1$, 
namely on all $q=1,\ldots,Q$ and on all $h=1,\ldots,m_q$. Actually observe that
the contribution of a strip $q>1$ to the right hand side is zero so that sum
is independent of $Q$. The sum is reorganised
\bea\label{sumzeros}
\frac2{\pi} \sum_{h} e^{-\hy_h^{(1)}}&=&\frac12 \sum_{q=1,h}^{Q} n_h^{(q)}+
\sum_{q=1,h,r,k}^{Q}A_{q,r}\B_{qh,rk}+\sum_{q=1,h,r}^{Q} A_{q,r} G_{r,qh}=\\
&=&\frac12 \sum_{q=1,h}^{Q} n_h^{(q)}+\sum_{q=2,h,k}^{Q}\B_{qh,q-1 k}+
\sum_{q=1,h,k}^{Q-1}\B_{qh,q+1 k}+\sum_{h,k}\B_{Qh,Q+1 k}+\nonumber\\
&+&\sum_{q=1,h}^{Q-1}G_{q+1,qh}+\sum_{q=2,h}^{Q}G_{q-1,qh}+\sum_{h}G_{Q+1,Qh}
\nonumber
\eea
In the second line, the sum on the quantum numbers is evaluated with 
(\ref{sc_psi_xxx})
and gives 
\be
\frac12 \sum_{q=1,h}^{Q} n_h^{(q)}=\sum_{q=1,h}^{Q}I_h^{(q)}+
\frac12 \sum_{q=1}^{Q} m_q^2 
\ee
the second and third terms are evaluated with (\ref{bid}) and yield
\be
-\frac12 \sum_{q=1}^{Q-1} m_{q}m_{q+1}
\ee
The sum evaluated in (\ref{sumzeros}) can now be inserted in the energy 
expression (\ref{sc_energ_xxx}) so we write
\bea\label{sumzeros2}
E&=&\sum_{q=1,h}^{Q}I_h^{(q)}+\frac{m_1^2}2+\frac12\sum_{q=1}^{Q-1} m_{q+1}
(m_{q+1}-m_q)+\\
&+&\sum_{h,k}\B_{Qh,Q+1 k}+\sum_{q=2,h}^{Q}G_{q-1,qh}+\sum_{q=1,h}^{Q-1}G_{q+1,qh}
+\sum_{h}G_{Q+1,Qh}-\int_{-\infty}^{\infty}\frac{dy}{\pi^2}e^{-y}\,L_1(y)=
\nonumber\\
&=&\sum_{q=1,h}^{Q}I_h^{(q)}+\frac{m_1^2}2+\frac12\sum_{q=1}^{Q-1} m_{q+1}
(m_{q+1}-m_q)-\frac{S^Q}{8\pi^2}+T^Q \nonumber
\eea
and the newly introduced quantities are defined as
\bea
S^Q&=&8\int_{-\infty}^{\infty}dy\,e^{-y}\,L_1(y)-8\pi^2
\sum_{q=2,h}^{Q}G_{q-1,qh}-8\pi^2\sum_{q=1,h}^{Q-1}G_{q+1,qh}\\
T^Q&=&\sum_{h,k}\B_{Qh,Q+1 k}+\sum_{h}G_{Q+1,Qh}
\eea
We need to sum up to $Q$ the values
\be\label{S_val}
S_q=\int_{-\infty}^{\infty}dy\Big[-\re(\epsilon_q(y))'\,L_q(y)+
\re(\epsilon_q(y))\,L_q'(y)\Big] 
\ee
where the real part is taken to remove all imaginary contributions coming
from the zero terms in (\ref{sc_tba_xxx}), namely
\be
\re \log \tanh (\frac{y-\hy_k^{(r)}}{2})=\log |\tanh (\frac{y-\hy_k^{(r)}}{2})|
\ee
With some integrations by parts in (\ref{S_val}) we get the following expression
\bea
S_q=\delta_{1,q}\,8 \int_{-\infty}^{\infty}dy\ e^{-y}\, L_1(y)-8\pi^2\sum_{rk}
A_{q,r}G_{q,rk}+\sum_{r}A_{q,r}\big(L_q\cdot K*L_r'-L_q'\cdot K*L_r\big)
\eea
where the dot product $\cdot$ indicates the bilinear form
\be\label{bilin}
f\cdot g=\int_{-\infty}^{\infty}dy f(y)\,g(y)
\ee
and the convolution ``$*$" has higher priority than the bilinear form ``$\cdot$". 
Actually, the two types of product can be exchanged because the kernel is even, for example 
\be\label{exchange}
L_q'\cdot K*L_r=L_r\cdot  K*L_q'=L_q'*K\cdot L_r
\ee
This property is extremely useful in the following sum  
because all bilinear terms cancel except those with largest indexes
\bea
\sum_{q=1}^{Q} S_q&=& 8 \int_{-\infty}^{\infty}dy\ e^{-y}\, L_1(y)
-8\pi^2\sum_{q=2,k}^{Q} G_{q-1,qk}-8\pi^2\sum_{q=1,k}^{Q-1} G_{q+1,qk}+\nonumber\\
&-&8\pi^2\sum_{k}G_{Q,Q+1\,k}+ K*L_{Q+1}'\cdot L_{Q}-K*L_{Q+1}\cdot L_Q' =\nonumber \\
&=& S^Q -8\pi^2\sum_{k}G_{Q,Q+1\,k}+ K*L_{Q+1}'\cdot L_{Q}-K*L_{Q+1}\cdot L_Q'
\label{sumq}
\eea
and we see that the quantity $S^Q$ appears in this sum; in this way, the sum of 
(\ref{S_val}) enters into the energy expression. 
The evaluation of (\ref{S_val}) with dilogarithms has been done in section~\ref{sec:dilogt} and holds in the present case; in particular we will use
(\ref{sdil}). 
From the asymptotic values (\ref{asy1}) and (\ref{asy2}), using also the properties given 
in Appendix~A, we get 
\bea
L_{+}(\hd^q(+\infty))=\mathcal{L}\Big(\frac{\hd^q(+\infty)}{1+\hd^q(+\infty)}\Big)=
\mathcal{L}\Big(\frac{q(q+2)}{(q+1)^2}\Big)=\frac{\pi^2}6-\mathcal{L}\Big(\frac{1}{(q+1)^2}\Big)   \\
L_{+}(\hd^q(-\infty))=\mathcal{L}\Big(\frac{\hd^q(-\infty)}{1+\hd^q(-\infty)}\Big)=
\mathcal{L}\Big(\frac{q^2-1}{q^2}\Big)=\frac{\pi^2}6-\mathcal{L}\Big(\frac{1}{q^2}\Big) 
\eea
and finally
\be
S_q=2\Big[\mathcal{L}\Big(\frac{1}{q^2}\Big)-\mathcal{L}\Big(\frac{1}{(q+1)^2}\Big)\Big]
\ee
In the sum on $q$ all terms disappear but two 
\be\label{dilsum}
\sum_{q=1}^{Q} S_q=2\sum_{q=1}^{Q} \Big[\mathcal{L}\Big(\frac{1}{q^2}\Big)
-\mathcal{L}\Big(\frac{1}{(q+1)^2}\Big)\Big]=2\mathcal{L}(1)-
2\mathcal{L}\Big(\frac{1}{(Q+1)^2}\Big)=\frac{\pi^2}3-
2\mathcal{L}\Big(\frac{1}{(Q+1)^2}\Big)
\ee
and (\ref{sumq}) reads
\be
S^Q=8\pi^2\sum_{k}G_{Q,Q+1\,k}-K*L_{Q+1}'\cdot L_{Q}+K*L_{Q+1}\cdot L_Q'+
\frac{\pi^2}{3}-2 \mathcal{L}\Big(\frac{1}{(Q+1)^2}\Big)
\ee
Now we are ready to express the energy. From (\ref{sumzeros2}), using also (\ref{sumq})
and (\ref{dilsum}), we have
\bea\label{sumzeros3}
E&=&-\frac{1}{24}+\sum_{q=1,h}^{Q}I_h^{(q)}+\frac{m_1^2}2+\frac12\sum_{q=1}^{Q-1} m_{q+1}
(m_{q+1}-m_q)+U^Q\\
U^Q&=&-\frac{S^Q}{8\pi^2}+T^Q+\frac{1}{24}=
\frac{1}{4\pi^2}\mathcal{L}\Big(\frac{1}{(Q+1)^2}\Big)-
\sum_{k}G_{Q,Q+1\,k}+\sum_{k}G_{Q+1,Q\,k}+\\
&+&\frac{1}{8\pi^2} \Big(K*L_{Q+1}'\cdot L_{Q}-K*L_{Q+1}\cdot L_Q'\Big)
+\sum_{h,k}\B_{Qh,Q+1\,k} \nonumber
\eea
By construction, the energy expression (\ref{sumzeros3}) is actually independent of $Q$
because in (\ref{quanz}) the contributions with $q>1$ actually vanish. 
We will show that in the limit $Q\rightarrow \infty$ the energy (\ref{sumzeros3})
will take a completely explicit form; to argue this, we need to know the behaviour of 
the transfer matrix eigenvalues for very large strip index $q$.

The number of 2-strings $n_1$ grows thermodynamically as $N$ while all the other 
one- and 2-strings numbers remain finite therefore the first line $a=1$ of
(\ref{Tmnsystem}) becomes the trivial identity $\infty=\infty$.
This mechanism applies to all states; suppose a state is described by a lattice path
of lenght $N_0$ and maximal height $L_0=\lfloor N_0/2 \rfloor$+1. As we increase $N$, 
the same path fits in the extended lattice with just the addition of particles of 
type $n_1$ (Dirac sea) on the right-most part of the lattice. All the other 
particle numbers
are unchanged. Explicitly, the following particle contents do not vary with $N$
\newcommand{\mtow}{m_{\infty}}
\bea
m_a &\mbox{ for }& a=1,2,\ldots,L_0-1 \nonumber\\
n_a &\mbox{ for }& a=2,3,\ldots,L_0-1 \label{oldpart}\\
n_0 \nonumber
\eea
while the newly-arrived particles $m_a,n_a$, $a=L_0,L_0+1,\ldots$ are fixed by 
consistency as
\bea\label{newcomers}
m_a=n_0 &\mbox{ for }& a=L_0,\ldots \\
n_a=0 &\mbox{ for }& a=L_0,\ldots \nonumber
\eea
where we do not specify an upper limit because these equations hold true for
all the following strips. It is also useful to define 
\be\label{mntower}
\mtow=\lim_{L\rightarrow \infty} m_L=n_0
\ee
If we suppose that (\ref{Tmnsystem}) is verified for a size $N_0=2(L_0-1)$, 
the highest tadpole equation $a=L_0-1$ becomes (remember that $m_{L_0-1}=n_0=m_{L_0}$)
\be
m_{L_0-1}+n_{L_0-1}=\frac{m_{L_0-2}+m_{L_0-1}}{2}=
\frac{m_{L_0-2}+m_{L_0}}{2}
\ee
therefore the last member takes the form of an A-like adjacency rule. 
With the given definitions (\ref{newcomers}), for all $L\geqslant L_0$
the following A-like equality holds
\be
m_{L}+n_{L}=\frac{m_{L-1}+m_{L+1}}{2}=n_0
\ee
We unify all this in the $(m,n)$-system notation
\bea
\mybox{m_a+n_a=\half\sum_{b=1}^{\infty} A_{a,b} m_b\qquad a=2,3,\ldots}
\label{Tmnsystem_inf}
\eea
where the adjacency matrix is now an infinite $A_L$-like matrix. 
In summary, the identity $m_{L_0-1}=n_0=m_L $ for all $L\geqslant L_0$
embeds the tadpole diagram in an infinite $A_{\infty}$ diagram.
The property (\ref{newcomers}) is extremely important: it states that high-numbered strips  
have the same particle content, from a given $L_0$ on. The rectangle 
$N_0=2 (L_0-1)$ times $L_0$ represent the smallest lattice at which a generic 
state (\ref{oldpart}) exists.

With (\ref{newcomers}) we can safely take the limit $Q\rightarrow \infty$ in the energy
(\ref{sumzeros3}). In that limit, the sum on quantum numbers is finite because 
strips (labelled by $a$) do not contain 2-strings for large $a$; the sum on the $m$-family particles 
is also finite because $m_{q+1}-m_q=0$ for all $q\geqslant L_0-1$. 
The dilogarithm vanishes because its argument vanishes.

We are left with the following sum of terms to be evaluated in the limit 
$Q\rightarrow \infty$
\be\label{sumQQ}
\Big(-\sum_{k}G_{Q,Q+1\,k}+\sum_{k}G_{Q+1,Q\,k}\Big)
+\frac{1}{8\pi^2} \Big(K*L_{Q+1}'\cdot L_{Q}-K*L_{Q+1}\cdot L_Q'\Big)
+\sum_{h,k}\B_{Qh,Q+1\,k} 
\ee
We assume that the full information content of the high strips is the same namely that 
the following limits exist at large $q$
\begin{gather}\label{highstrip}
\lim_{q\rightarrow \infty} y^{(q)}_h = y^{(\infty)}_h \\
\lim_{q\rightarrow \infty} \hd^q(x)=\hd^{\infty}(x) \nonumber
\end{gather}
The two conditions are not fully independent because of the TBA and auxiliary equations.
This limit says that the high strips coincide, they carry the same information (see the 
section on numerical evaluations for a justification). 
With this assumption, we can evaluate (\ref{sumQQ}) by substituting the full
expression (\ref{gint})
\bea
&&\lim_{Q\rightarrow\infty} \Big[-\sum_{k}G_{Q,Q+1\,k}+\sum_{k}G_{Q+1,Q\,k}\Big]=\nonumber \\
&&=\lim_{Q\rightarrow\infty}\sum_{k}\int \frac{dy}{4\pi^2}\Big[
\frac{-L_Q(y)}{\sinh(y^{(Q+1)}_k-y)}+\frac{L_{Q+1}(y)}{\sinh(y^{(Q)}_k-y)}\Big]=0
\eea
The convolution term is evaluated by remembering the commutativity
(\ref{exchange})
\be
\lim_{Q\rightarrow\infty} \frac{1}{8\pi^2} 
\Big(K*L_{Q+1}'\cdot L_{Q}-K*L_{Q+1}\cdot L_Q'\Big)=
\frac{1}{8\pi^2} \Big(K*L_{\infty}'\cdot L_{\infty}-K*L_{\infty}\cdot L_{\infty}'\Big)
=0
\ee
The last term must be written explicitly and, according to (\ref{mntower}) and to
(\ref{newcomers}), we will use $m_Q=\mtow$ because $Q$ is sufficiently large
\bea
\lim_{Q\rightarrow\infty} \sum_{h,k}\B_{Qh,Q+1\,k}&=&\lim_{Q\rightarrow\infty} - \frac{i}{2\pi} 
\sum_{h,k=1}^{\mtow}\log \tanh (\frac{\hy_h^{(Q)}-\hy_k^{(Q+1)}}{2}-i\frac{\pi}{4})\nonumber\\
&=& - \frac{i}{2\pi} \sum_{h,k=1}^{\mtow}
\log \tanh (\frac{\hy_h^{(\infty)}-\hy_k^{(\infty)}}{2}-i\frac{\pi}{4})
=-\frac{i}{2\pi} \sum_{h=1}^{\mtow} \log \tanh (-i\frac{\pi}{4})+ \nonumber\\
&-&\frac{i}{2\pi} \mathop{\sum_{h,k=1}^{\mtow}}_{h<k}\left[
\log \tanh (\frac{\hy_h^{(\infty)}-\hy_k^{(\infty)}}{2}-i\frac{\pi}{4})
+\log \tanh (\frac{\hy_k^{(\infty)}-\hy_h^{(\infty)}}{2}-i\frac{\pi}{4})\right]\nonumber\\
&=&-\frac{\mtow}4 -\frac14 \mtow(\mtow-1)=-\frac14 \mtow^2
\eea
where we have used the identity (\ref{logid}). Notice that we always use 
the logarithms in their fundamental determination.
In (\ref{sumzeros3}) all terms have been evaluated in the limit 
$Q\rightarrow\infty$ so we write the final result
\bea
E&=&-\frac{1}{24}+\sum_{q=1}^{\infty}\sum_{h=1}^{m_q} I_h^{(q)}+
\frac{m_1^2}2+\frac12\sum_{q=1}^{\infty} m_{q+1}(m_{q+1}-m_q)-\frac14 \mtow^2
\nonumber \\
&=&-\frac{1}{24}+\sum_{q=1}^{\infty}\sum_{h=1}^{m_q} I_h^{(q)}+
\frac{1}{4} \m^T\,C\,\m -\frac14 \mtow^2 \label{energy_xxx}
\eea
where $C$ is now the infinite $A_L$-like Cartan matrix, in agreement with 
the $(m,n)$ system (\ref{Tmnsystem_inf}). 
We have to compare the energy expression (\ref{energy_xxx}) with the 
finitized character expression (\ref{xxxvac}). 
As noticed in (\ref{oldpart}), every state is characterized by $L_0$ namely 
the minimal lattice that contains it. Considering a finite lattice and 
matrices and vectors of length $L_0$, it is easy to prove that
\be
\frac{1}{4} \m^T\,C_{\text{tad}}\,\m=\frac{1}{4} \m^T\,C\,\m-\frac14 \mtow^2
\ee
where $C_{\text{tad}}$ is the tadpole $T'_{L_0-1}$ Cartan matrix and $C$
is the $A_{L_0-1}$ Cartan matrix.
Observe that, a priori, on the left hand side the limit $L_0\rightarrow \infty$ is not
defined because we don't know how to treat the entry corresponding to the loop
in the diagram, namely the entry $(L_0,L_0)$ of the Cartan matrix, 
while in the right hand side the limit 
is perfectly defined thanks to the properties (\ref{newcomers}). 
More explicitly, in the right hand side the general term takes the form
\be
\frac{m_q(m_q-m_{q-1})}2 
\ee
and it is trivially zero from $L_0$ on because of (\ref{newcomers}).

\subsection{Tadpole truncation for numerical TBA}
In (\ref{highstrip}) we have formulated an assumption on the behaviour of the transfer matrix 
eigenvalues in strip $q$, with $q$ growing to infinity. 
Here we try to motivate this assumption on the basis of numerical solutions of the infinite TBA 
system of (\ref{sc_tba_xxx}), (\ref{sc_psi_xxx}). 

It is important to remember that a strip
is directly related to the neighboring strips only, with an $A_{\infty}$ adjacency rule,
therefore we face the problem of choosing a suitable truncation. Inspired by 
the argument
given near (\ref{Tmnsystem_inf}) and also by the assumption in (\ref{highstrip}), we use a 
{\em tadpole truncation}, in which the amount of information that should come in the last 
strip $Q$ from the strip $Q+1$ is replaced by the strip content of $Q$ itself 
(this $Q$ is not the same of Section~\ref{exact_xxx}). In other words,
(\ref{sc_tba_xxx}), (\ref{sc_psi_xxx}) are unchanged for all strips $q<Q$ but for the 
last strip we replace them with
\bea
\log \hd^{Q}(x)&=& \log \prod_{j=Q-1}^{Q}\prod_{k=1}^{m_j} \tanh\frac{x-\hy_k^{(j)}}{2} +
\sum_{j=Q-1}^{Q} \big(K*\log(1+\hd^{j})\big)(x) \label{num_tba}\\
\hat{\Psi}^{Q}(x)&=&i\sum_{r=Q-1}^{Q} \sum_{k=1}^{m_r} 
\log\tanh(\frac{x-\hy_k^{(r)}}{2}-i\frac{\pi}{4}) 
-\sum_{r=Q-1}^{Q} \pv dy\, \frac{\log(1+\hd^{r}(y))}{2\pi\sinh(x-y)}\label{num_psi}
\eea
Quantization conditions and (\ref{sc_energ_xxx}) are unchanged.
This truncation can be equivalently formulated by saying that the adjacency matrix of $A_{\infty}$ in
(\ref{sc_tba_xxx}), (\ref{sc_psi_xxx}) is replaced by that of the tadpole $T'_{Q}$ as in  
Section~\ref{sec:quasi} and expecially in (\ref{Tmnsystem}).
Notice that the counting in (\ref{tba_mn}) makes inconsistent a naive truncation in which the last 
strip $Q$ contains the previous one $Q-1$ but not the $Q$ itself: that would make no space to allocate 
the required number of 1-strings.

We have implemented numerical TBA equations for the truncations $T'_2,\,T'_3,\,T'_4$ and 
the actual numerical results fully confirm our assumption. For the second state in 
Figure~\ref{fig:wzw1}, characterized by two zeros in each strip, with quantum numbers 
\be
I_1^{(q)}=I_2^{(q)}=0 \qquad \forall \qquad q\,,
\ee
we obtain the zeros positions in Table~\ref{tab:zeri}. 
\begin{table}[b]\caption{\label{tab:zeri}Zeros and energies from the numerical solution of the TBA 
system at three different levels of the tadpole truncation are given here. }
$$
\begin{array}{c|c|cccccccc}
  & E &y_1^{(1)}&y_2^{(1)} & y_1^{(2)} & y_2^{(2)} & y_1^{(3)} &y_2^{(3)} & y_1^{(4)} & y_2^{(4)}\\\hline
T'_2 & 1.00714 & -0.2783 & 1.6430 & -0.6732 & 2.2468 \\
T'_3 & 1.00397 & -0.2874 & 1.7675 & -0.8257 & 2.6191 & -1.1155 & 3.0052 \\
T'_4 & 1.00253 & -0.2911 & 1.8293 & -0.8989 & 2.8152 & -1.3431 & 3.4124 & -1.5754 & 3.7014 
\end{array}
$$
\end{table}
The distance between corresponding zeros in different strips $|y_k^{(j)}-y_k^{(j+1)}|$ can 
be evaluated from Table~\ref{tab:zeri} and suggests a vanishing behaviour
\be\label{converg}
\begin{array}{c|ccc}
 & j=2 &j=3 & j=4 \\\hline
 |y_2^{(j)}-y_2^{(j+1)}| &  0.6038 & 0.3861 & 0.2890 \\
 |y_1^{(j)}-y_1^{(j+1)}| &  0.3949 & 0.2898 & 0.2323
\end{array}
\ee
consistent with the existence of the limit (\ref{highstrip}). The actual positions grow with 
$j$ but we expect this growth will slow down later on. 
From Table~\ref{tab:zeri} we also observe that the energy value decreases toward 1, as expected for 
this state, see Figure~\ref{fig:wzw1}.

The result in (\ref{converg}) is precisely as expected: different strips tend to resemble each other,
with the same zeros and the same eigenvalues (plots of the transfer matrix eigenvalues are not given here).
We obtained similar results for various other states, for example for the first of Figure~\ref{fig:wzw2}.

%%%%%%%%%%%%%%%%%%%%%%%%%%%%%%%%%%%%%%%%%%%%%%%%%%%%%%%%%%%%%%%%%%%%%%%
\section{Discussion}

In this paper we have introduced a combinatorial formalism, based on paths and quasiparticles (particles and dual-particles), 
to classify the eigenvalues and eigenstates of the transfer matrices of critical two-dimensional Yang-Baxter integrable lattice models. This is achieved by implementing a conjectured energy-preserving bijection between patterns of zeros (in the plane of the complex spectral parameter $u$) of the transfer matrix eigenvalues and lattice paths coinciding with one-dimensional configurational sums. 
The TBA equations are solved for the conformal spectra of energies using the analyticity information encoded in the patterns of zeros, that is, the relative locations of the 1-strings (dual-particles) and 2-strings (particles).
This program has been carried to completion for the ABF RSOS and XXX models in their vacuum sectors. 
However, we expect our methods to generalize to other sectors and to other models including the $D$ and $E$ RSOS models as well as $Z_k$ parafermions.
Finally, our methods should extend to boundary conditions other than fixed boundary conditions such as periodic and toroidal boundaries.

%%%%%%%%%%%%%%%%%%%%%%%%%%%%%%%%%%%%%%%%%%%%%%%%%%%%%%%%%%%%%%%%%%%%%%%%%%%%%%%
\section*{Acknowledgments}
GF and PAP thank the APCTP, Pohang, South Korea, where parts of this work were carried out 
during the 2005 and 2008 programs on 
{\em Finite-size technology in low dimensional quantum systems} and the GGI, Florence, Italy for hospitality during the program on
{\em Low-dimensional quantum field theories and applications}, 2008. 
This work is supported by the Australian
Research Council.
We thank the referee for many helpful comments and additional references.

%%%%%%%%%%%%%%%%%%%%%%%%%%%%%%%%%%%%%%%%%%%%%%%%%%%%%%%%%%%%%%%%%%%%%%%%%%%
\setcounter{equation}{0}
\renewcommand{\theequation}{A.\arabic{equation}}
\section*{A. Dilogarithm Identities}
\addcontentsline{toc}{section}{A. Dilogarithm Identities}

The Rogers dilogarithms are defined as
\be\label{dilog}
\calL(x)=-\frac12 \int_{0}^{x}dy\:
\Big(\frac{\log(1-y)}{y}+\frac{\log\: y}{1-y}\Big), \qquad 
0\leqslant x\leqslant 1.
\ee
Following \cite{KirillovDilog95a, KirillovDilog95b}, we extend the definition to the whole real axis by
\bea
\calL(x)&=&{\pi^2\over 3}-\calL(x^{-1}), \quad {\rm if} \quad x>1,
\label{dilp}\\
\calL(x)&=&\calL\left({1\over 1-x}\right) -{\pi^2\over 6}, \quad {\rm if} 
\quad x<0, \\
\calL(0)&=&0,\quad \calL(1)={\pi^2\over 6}, \quad \calL(+\infty )
={\pi^2\over 3}, \quad \calL(-\infty )= -{\pi^2\over 6}.
\eea
We will use the following identity that remains true for all $x$
\be \label{identity}
\calL(x)+\calL(1-x)=\frac{\pi^2}6
\ee
In \cite{KirillovDilog95a}, an important identity is stated and proven. We write it with 
special reference to our case. 
Following eq. (2.1) of Kirillov, we define
\be\label{skir}
s(j,2,k)=\frac6{\pi^2}\sum_{l=1}^k\calL\Big(\frac{\sin^2\varphi}
{\sin^2(l+1)\varphi}\Big),\qquad \varphi=\frac{(j+1)\pi}{k+2}.
\ee
Corollary 2.5 of Kirillov states that
\be\label{coroll}
s(r-1,2,L-2)+1-s(s-1,2,L-1)=c_L-24 h_{r,s}+6(r-s)(r-s+1)
\ee
where
\be\label{coroll1}
c_L=1-\frac6{L(L+1)},\qquad h_{r,s}=\frac{[(L+1)r-Ls]^2-1}{4L(L+1)}
\ee
and $L,r,s$ are certain positive integers.

%%%%%%%%%%%%%%%%%%%%%%%%%%%%%%%%%%%

%:Bibliography
\goodbreak

\end{document}